\documentclass[aps,prd,twocolumn,superscriptaddress,preprintnumbers,nofootinbib,floatfix,10pt]{revtex4-2}
\usepackage[utf8]{inputenc}
\usepackage{amsmath,amsthm,amssymb,mathtools,physics,cancel,tikz,bm}
\usepackage{graphicx,flexisym,mwe,float}
\usepackage{dcolumn}
\usepackage[hidelinks]{hyperref}
\hypersetup{colorlinks=true,
    linkcolor=blue,
    filecolor=blue,
    urlcolor=blue,
    citecolor=blue,
    }

\usepackage[version=4]{mhchem}
\usepackage{caption,subcaption}
\usepackage{adjustbox}
\usepackage{gensymb}
\usepackage{breqn}
\usepackage{braket}
\usepackage{enumitem}
\usepackage{gensymb}
\usepackage{comment}
\usepackage{xcolor}
\usepackage{soul}
\usepackage{amsfonts,slashed,dsfont,multirow}
\usepackage{psfrag,nicefrac}
\usepackage{bold-extra}

\graphicspath{{./figures/}}

\begin{document}

\author{Raghda Abdel Khaleq}
\email{Raghda.AbdelKhaleq@anu.edu.au }
\affiliation{Department of Fundamental and Theoretical Physics$,$~ \\~Research~ School~ of~ Physics$,$~ Australian~ National~ University$,$~ ACT$,$~ 2601$,$~ Australia}
\affiliation{Department of Nuclear Physics and Accelerator Applications$,$~ Research~ School~ of~ Physics$,$~ Australian~ National~ University$,$~ ACT$,$~ 2601$,$~ Australia}
\affiliation{ARC Centre of Excellence for Dark Matter Particle Physics$,$~ Australia}

\author{Jayden L.~Newstead}
\email{jnewstead@unimelb.edu.au}
\affiliation{ARC Centre of Excellence for Dark Matter Particle Physics$,$ \\~School of Physics$,$~ The~ University~ of~ Melbourne$,$~ Victoria~ 3010$,$~ Australia}

\author{Cedric Simenel}
\email{cedric.simenel@anu.edu.au}
\affiliation{Department of Fundamental and Theoretical Physics$,$~ \\~Research~ School~ of~ Physics$,$~ Australian~ National~ University$,$~ ACT$,$~ 2601$,$~ Australia}
\affiliation{Department of Nuclear Physics and Accelerator Applications$,$~ Research~ School~ of~ Physics$,$~ Australian~ National~ University$,$~ ACT$,$~ 2601$,$~ Australia}
\affiliation{ARC Centre of Excellence for Dark Matter Particle Physics$,$~ Australia}

\author{Andrew E. Stuchbery}
\email{andrew.stuchbery@anu.edu.au}
\affiliation{Department of Nuclear Physics and Accelerator Applications$,$~ Research~ School~ of~ Physics$,$~ Australian~ National~ University$,$~ ACT$,$~ 2601$,$~ Australia}
\affiliation{ARC Centre of Excellence for Dark Matter Particle Physics$,$~ Australia}

\title{Detailed nuclear structure calculations for coherent elastic neutrino-nucleus scattering}

\begin{abstract}
Any discovery of `new physics' in the neutrino sector will first require a precise baseline prediction of the expected Standard Model cross section. While Coherent Elastic neutrino-Nucleus Scattering (CEvNS) experiments are currently statistics limited, upcoming and future experiments at larger scales will be systematics limited. At this point it will be necessary to improve the theoretical predictions to improve the experimental sensitivity. Here we review the calculation of the CEvNS cross section in a consistent theory of hadronic currents and compute the relevant nuclear form factors using the nuclear shell model. The uncertainty on the form factors is explored by repeating the calculation for various shell model interactions and with Skyme-Hartree-Fock evaluations of the Weak-charge radii. We then refine the Standard Model predictions for the recent experimental results of the COHERENT experiment. We find that our cross sections are in good agreement with previous predictions, but with significantly smaller uncertainties. Near-future CEvNS experiments will meaningfully benefit from improved predictions through an increased sensitivity to new-physics signals.
\end{abstract}

\maketitle

\section{Introduction}

The elastic scattering of neutrinos from nuclei via the weak neutral current is a process that can experience a coherent enhancement at sufficiently small momentum transfer. Dubbed CEvNS, the coherent enhancement of this process boosts the cross section by orders of magnitude, making it much larger than other processes involving neutrinos. This enhancement would suggest that the process should be relatively easy to observe, however the observational signature, a low-energy nuclear recoil, is very difficult to detect. Steady advancement in detector technology enabled CEvNS to finally be observed in 2017.

The first detection of neutrinos via CEvNS was performed by the COHERENT experiment using a small (14.6 kg) cesium-iodide detector \cite{COHERENT:2017ipa}. This measurement was quickly followed up by a detection using a liquid argon detector~\cite{COHERENT:2020iec}. The next target in the COHERENT program is an array of germanium detectors (total target mass of 18 kg), which recently made an observation rejecting the no-CEvNS hypothesis at 3.9$\sigma$~\cite{Adamski:2024yqt}. All of these measurements were carried out at the Spallation Neutron Source (SNS) at Oak Ridge National Laboratory. There, neutrinos with energies up to $\sim$50 MeV are created from pions decaying at rest within the mercury spallation target. Together, these measurements have observed hundreds of CEvNS events and are broadly consistent with the Standard Model expectation for the process.

CEvNS has many potential applications, both practical and fundamental. While the practical applications, such as nuclear reactor monitoring~\cite{Bowen:2020unj}, require further detector development before they can be realized, advances in nuclear \cite{Cadeddu:2017etk,Papoulias:2019lfi,Coloma:2020nhf} and fundamental physics are already being realized with current experiments \cite{Bisset:2023oxt,Denton:2018xmq,Coloma:2017ncl,Dent:2017mpr}. Notably, CEvNS experiments are the leading probe of some realizations of sterile neutrinos and non-standard neutrino interactions~\cite{PhysRevD.96.063013}. CEvNS will also be detected in dark matter experiments, where it is an irreducible background to dark matter searches, giving rise to the so-called neutrino floor~\cite{Billard:2013qya} or fog~\cite{OHare:2021utq}. Recently, XENONnT and PandaX reported the first indication of CEvNS due to $^8$B solar neutrinos~\cite{XENON:2024ijk,PandaX:2024muv}. Key to maximizing the effectiveness of these applications is having precise predictions for the scattering rate. That is, if one is interested in detecting deviations from the Standard Model at the sub-percent level, then a commensurate level of precision is required for the theoretical predictions.

The dominant systematic uncertainty in CEvNS rate predictions comes from the neutrino flux, which is presently $\sim10\%$ for the pion-decay-at-rest neutrinos produced at the SNS. However, future dedicated measurements by COHERENT aim to reduce this significantly, to around 2-3\%~\cite{COHERENT:2021xhx}. With this improvement, cross section uncertainties, currently around 2\%~\cite{COHERENT:2020iec}, will provide a significant contribution to the overall systemic uncertainty in the predicted event rates. Relatively little attention has been given to the theoretical uncertainties of the cross section. With CEvNS being a low-energy process, it is expected that predictions that treat the nucleus as an elementary particle are sufficient. The nuclear structure is then only accounted for through a simple charge form factor. Recently, the cross section calculation has been revisited and a modern accounting of nucleon currents applied \cite{PhysRevD.102.074018}; however, this formalism has not yet been adopted for use in predicting experimental observables. That said, cross section calculations using alternative approaches, using coupled cluster theory~\cite{Payne:2019wvy} (for argon only) and Skyrme-Hartree-Fock theory~\cite{VanDessel:2020epd} (for carbon, oxygen, argon, iron and lead) have been performed.

In this work we present theoretical predictions for the CEvNS cross sections, based on detailed nuclear structure calculations, for targets used by the COHERENT experiment (argon, germanium, cesium and iodine). In section \ref{sec:crossSection} we review the standard calculation of the CEvNS cross section and present two methods to derive the nuclear form factors. In section \ref{sec:shellModel} we revise the full CEvNS cross section calculation (following reference~\cite{PhysRevD.102.074018}) and present our computations of the relevant nuclear form factors based on the nuclear shell model. In section~\ref{sec:obs} we use our form factors to improve the theoretical predictions of the CEvNS flux averaged cross sections and event rates for the COHERENT experiment. Finally, in section~\ref{sec:conclusion} we conclude.


\section{CEvNS cross section}
\label{sec:crossSection}
Shortly after the discovery of Weak neutral currents in 1973, it was suggested that this interaction should coherently enhance the scattering of neutrinos from the nucleus~\cite{PhysRevD.9.1389}. At very low momentum transfer, $q$, where $q$ is less than the inverse size of the nucleus, the cross section approximately scales with the number of neutrons in the nucleus. For neutrinos with energies of tens of MeV, the momentum transfer can be large enough to probe the internal structure of the nucleus, resulting in a loss of coherence.

The CEvNS process can be represented by the reaction:
\begin{equation}
    \nu(p) +\mathcal{N}(k) \rightarrow \nu(p') + \mathcal{N}(k')
\end{equation}
where $p,k$ are the incoming momenta of the neutrino and nucleus, respectively, and $p',k'$ are the outgoing counterparts. The momentum transfer is then given by $t=(p-p')^2=(k'-k)^2$. When evaluated in the lab frame (with the nucleus beginning at rest), this gives $t=-q^2=-2m_T E_R$, where $m_T$ and $E_R$ are the mass and kinetic energy of the recoiling nucleus. With the mass of the neutrinos safely neglected, the maximum recoil an incoming neutrino of energy $E_\nu$ can impart is then
\begin{equation}
    E_R^{\rm{max}} = \frac{2E_\nu^2}{m_T+2E_\nu}.
\end{equation}

The cross section is typically calculated by treating the nucleus as composed of point-like nucleons and accounting for their spatial extent within the nucleus via a form factor. This results in a standard expression for the CEvNS cross section
\begin{equation}
\frac{d\sigma}{dE_R} = \frac{G_F^2 m_T}{4\pi}\left(1-\frac{E_R m_T}{2E_\nu^2}\right)Q_{\text{w}}^2F^2(q^2),
\label{eq:crossSection}
\end{equation}
where $Q_{\text{w}} = Q_{\text{w}}^p Z + Q_{\text{w}}^n (A-Z)$ is the weak nuclear charge in terms of the proton and neutron weak charges and $F(q)$ is the nuclear form factor, typically taken from the Helm~\cite{Lewin:1995rx} or Klein-Nystrand~\cite{PhysRevC.60.014903} model.\\

The above is a good approximation to the CEvNS cross section, however, several improvements are variously made in the literature, which can have percent-level corrections. For example, in the above equation, the sub-leading kinematic factor $E_R/E_\nu$, has been dropped. Additionally, one can take into account the different proton and neutron distributions within the nucleus, assigning them separate form factors~\cite{AristizabalSierra:2019zmy}. Furthermore, while radiative corrections are typically included in the nucleon weak charges, there are additional flavor-dependent radiative corrections that can be included~\cite{Tomalak:2020zfh}. Lastly, some effort has been made to include the spin-dependent effects of the neutrinos coupling to odd-$A$ nuclei~\cite{Barranco:2005yy,Dent:2016wcr}.

Putting these together, but neglecting the spin-dependent effects (which are small but will nevertheless be treated with more rigor in the following section) we can more accurately write the cross section as
\begin{eqnarray}
\label{eq:improved}
\frac{d\sigma_{T\nu_i}}{dE_R} = \frac{G_F^2 m_T}{4\pi}\left(1-\frac{E_R m_T}{2E_\nu^2}-\frac{E_R}{E_\nu}\right)\nonumber\\
\times \left(Q_{\text{w}}^{p,\nu_i} F_p(q^2) + Q_{\text{w}}^n F_n(q^2)\right)^2.
\end{eqnarray}
Here we take the flavor dependent charges from \cite{Tomalak:2020zfh}:
\begin{eqnarray}
    Q_{\text{w}}^{n} &=& -1.02352(25)\nonumber\\
    Q_{\text{w}}^{p,\nu_e} &=&  0.0747(34), \,\,\,\,    Q_{\text{w}}^{p,\nu_\mu} = 0.0582(34),\label{eq:charges}\\\nonumber
\end{eqnarray}
and use them in all further calculations. Using Eq.~\eqref{eq:improved} requires knowledge of both the proton and neutron form factors. While the former can be inferred from the electric charge distribution, the latter is not well constrained experimentally. Given that the weak charge of the neutron is much larger than the proton's, we can deduce that the dominant contribution to the uncertainty of the CEvNS cross section comes from the neutron form factor.

\begin{table*}[tbp]
\caption{Experimental charge rms radii $r_{\text{rms}}^p$ values (in fm) for $^{40}$Ar, $^{70,72,73,74,76}$Ge, $^{127}$I and $^{133}$Cs,  from \cite{Angeli:2013epw}. \label{tab:proton rms}}
\begin{tabular}{l|c|cccccc|c|c}
\hline \hline 
& $^{40}$Ar & \multicolumn{5}{c}{Ge} & & $^{127}$I & $^{133}$Cs \\
\cline{2-10}
 & & 70 & 72 & 73 & 74 & 76 & & &  \\
\hline
$r_{\text{rms}}^p$ & 3.4274(26) & 4.0414(12) & 4.0576(13) & 4.0632(14) & 4.0742(12) & 4.0811(12) &  & 4.7500(81) & 4.8041(46) \\
\hline \hline
\end{tabular}
\end{table*}

\subsection{Nuclear form factors}\label{standard FF}

Nuclear form factors are defined as the Fourier transform of density distributions of the nucleus:
\begin{equation}
    F(q^2) = \int e^{i\vec{q}.\vec{r}} \rho(r) \ d^3\vec{r},
\end{equation}
where $\rho(r)$ is the relevant density distribution, which is commonly assumed to be a spherically symmetric function. The root-mean-square (rms) radius for a density distribution is then given by
\begin{equation}
    r^2_{\text{rms}} \equiv \langle r^2 \rangle = \frac{\int \rho(r) \ r^2 \ d^3\vec{r}}{\int \rho(r) \ d^3\vec{r}}.
\end{equation}
The density distribution used, i.e. electric charge-, weak charge- or mass-density, results in different form factors (and rms radii) which are appropriate for use in different scattering processes. When modelling the nucleus as a bound state of point-like nucleons, the neutron and proton point radii, $R_{n,p}$, are obtained from their respective distributions via
\begin{equation}
     R_n^2=\frac{1}{A-Z}\int d^3r \, r^2\rho_n(\mathbf{r}),
\end{equation}
and
\begin{equation}
     R_p^2=\frac{1}{Z}\int d^3r \, r^2\rho_p(\mathbf{r}).
\end{equation}

Neutral-current neutrino scattering proceeds through the coupling of the Z-boson to the weak charge of the nucleus. Therefore CEvNS requires the use of weak-charge form factors which in turn require knowledge of the weak-charge density. It is possible to measure the weak charge distribution directly through parity-violating electron scattering. This has been performed at Jefferson lab by the PREX and CREX experiments, on 
$^{208}$Pb and $^{48}$Ca targets, respectively. However, these experiments have only measured the weak charge at a single momentum transfer: $q = 0.76$ fm$^{-1}$ and 1.28 fm$^{-1}$, respectively~\cite{PREX:2021umo,Kumar:2020ejz}. Additionally, these experiments are difficult and unlikely to be performed on every isotope relevant to CEvNS. Therefore, it is necessary to use indirect knowledge of the weak-charge distribution from e.g. the electric charge distributions, or theoretical modelling.  While electric-charge rms radii are measured to high-precision, they can differ significantly from the weak-charge radii~\cite{Reed:2020fdf} and form factors~\cite{VanDessel:2020epd}. Therefore they cannot be used to produce accurate CEvNS cross sections.

In this work we will compare several methods for modelling the CEvNS form factors and determine their effect on the resulting cross sections. This includes theoretical modelling using both Skyrme-Hartree-Fock (SHF) calculations and the nuclear shell model. The SHF results will allow us to model the nucleon distributions accurately, from which rms radii are computed. The shell model calculations will allow us to include sub-leading nuclear effects through the chiral EFT formalism of~\cite{PhysRevD.102.074018}. For comparison with previous work in the literature, we will also perform a data-driven calculation based on electric-charge radii.


\subsection{Empirically derived form factor}

The most reliable measurements of nuclear size come from probes of the nuclear electric-charge distribution, e.g., from electron scattering data. Protons, being charged, contribute more to the  distribution than the neutrons do. However, the reverse is true for the weak-charge distribution (where the neutron contribution dominates), making the electric-charge distribution an imperfect surrogate for the weak-charge distribution. Therefore, while nuclear charge-radius measurements can help define the proton distribution, additional information (or assumptions) are required to define the neutron distribution.

While it is difficult to probe the neutron distribution directly, it is possible to measure the proton-neutron radius difference, $\Delta r_{np} = r_{rms}^n-r_{rms}^p$, via nuclear spectroscopy of antiprotonic atoms~\cite{PhysRevLett.87.082501}. Presently, these data have large uncertainties, however, an approximately linear relationship between $\Delta r_{np}$ and $I\equiv(N-Z)/A$ is apparent. A global fit to the available data is given in reference \cite{ORRIGO2016414}:
\begin{equation}
    \Delta r_{np} = [(0.82\pm0.54)I-(0.02\pm0.08)]\,\rm{fm}.
    \label{eq:DeltaR}
\end{equation}
Here we disregard the small uncertainty values given by Eq.~\eqref{eq:DeltaR}, as we don't believe the simple linear relationship captures the true scatter that exists in the data. However, we can use this formula to estimate that $\Delta r_{np} = 0.050\,\rm{fm}$ for $^{70}$Ge ($I = 0.09$) and $\Delta r_{np} = 0.15\,\rm{fm}$ for $^{136}$Xe ($I = 0.21$). All other nuclei in our study lie within this range of $I$ values. Therefore, following \cite{AristizabalSierra:2019zmy}, we conservatively assume that the neutron rms radius lies within the range $r^p_{rms}< r^n_{rms} < r^p_{rms}+0.3$ fm.

To define our empirical form factors we take $r^p_{rms}$ directly from electric-charge radii measurements, and use the range of $r^n_{rms}$ to define an uncertainty band. For the form factor we use the Klein-Nystrand (KN)~\cite{PhysRevC.60.014903} model, which has been previously used in CEvNS research (including by the COHERENT collaboration). The KN model is based on a hard sphere distribution with radius $r_A$ and a Yukawa potential of range $a_k$. The form factor is given by:
\begin{equation}
    F_{\text{KN}} (q) = 3 \frac{\sin(qr_A) -qr_A \cos(qr_A)}{(q r_A)^3} \left[1 + (q a_k)^2 \right]^{-1}.
\end{equation}
Here we take $a_k= 0.7$~fm and obtain $r_A$ by equating the rms radius for the distribution with the proton rms charge radius: 
\begin{equation}
   \langle r_p^2 \rangle \equiv \langle r^2 \rangle_{KN} = \frac{3}{5}r_A^2 + 6a_k^2.
\end{equation}
The proton rms charge radii for the target nuclei are shown in table~\ref{tab:proton rms}.  The neutron rms radii are then taken to lie in the range $r_{\text{rms}}^p \leq r_{\text{rms}}^n \leq r_{\text{rms}}^p + 0.3$ fm for Ge, I and Cs, whilst for Ar the upper limit is taken to be $r_{\text{rms}}^p + 0.2$ fm.

\subsection{Skyrme-Hartree-Fock distributions}

To model the weak-charge distribution of the nucleus more directly, we perform Skyrme-Hartree-Fock calculations to find the point-like nucleon distributions within the nucleus. The weak radius of the nucleus can then be calculated from:
\begin{equation}
    R_W^2 = \frac{1}{Q_{\text{w}}}\left[Z Q_{\text{w}}^p (R_p^2+r_{\text{w},p}^2)+(A-Z) Q_{\text{w}}^n (R_n^2+r_{\text{w},n}^2)\right]
    \label{eq:Rweak}
\end{equation}
where $r_{\text{w},p} = 1.545(17)$~fm and $r_{\text{w},n}=0.8434(12)$~fm are the weak radii of the proton and neutron, respectively. In this equation we have dropped the flavor dependence for visual clarity. We take it as understood that all charges and radii are neutrino-flavor dependent and retain that dependence in all calculations.

The nucleon densities were evaluated with the \textsc{skyax} code \cite{reinhard2021} assuming axial symmetry for several parametrizations of the Skyrme functional, namely SkM$^*$, SLy6, SV-min and UNEDF1 as defined in \textsc{skyax}. Pairing correlations are treated at the BCS level with a ``density-dependent delta interaction'' (DDDI) for SV-min and  SkM$^*$, and with a ``volume delta interaction'' (VDI) for SLy6 and UNEDF1. The final theoretical values of $R_{p,n}$  are obtained from the average of the results for each interaction, while the standard deviations are used to evaluate the uncertainties. The full results for each interaction are compiled in the appendix (Table~\ref{tab:SHF}). From this we find the average and standard deviation for each isotope and use it to compute weak radii from Eq.~(\ref{eq:Rweak}). The results are summarised in Table~\ref{tab:SHFsummary}.

For simplicity we then parameterize the distribution via the symmetrized 2-parameter Fermi model~\cite{DWLSprung_1997}, which has been shown to provide a reasonable fit to similar calculations~\cite{Reed:2020fdf}. The Fermi distribution is defined by:
\begin{equation}
    \rho (r)  = \rho_0 \frac{\sinh(c/a)}{\cosh(r/a)+\cosh(c/a)},
\end{equation}
where its two parameters $c$ and $a$ relate to the size and thickness of the skin, respectively. The rms radius of this distribution is:
\begin{equation}
    \langle r^2 \rangle = \frac{3}{5} c^2 + \frac{7}{5}\pi^2a^2,
    \label{eq:rms}
\end{equation}
and the normalization constant is
\begin{equation}
    \rho_0 = \frac{3 Q_{\text{w}}}{4\pi c(c^2+\pi^2a^2)}.
\end{equation}
The (normalized) form factor is then:
\begin{eqnarray}
    F_F(q) &= \frac{3}{q c}\left(\frac{\sin(qc)}{(qc)^2}\frac{\pi q a}{\tanh(\pi q a)}-\frac{\cos(qc)}{qc}\right)\nonumber\\
    &\times\frac{\pi q a}{\sinh(\pi q a)}\frac{1}{1+(\pi a/c)^2}.
\end{eqnarray}
Motivated by the results for the weak skin of calcium and lead in~\cite{Reed:2020fdf}, in this work we set $a = 0.55\pm 0.05$ fm. We then derive the parameter $c$ by equating $R_W = \langle r^2 \rangle$, i.e. Eq.~(\ref{eq:rms}) with Eq.~(\ref{eq:Rweak}).

\begin{table}[htbp]
\caption{Skyrme-Hartree-Fock predictions of rms point-like proton and neutron radii and the resulting predicted rms Weak radius (for $\nu_e$). \label{tab:SHFsummary}}
\begin{tabular}{lccc}
\hline \hline 
Nucleus     &$R_p$ (fm) &$R_n$ (fm) & $R_W$ (fm)\\
\hline
 $^{40}$Ar  &3.326(25)  &3.420(17) &3.514(17) \\
\hline
 $^{70}$Ge  &3.930(21)  &3.986(12) &4.065(12) \\
\hline
 $^{72}$Ge  &3.946(20)  &4.036(12) &4.116(12) \\
\hline
 $^{73}$Ge  &3.954(19)  &4.059(11) &4.140(12) \\
\hline
 $^{74}$Ge  &3.963(18)  &4.084(10) &4.165(11) \\
\hline
 $^{76}$Ge  &3.993(20)  &4.143(14) &4.224(15) \\
\hline
 $^{127}$I  &4.688(14)  &4.824(9)  &4.895(10) \\
\hline
 $^{133}$Cs &4.750(12)  &4.887(8)  &4.957(9) \\
\hline \hline
\end{tabular}
\end{table}


\section{Detailed nuclear treatment}
\label{sec:shellModel}
The formalism discussed thus far has been derived top-down, dealing with the most dominant contributions of neutrino-nucleus couplings. However, this does not systematically account for the nuclear effects stemming from the neutrino coupling to both vector (spin-independent) and axial-vector (spin-dependent) currents within the nucleus. These effects are generally expected to be small and therefore are typically ignored.

The full, bottom-up, calculation starts with the hadronic currents within the nucleus and using effective field theory, derives the Wilson coefficients for the nuclear couplings. This task has been performed for the CEvNS process, in full generality, in \cite{PhysRevD.102.074018}. The resulting cross section is similar to Eq.~(\ref{eq:crossSection}), with an additional term for the axial form factor:

\begin{equation}
\begin{split}
    \frac{d\sigma_A}{dE_R} &= \frac{G_F^2 m_T}{4\pi} \left(1-\frac{m_T E_R}{2 E_\nu^2} -\frac{E_R}{E_\nu} \right) Q_{\text{w}}^2 |F_{\text{w}}(\mathbf{q}^2)|^2\\
    &+ \frac{G_F^2 m_T}{4\pi} \left(1+\frac{m_T E_R}{2 E_\nu^2} -\frac{E_R}{E_\nu} \right) F_A(\mathbf{q}^2),
\end{split}
\label{eq:fullcs}
\end{equation}

However, the full nuclear form factor goes beyond the density distributions discussed previously.  The Weak form factor becomes 

\begin{equation}\label{weak F}
\begin{split}
F_{\text{w}}(\mathbf{q}^2)&=\frac{1}{Q_{\text{w}}}\bigg[\bigg(Q_{\text{w}}^p\Big(1+\frac{\langle r_E^2\rangle^p}{6} t +\frac{1}{8 m_N^2} t \Big)\\
& + Q_{\text{w}}^n \frac{\langle r_E^2\rangle^n+\langle r_{E,s}^2\rangle^N}{6} t\bigg) \mathcal{F}^M_p(\mathbf{q}^2) \\
&+\bigg(Q_{\text{w}}^n\Big(1+\frac{\langle r_E^2\rangle^p+\langle r_{E,s}^2\rangle^N}{6} t +\frac{1}{8 m_N^2} t\Big)\\
& +Q_{\text{w}}^p \frac{\langle r_E^2\rangle^n}{6} t\bigg) \mathcal{F}^M_n(\mathbf{q}^2) \\
&- \frac{Q_{\text{w}}^p(1+2\kappa^p)+2Q_{\text{w}}^n(\kappa^n+\kappa_s^N)}{4 m_N^2} t \mathcal{F}^{\Phi''}_p(\mathbf{q}^2)\\
&- \frac{Q_{\text{w}}^n(1+2\kappa^p+2\kappa_s^N)+2Q_{\text{w}}^p\kappa^n}{4 m_N^2} t \mathcal{F}^{\Phi''}_n(\mathbf{q}^2)\bigg],
\end{split}
\end{equation}

\noindent where $m_N$ is the nucleon mass and the remaining constants are provided in \cite{PhysRevD.102.074018}. The weak form factor hence depends on the coherent and semi-coherent responses $\mathcal{F}^M_N$ and $\mathcal{F}^{\Phi''}_N$, respectively, where $N = \{p, n\}$ denotes protons and neutrons. The axial form factor is given by

\begin{equation}\label{axial-vector F}
\begin{split}
    F_A & (\mathbf{q}^2)  = \frac{8\pi}{2J+1} \\
     \times & \left((g_A^{s,N})^2  S_{00}^{\mathcal{T}}(\mathbf{q}^2) - g_A g_A^{s,N} S_{01}^{\mathcal{T}} (\mathbf{q}^2) + (g_A)^2 S_{11}^{\mathcal{T}}(\mathbf{q}^2) \right)
\end{split}
\end{equation}

\noindent where, in the isospin-basis, we have the spin structure functions in terms of the nuclear responses $\mathcal{F}_{\pm}^{\Sigma'_L} = \mathcal{F}_{p}^{\Sigma'_L} \pm \mathcal{F}_{n}^{\Sigma'_L}$:

\begin{equation}
\begin{split}
    S_{00}^{\mathcal{T}} &= \sum_L \left[\mathcal{F}_+^{\Sigma'_L}(\mathbf{q}^2)  \right]^2,\\
    S_{11}^{\mathcal{T}} &= \sum_L \left[[1+\delta' (\mathbf{q}^2)] \mathcal{F}_-^{\Sigma'_L}(\mathbf{q}^2)  \right]^2,\\
    S_{01}^{\mathcal{T}} &= \sum_L 2 \left[ 1+\delta' (\mathbf{q}^2) \right] \mathcal{F}_+^{\Sigma'_L}(\mathbf{q}^2) \mathcal{F}_-^{\Sigma'_L}(\mathbf{q}^2),
\end{split}
\end{equation}

The correction to the leading spin-dependent (SD) coupling for the transverse SD response $\delta' (\mathbf{q}^2)$ is given in \cite{PhysRevD.102.074018}.  In this more-detailed treatment of the nucleus, the full weak radius receives additional contributions from spin-orbit effects and the strange electric-charge distribution:
\begin{equation}\label{weak radius}
\begin{split}
R_{\text{w}}^2 &=\frac{Z Q_{\text{w}}^p}{Q_{\text{w}}}\bigg(R_p^2+ \langle r_E^2\rangle^p
 + \frac{Q_{\text{w}}^n}{Q_{\text{w}}^p}\left(\langle r_E^2\rangle^n+\langle r_{E,s}^2\rangle^N \right)\bigg)\\
 &+\frac{N Q_{\text{w}}^n}{Q_{\text{w}}}\bigg(R_n^2+ \langle r_E^2\rangle^p+\langle r_{E,s}^2\rangle^N +  \frac{Q_{\text{w}}^p}{Q_{\text{w}}^n}\langle r_E^2\rangle^n\bigg)\\
 & +\frac{3}{4 m_N^2}+\langle \tilde r^2\rangle_{\text{so}},
\end{split}
\end{equation}
where the spin-orbit contribution is given by:
\begin{equation}
 \begin{split}
\langle \tilde r^2\rangle_{\text{so}}&=-\frac{3Q_{\text{w}}^p}{2 m_N^2Q_{\text{w}}}\Big(1+2\kappa^p+2\frac{Q_{\text{w}}^n}{Q_{\text{w}}^p}(\kappa^n+\kappa_s^N)\Big) \mathcal{F}^{\Phi''}_p(0) \\
 &-\frac{3Q_{\text{w}}^n}{2 m_N^2Q_{\text{w}}}\Big(1+2\kappa^p+2\kappa_s^N+2\frac{Q_{\text{w}}^p}{Q_{\text{w}}^n}\kappa^n\Big) \mathcal{F}^{\Phi''}_n(0).
\end{split}
 \end{equation}

To make use of this formalism we need to calculate the nuclear responses $\mathcal{F}_{N}^{M}$, $\mathcal{F}_{N}^{\Phi''}$, and $\mathcal{F}_{N}^{\Sigma'}$ for our chosen targets -- for this we employ the nuclear shell model.


\subsection{Evaluating the nuclear responses}

\begin{figure*}[th]
    \centering
    \includegraphics[height=4.15cm]{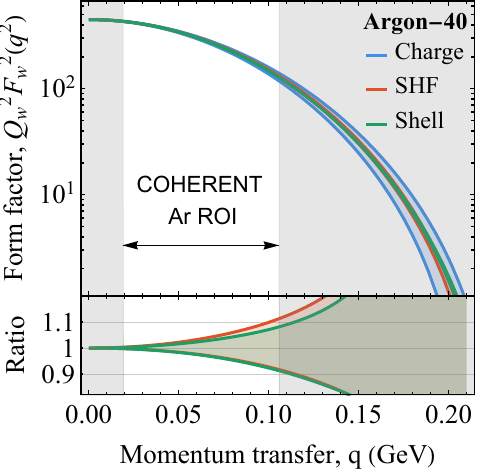}
    \includegraphics[height=4.15cm]{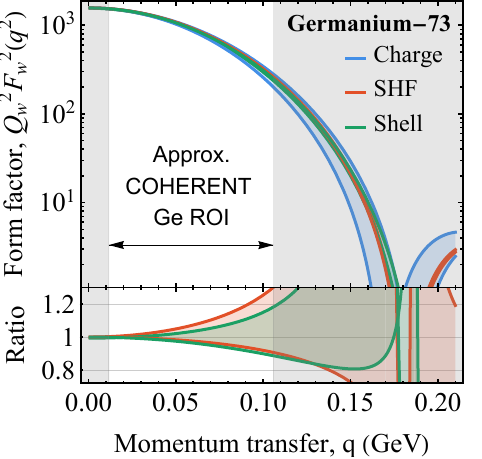}
    \includegraphics[height=4.15cm]{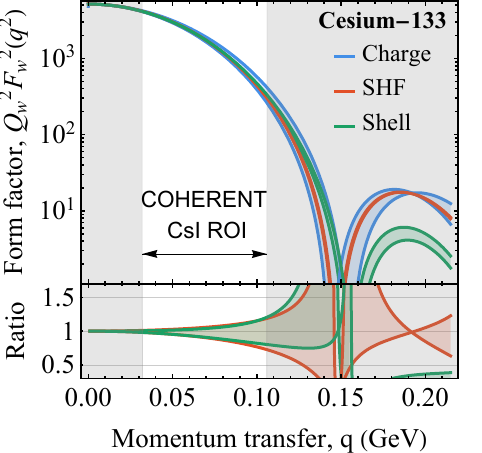}
    \includegraphics[height=4.15cm]{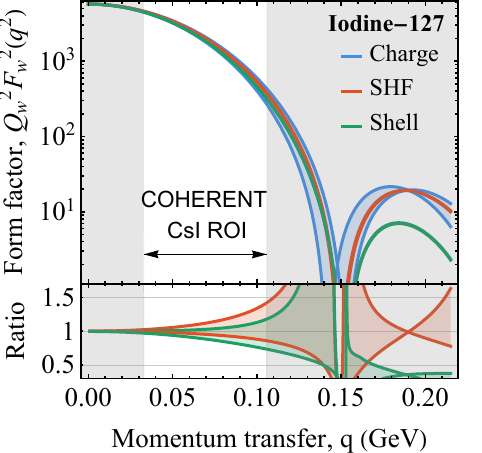}
        \caption{Spin-independent weak form factors for CEvNS in the four targets relevant to current experiments. Bands show the error due to uncertainties in the neutron distribution (Charge) and from using different 
        nuclear structure models
        (SHF and Shell model). The arrows indicate the COHERENT regions of interest (ROI).}
    \label{fig:FFSI}
\end{figure*}

The nuclear responses originate from a multipole decomposition of the lepton-nucleus amplitude, developed in~\cite{PhysRevC.6.719,doi:10.1080/00018736600101254}. Here, they are defined as:

\begin{eqnarray}\label{FF eqn}
    |\mathcal{F}^{X}_{N} (q^2)|^2 \equiv &\frac{4\pi}{2J_i+1} \sum\limits_{J=0} \langle J_i || X_J^{(N)} || J_i \rangle \langle J_i || X^{(N)}_J || J_i \rangle, \nonumber\\
    |\mathcal{F}^{\Sigma'}_{N} (q^2)|^2  \equiv &\sum\limits_{J=1} \langle J_i || \Sigma_J^{\prime(N)} || J_i \rangle \langle J_i || \Sigma^{\prime(N)}_J || J_i \rangle, \nonumber\\
\end{eqnarray}
where $J_i$ is the nuclear ground state angular momentum, and $X =\{ M,\,\,\Phi''\}$ and $\Sigma'$ are nuclear operators whose full form is given in Appendix \ref{operator appendix}.  

The nature of these operators can be understood from their long-wavelength ($q \rightarrow 0$) limit (provided in Table~\ref{long wavelength limit}). The $M$ operator is spin-independent (SI) and adds coherently from all nucleons, giving rise to the standard coherent enhancement of CEvNS. The $\Phi''$ term in the $q \rightarrow 0$ limit is proportional to a spin-orbit operator, and is sensitive to nucleon occupations of the spin-orbit partner levels. This partially coherent operator is highest when one spin-orbit partner is filled and the other is empty. The $\Sigma'$ operator is spin-dependent (SD), picking out the transverse projection of the spin current. All nuclei have non-zero responses to the $M$ and $\Phi''$ operators, while only odd-nuclei with a $J_i\neq0$ ground state have a non-zero $\Sigma'$ response.

{\centering
\begin{table*}[htbp] 
\caption{Leading order terms of the nuclear operators in the long-wavelength limit $q \rightarrow 0$ \cite{Fitzpatrick:2012ix}. 
\label{long wavelength limit}}
\begin{tabular}{lcc}
 \hline \hline
 Response Type & Leading Multipole  & Long-wavelength Limit  \\
 \hline 
    $M_{JM}$ : Charge  &  $M_{00} (q \vec{x}_m)$   &  $\frac{1}{\sqrt{4\pi}} 1(m)$ \\
    $T^{\text{el5}}_{JM}$ : Axial Transverse Electric  &  $\Sigma'_{1M} (q \vec{x}_m)$ & $\frac{1}{\sqrt{6\pi}} \sigma_{1M} (m)$ \\
    $L_{JM}$ : Longitudinal &   $\frac{q}{m_N} \Phi''_{00} (q \vec{x}_m)$  &  $-\frac{q}{3 m_N \sqrt{4\pi}} \vec{\sigma} (m). \vec{l} (m)$ \\
    & $\frac{q}{m_N} \Phi''_{2M} (q \vec{x}_m)$  &  $- \frac{q}{m_N} \frac{1}{\sqrt{30 \pi}} \left[x_m \otimes \left(\vec{\sigma} (m) \times \frac{\vec{\nabla}}{i} \right)_1 \right]_{2M}$  \\ 
 \hline \hline
\end{tabular}
\end{table*}}

Using the Wigner-Eckhart theorem the nuclear matrix elements in Eq.~(\ref{FF eqn}) can be reduced in angular momentum and isospin, giving:

\begin{eqnarray}
\label{reduced nuclear matrix element}
     \langle J_i; T M_T \ ||  \sum_{m=1}^{A} \hat{O}_{J, \tau} (q \vec{x}_m) ||  \ J_i ; T M_T \rangle = (-1)^{T-M_T}  \nonumber\\
     \begin{pmatrix}
         \vspace{2mm} T  &  \tau  &  T \\
         \vspace{2mm} -M_T &  0 &  M_T  \hspace{2mm}
      \end{pmatrix}  
         \langle J_i; T \ \vdots \vdots \sum_{m=1}^{A} \hat{O}_{J, \tau} (q \vec{x}_m) \vdots \vdots \ J_i ; T \rangle. \nonumber\\
\end{eqnarray}
Here, $\hat{O}_{J,\tau}$ signify the nuclear operators, with $ \hat{O}_{J, \tau} =  \hat{O}_{J} \ \tau_3^\tau$ and $\tau_3$ being the nucleon isospin operator ($\tau=\{0, 1\}$). The isospin-independent(dependent) component is given by the $\tau=0$ ($\tau=1$) term. 

This can then be written as a product of single-nucleon matrix elements and one-body density matrix elements (OBDMEs) $\Psi^{J;\tau}_{|\alpha|, |\beta|}$ in the  harmonic oscillator basis~\cite{DONNELLY1979103}:
\begin{eqnarray} \label{reduced nuclear matrix element_obdm}
     \langle J_i; T M_T \ ||  \sum_{m=1}^{A} \hat{O}_{J, \tau} (q \vec{x}_m) ||  \ J_i ; T M_T \rangle  =    (-1)^{T-M_T} \nonumber\\
     \times\begin{pmatrix}
         \vspace{2mm} T  &  \tau  &  T \\
         \vspace{2mm} -M_T &  0 &  M_T  \hspace{2mm}
     \end{pmatrix}
         \sum_{|\alpha|, |\beta|} \Psi^{J;\tau}_{|\alpha|, |\beta|}  \langle |\alpha| \ \vdots \vdots \hat{O}_{J, \tau} (q \vec{x}) \vdots \vdots \ |\beta| \rangle.\nonumber\\
\end{eqnarray}
These OBDMEs hold all the nuclear structure information for any nucleus of interest, and can be conveniently calculated using the nuclear shell model. The harmonic-oscillator states for single nucleons, $\alpha$ and $\beta$, are defined by the quantum numbers $\beta = \{n_\beta, l_\beta, j_\beta, m_{j_\beta}, m_{t_\beta} \}$, with the reduced state notation $|\beta|= \{n_\beta, l_\beta, j_\beta\}$. The nucleon isospin state $t_\beta=t_\alpha=1/2$ is implicit in the notation. The nuclear isospin is denoted by $T$ and its projection $M_T$. The notation $\vdots \vdots$ denotes a matrix element reduced in both angular momentum and isospin. The explicit form of the reduced matrix elements is given in Appendix~\ref{operator appendix}. The OBDMEs are found via:
\begin{equation}
\begin{split}
    \Psi^{J;\tau}_{|\alpha|, |\beta|} & \equiv \frac{\langle J_i; T \ \vdots \vdots \left[a^\dagger_{|\alpha|} \otimes \tilde{a}_{|\beta|} \right]_{J;\tau} \vdots \vdots \ J_i ; T \rangle}{\sqrt{(2J+1)(2\tau+1)}},
\end{split}
\end{equation}
where $\tilde{a}_{|\beta| }=(-1)^{j_\beta -m_{j_\beta}+1/2-m_{t_\beta}} \ a _{|\beta|;-m_{j_\beta}, -m_{t_\beta}}$ and $\otimes$ denotes a tensor product.  This basis gives response functions of the form $e^{-2y}p(y)$, where $p(y)$ is a polynomial of the dimensionless parameter $y=(qb/2)^2$ and $b= {1}/{\sqrt{m_N \omega}} \approx \sqrt{41.467/(45A^{-1/3} - 25A^{-2/3})}$~fm is the harmonic oscillator length parameter with $\omega$ as the oscillator frequency.  

\begin{figure*}[th]
    \centering
    \includegraphics[height=5.cm]{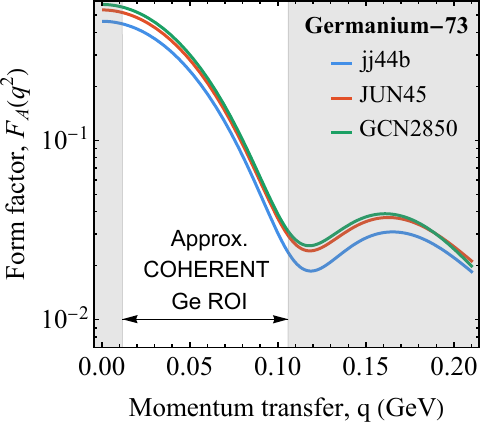}
    \includegraphics[height=5.cm]{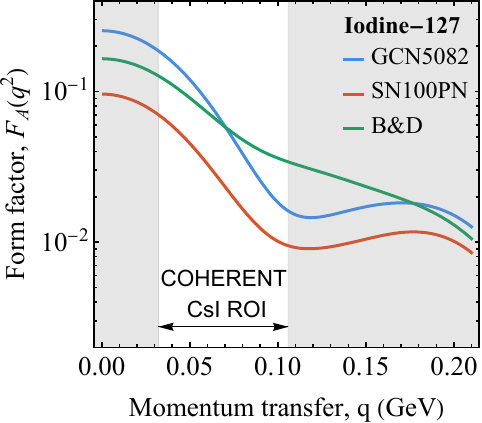}
    \includegraphics[height=5.cm]{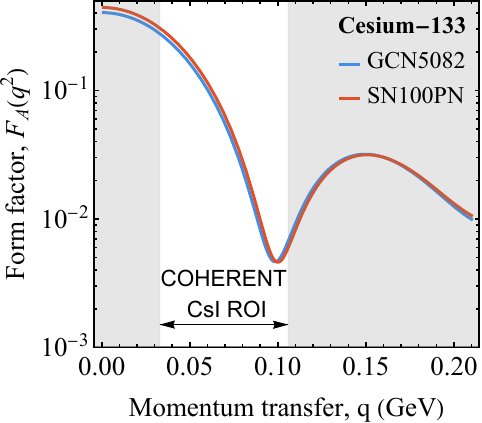}
        \caption{Spin-dependent form factors relevant to CEvNS for the three $J_i\neq0$ targets we consider. The curves denote the different shell model interactions as described in Appendix~\ref{nuclear observable comparison}.}
    \label{fig:FFSD}
\end{figure*}   

\subsection{Form factors from shell model calculations}\label{shell model FF}

Large-scale nuclear shell model calculations were performed using NuShellX \cite{Brown2014TheNuShellXMSU}, which can produce level energies, wave functions and spectroscopic overlaps for a wide range of nuclei. The program's inputs include the relevant valence space for the nucleus of interest -- which constitutes a set of nucleon orbits/levels above a completely filled nuclear core -- as well as its corresponding list of two-body interactions. Each set of two-body interactions, also referred to as the `interaction', is fitted to selected experimental nuclear data. The contemporary approach is to derive the interactions from the nucleon-nucleon interaction using effective field theory. The interactions are then fine tuned by fitting a limited number of parameters to a body of nuclear data -- usually excitation energies in a range of nuclei relevant to the valence space. The aforementioned OBDMEs are extracted for the valence space levels, with the values dependent on the particular shell model interactions used. 



In this work we explore a variety of shell model interactions and truncations for our chosen targets: $^{40}$Ar, $^{70,72,73,74,76}$Ge, $^{127}$I, and $^{133}$Cs. Details of the shell model calculations, as well as comparisons with experimental nuclear structure data can be found in Ref.~\cite{AbdelKhaleq:2023ipt} and in Appendix~\ref{nuclear observable comparison}. The resulting OBDMEs are then used to compute the nuclear responses with the Mathematica package described in \cite{Anand:2013yka}. The form factors for each of the nuclear responses $M$, $\Phi''$ and $\Sigma'$ are plotted as functions of $q$ in Appendix~\ref{form factor appendix}. Analytic expressions of the form factors
for $^{40}$Ar, $^{70,72,73,74,76}$Ge and $^{127}$I are provided in \cite{AbdelKhaleq:2023ipt}, while those for $^{133}$Cs are provided in Appendix~\ref{Cs FF expressions}. For comparison with our SHF calculations, the form factors are used to derive point-proton, point-neutron and weak radii for all isotopes, and are given in Table~\ref{tab:SMradii} of Appendix~\ref{app:point_nucleon_dist}.

The shell-model derived weak form factors are shown in Fig.~\ref{fig:FFSI}, compared to the form factors based on the charge distribution and SHF calculations. The uncertainty band denotes the range of values obtained given the different shell model interactions. The shell model results for each isotope show good agreement with each other (low overall uncertainty) and fall within the large uncertainty band of the form factors derived from the charge distribution. They are also generally in very good agreement with the SHF form factors within the experimentally relevant regions of interest (i.e. the momentum transfer, $q$, which corresponds to observable nuclear recoils in the COHERENT detectors). The axial form factors for each target and interaction are shown in Fig.~\ref{fig:FFSD}. The germanium and cesium axial form factors exhibit good agreement across the different interactions, while iodine has more disparate results. However, inspecting Eq.~\eqref{eq:fullcs} and Figs.~\ref{fig:FFSI} and \ref{fig:FFSD}, we see that the axial term is a factor of (at least) $10^{3}$ smaller than the spin independent term, and is thus a subdominant contribution. This leads us to conclude that the axial form factor uncertainty will be a negligible contribution to the total uncertainty.

\begin{table}[hbt]
    \centering
    \caption{Our CEvNS cross section predictions, compared with COHERENT's predictions and measurements (in units of 10$^{-40}$cm$^2$)}
    \begin{tabular}{cccc|cc}
    \hline
    \hline
         &  &  &                                & \multicolumn{2}{c}{COHERENT} \\
    Target     & Charge     &  SHF   &  Shell Mod.   & Prediction & Exp. \\
    \hline
       Ar      & 18.65(22) & 18.68(6) & 18.64(5) & 18.0(4) & 23(7)\\
       Ge      & 59(1)     & 59.8(2)  & 59.2(2)  & - & - \\
      CsI      & 189(4)    & 189.3(5) & 186.3(5) & 189(6) & 165(30)\\
      \hline
      \hline
    \end{tabular}
    \label{tab:CS}
\end{table}

\section{Experimental observables}
\label{sec:obs}

To date, CEvNS has been observed by the COHERENT collaboration using the stopped pion source of neutrinos from the SNS, with both cesium-iodide~\cite{COHERENT:2021xmm} and argon~\cite{COHERENT:2020iec} detector targets. Presently, COHERENT have deployed a germanium detector which has observed a small excess of events, with statistical significance of 3.9$\sigma$~\cite{Adamski:2024yqt}. There are also experiments aiming to measure CEvNS from reactor neutrinos, including CONUS~\cite{Ackermann:2024kxo}, RICOCHET~\cite{Ricochet:2023yek} and NUCLEUS~\cite{NUCLEUS:2019igx}. The momentum transfer from reactor neutrinos is very small and thus the form factor has little effect on the total cross section. Therefore we choose to focus the present analysis on the cesium-iodide, argon and germanium targets of COHERENT and leave reactor based experiments to a future work.

\begin{figure}[t]
    \centering
    \includegraphics[width=\columnwidth]{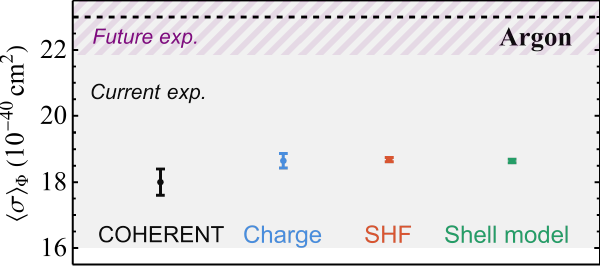}\\
    \includegraphics[width=\columnwidth]{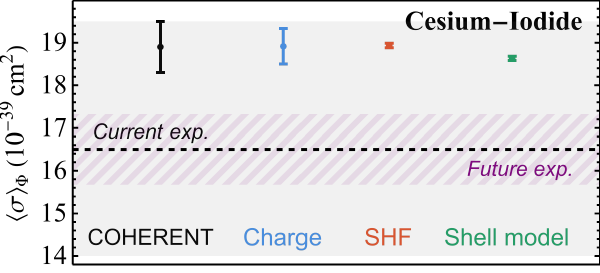}\\
    \includegraphics[width=\columnwidth]{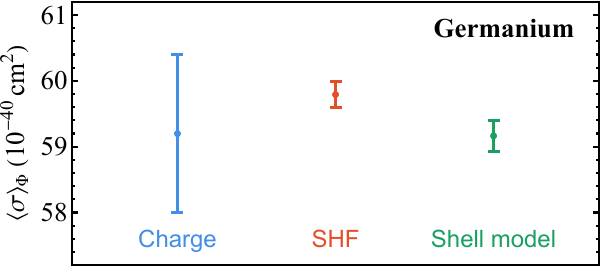}\\
    \caption{Theoretical predictions of SNS-flux averaged cross sections for the three targets considered in this work. Shown are our main results using the shell model and Skyrme-Hartree-Fock (SHF) and comparisons with the Klein-Nystrand form factor (using electric charge radii) and COHERENT's prediction. Also shown is COHERENT's experimental result (black dashed, with gray region showing uncertainty), and a potential future measurement with 5\% error (purple hatched).}
    \label{fig:cs}
\end{figure}

\subsection{Flux-averaged cross sections}
First we calculate the flux-averaged cross sections, by performing a weighted average of the cross sections over the three incoming neutrino flavor spectra and the isotopes in the target:
\begin{equation}
    \langle\sigma\rangle_\Phi = \int dE_\nu  \frac{d\phi}{dE_\nu} \int dE_R \frac{d\sigma}{dE_R}.
\end{equation}
Here the neutrino spectra, $\frac{d\phi}{dE_\nu}$, are normalized to unity. For this calculation we use the neutrino spectra provided in reference \cite{COHERENT:dataCsI:2023aln}, which were derived from detailed GEANT4 simulations of the proton beam colliding in the mercury spallation target. The KN and SHF models are calculated using Eq.~(\ref{eq:improved}), while the shell model result uses the full cross section in Eq.~(\ref{eq:fullcs}).

The uncertainties in our cross section calculations are determined from two sources: the form factors and the nucleon charges. We neglect uncertainties in the modeling of the neutrino spectra, which are unavailable, but note that errors in the spectra should affect all of the compared cross sections approximately the same. The form factor uncertainties are found by computing the flux averaged cross section with our two extremal form factors for each model (i.e. the upper and lower limits of the bands shown in Fig.~\ref{fig:FFSI}). The central value and uncertainty are then taken as ($\sigma_{\rm{max}}+\sigma_{\rm{min}}$)/2 and ($\sigma_{\rm{max}}-\sigma_{\rm{min}}$)/2, respectively. The uncertainty in the nuclear weak charge is propagated from the values for the nucleon charges given in Eq.~\eqref{eq:charges} (taken from \cite{Tomalak:2020zfh}). The total uncertainty is determined by summing the two sources in quadrature.  We find that for our shell model and SHF calculations, the form factor uncertainties are sub-dominant to the nuclear charge uncertainties, with the latter contributing $\sim0.3\%$.

Figure~\ref{fig:cs} shows the main result of this work: the flux-averaged cross sections for argon, germanium, and cesium-iodide, calculated using our three approaches. These results are compared with COHERENT's theoretical prediction (in black) and experimental result (black dashed, with gray area showing the uncertainty band), from \cite{COHERENT:2021xmm,COHERENT:2020iec}. In the case of germanium, our prediction can be used to compare with future COHERENT results, when they become available. The results are also provided numerically in table~\ref{tab:CS}.

The argon cross sections are all in very close agreement with each other, and are slightly larger than the COHERENT prediction, but given the uncertainties, is still in reasonable agreement. Our SHF and shell model results have a much smaller uncertainty band than the COHERENT prediction and their concordance suggests the result is also accurate. The cesium-iodide cross sections are also in good agreement with the COHERENT prediction, falling within the uncertainty band, but again, ours have much smaller uncertainties. 

For germanium and cesium-iodide the shell model cross section is smaller than the SHF cross section, with both lying within the uncertainty band of the charge-derived result. There are several potential sources for this difference. Firstly, cesium-133, iodine-127, and germanium-73 are odd nuclei and thus the cross section receives a contribution from the axial current. However, in the previous section we found that the axial nuclear form factor, $F_A$, is small. Secondly, our shell model result includes the nuclear spin-orbit response, $\mathcal{F}^{\Phi''}$, however this enters the cross section $\propto q^2$, and thus, is also small. 

To investigate the size of the contributions from the additional nuclear responses, we computed the cross sections using only the shell model derived $F_M$ nuclear form factors. For argon, cesium-iodide and germanium, we found that including the axial and spin-orbit terms reduces the flux-averaged cross section by $0.05\%$, $0.05\%$ and $0.08\%$, respectively, well below our margin of error. This shows that the additional nuclear responses have very little effect on the cross section and can thus be neglected at the $<0.1\%$ level of precision. 

We therefore conclude that the difference between the SHF and shell model cross sections is due to the $F_M$ response. This can be seen visually in Fig.~\ref{fig:FFSI}, where the argon form factor is the same, while the germanium and iodine (but not cesium) shell model form factors decrease faster than the SHF ones. Note, while Fig.~\ref{fig:FFSI} only displays the result for germanium-73, the results are very similar for the other isotopes.

\begin{figure}[t]
    \centering
    \includegraphics[width=.95\columnwidth]{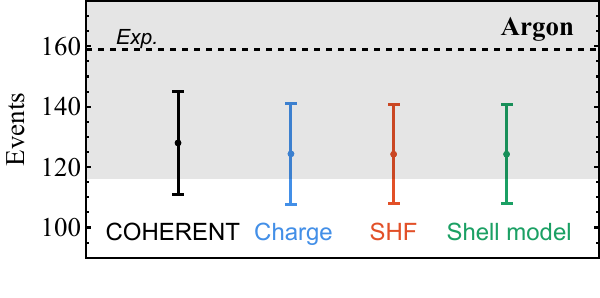}\\
    \vspace{-5mm}
    \includegraphics[width=.95\columnwidth]{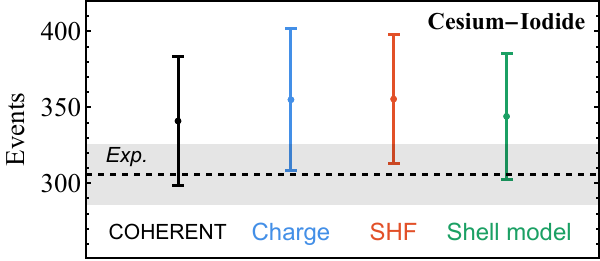}\\
    \caption{Total event count predictions for the COHERENT experiment's argon and cesium-iodide detectors, using our 3 cross section calculations. Also shown is COHERENT's theoretical prediction and experimental result.}
    \label{fig:events}
\end{figure}

Figure~\ref{fig:cs} also shows a potential future experimental measurement of the flux averaged cross section with a total error of 5\% (purple hatched area). We chose this level of error to represent what should be attainable in the coming years with additional statistics and reduced systematic uncertainties. Importantly, measurements of the SNS flux with a D$_2$O detector are expected to reduce flux uncertainties to 2-3\%~\cite{COHERENT:2021xhx}.  For presentation purposes we have assumed the same mean experimental measurement of $\langle\sigma\rangle_\Phi$ and simply reduced the errors. This highlights the importance of our results for increasing the sensitivity of COHERENT to potential new-physics signals.  Taking CsI as an example, if such a precise experimental measurement were made (i.e. $\langle\sigma\rangle_\Phi = (165\pm8)\times10^{-40}$ cm$^2$), the existing predictions would be merely in 2$\sigma$ tension with this result, while our prediction would be in tension at greater than 5$\sigma$.

We note that one must be cautious when interpreting deviations of experimental results from predictions. The experimentally determined flux-averaged cross section is typically calculated assuming a theoretical form of the cross section (e.g. through the calculation of an average efficiency factor). Where the energy dependence of the assumed cross section differs from the one used to produce a prediction, a fair comparison cannot be made. While this effect is likely small, for a direct comparison with the experimental results, we now compute the predicted number of events for the given experimental runs.

\subsection{Experimental event counts}

The differential scattering rate of neutrinos from nuclei, per target and incoming neutrino, is found by integrating over the incoming neutrino spectra and summing them:
\begin{equation}
    \frac{dR}{dE_R} = \sum_i \int_{E_{\nu}^{\mathrm{min}}} dE_\nu \frac{d\phi}{dE_\nu} \frac{d\sigma_{T\nu_i}}{dE_R}.
\end{equation}
Here, the neutrino flux is  integrated  from the minimum neutrino energy that can induce a recoil of energy $E_R$:
\begin{equation}
    E_{\nu}^{\mathrm{min}} = \sqrt{\frac{E_R m_T}{2}}.
\end{equation}

To go from the raw scattering rate to the expected number of counts in an experiment we follow the recipes provided by the COHERENT collaboration in reference \cite{COHERENT:dataCsI:2023aln} and \cite{COHERENT:dataAr:2020ybo}. The total number of detected CEvNS events is found by integrating over the detected energy taking into account the quenching of nuclear recoils, detector resolution effects and the energy dependent detection efficiency. For the cesium-iodide detector this is:
\begin{eqnarray}
    N_{\rm{events}}^{\rm{CsI}} &= N_T\,\Phi \int d{P\!E} \, \epsilon(P\!E)\nonumber\\  
    &\times \int dE_R P(P\!E|E_R)  \frac{dR}{dE_\nu},
\end{eqnarray}
while for the argon detector it is:
\begin{eqnarray}
    N_{\rm{events}}^{\rm{Ar}} &= N_T\,\Phi \int d{E_{ee}} \epsilon(E_{ee}) \nonumber\\
    &\times\int dE_R \, P(E_{ee}|E_R)  \frac{dR}{dE_\nu}.
\end{eqnarray}
Here $P\!E$ is the detected photoelectrons, $N_T$ is the number of targets in the detector and $\Phi$ is total number of neutrinos, per flavor, expected at the detector. The latter is computed as $\Phi = \frac{{\rm{POT}} f_{\pi}}{4\pi d^2}$, where POT is the total number of protons on target, $f_\pi$ is the average number of pions produced per POT, and $d$ is the distance of the detector from the beam-dump. The values for the experimental parameters are summarized in table \ref{tab:experiment}.

\begin{table}[htb]
    \centering
    \caption{Parameters for the COHERENT experimental runs that we considered in this work}
    \begin{tabular}{cccccc}
         \hline
         \hline
         Target       & Mass & Distance & POT & $\langle\pi/\rm{POT}\rangle$\\
        \hline
          Ar       & 24.4 kg & 27.5 m & $1.38\times10^{23}$ & 0.09\\
         CsI & 14.6 kg & 19.3 m & $3.20\times10^{23}$ & 0.0848\\
        \hline\hline
    \end{tabular}
    \label{tab:experiment}
\end{table}

The resulting expected counts for the two detectors' runs are shown in Fig.~\ref{fig:events}. The uncertainty in the number of counts is determined from two sources: the uncertainty in the cross section (calculated as above),  and from the experimental systematic uncertainty (which is dominated by the 10\% uncertainty in the total neutrino flux, but also includes detector effects such as efficiency cuts and quenching factors). We take our systematic uncertainty from COHERENT's analyses (not including their theoretical uncertainties): 13\% for argon and 12\% for cesium iodide~\cite{COHERENT:2021xmm,COHERENT:2020iec}. 

Overall we find good agreement with COHERENT's results, with all predictions well within the errors for both COHERENT's predictions and experimental results. However, without knowledge of their exact theoretical cross section, it is difficult to reproduce their result precisely. In this instance the result obtained using the form factor derived from the electric-charge distribution provides a useful benchmark to compare against. For the cesium-iodide detector our predictions were systematically larger, but shell model results were comparatively lower than from the electric-charge distribution form factors. This is likely due to the slightly larger charge radius predicted from the shell model, which causes coherence to be lost to a greater extent (for a given momentum transfer). For the argon detector our result is systematically lower than COHERENT's prediction, in contrast to our increased cross section prediction. As with the argon cross sections, the different form factors all produce very similar results because argon-40 is an even-even nucleus.

It is worth noting that the reduction in the theoretical uncertainty achieved for our cross section predictions has essentially no effect on the total uncertainty of our event rate predictions, since they are dominated by the systematic uncertainty in the total neutrino flux.

\section{Conclusion}
\label{sec:conclusion}

In this work we have computed precise flux-averaged cross section predictions for the CEvNS experiments being carried out by the COHERENT collaboration using argon, germanium and cesium-iodide targets. These predictions were made using both Skyme-Hartree-Fock calculations of weak radii, and shell model calculations of nuclear structure factors. These were compared with an empirically derived cross section prediction based on charge-radius measurements.  We find that the our calculated weak radii and $F_M$ nuclear responses showed reasonable agreement between the two methods and the charge radii data. While the shell model results included additional nuclear effects such as from spin-orbit coupling and axial currents, they were not found to alter the predictions significantly. We observe a small difference between the the SHF and shell model calculations, where the latter gives a slightly smaller cross section for germanium and iodine targets. Given that spin-orbit and axial effects are negligible, we conclude that this difference is entirely driven by the density distribution.

These results are in broad agreement with prior work, but have substantially smaller theoretical errors, even after accounting for the systematic difference between our SHF and shell model results. In the case of CsI, the difference between our SHF and shell model results is a factor of 2 times smaller than COHERENT's prediction. Further work should be done to understand the origin of this discrepancy and implement more advanced uncertainty estimates. Future advances on the experimental side will both reduce the uncertainty on the neutrino flux and increase the number of detected CEvNS events, reducing statistical errors. Combined with these future advancements, our results will pave the way for CEvNS to serve as a precision test of the Standard Model.

\section*{Acknowledgements}

We are grateful to Daniel Pershey, Jacob Zettlemoyer and Oleksandr Tomalak for useful discussions. This work was supported by the Australian Research Council through Discovery Project DP220101727. The work of JLN is supported by the Australian Research Council through the ARC Centre of Excellence for Dark Matter Particle Physics, CE200100008.


\appendix

\section{Point-like nucleon distributions}
\label{app:point_nucleon_dist}
Full results for the point-like distribution of nucleons in each nuclear target are given in tables~\ref{tab:SHF} and \ref{tab:SMradii}. The SHF results were used to construct Fermi-form factors for the Weak-charge distribution. The shell model results are given for comparison only and are not used for any subsequent calculations. The shell model point-nucleon radii were calculated using
 \begin{equation}\label{point p radius}
     R_p^2=-\frac{6}{Z}\frac{\text{d}F^M_p(\mathbf{q}^2)}{\text{d}\mathbf{q}^2}\bigg|_{\mathbf{q}^2=0},
\end{equation}
with an equivalent form for the neutrons.

\begin{table}[htbp]
\caption{Skyrme-Hartree-Fock predictions of root mean square point like proton, neutron, and weak radii. \label{tab:SHF}}
\begin{tabular}{lcccc}
\hline \hline 
Nucleus    &Interaction    &$R_p$ (fm) &$R_n$ (fm) & $R_w$ (fm)\\
\hline
 $^{40}$Ar  &SkM$^*$        &3.356      &3.445     &3.539 \\
            &SLy6           &3.330      &3.412     &3.507 \\
            &SV-min         &3.321      &3.418     &3.513 \\
            &UNEDF1         &3.296      &3.407     &3.503 \\
\hline
 $^{70}$Ge  &SkM$^*$        &3.949      &4.003     &4.083 \\
            &SLy6           &3.944      &3.988     &4.068 \\
            &SV-min         &3.922      &3.980     &4.061 \\
            &UNEDF1         &3.904      &3.975     &4.057 \\
\hline
 $^{72}$Ge  &SkM$^*$        &3.965      &4.053     &4.134 \\
            &SLy6           &3.959      &4.031     &4.112 \\
            &SV-min         &3.940      &4.031     &4.113 \\
            &UNEDF1         &3.922      &4.028     &4.111 \\
\hline
 $^{73}$Ge  &SkM$^*$        &3.973      &4.077     &4.159 \\
            &SLy6           &3.965      &4.054     &4.135 \\
            &SV-min         &3.947      &4.054     &4.136 \\
            &UNEDF1         &3.931      &4.053     &4.136 \\
\hline
 $^{74}$Ge  &SkM$^*$        &3.980      &4.099     &4.181 \\
            &SLy6           &3.976      &4.081     &4.163 \\
            &SV-min         &3.955      &4.077     &4.160 \\
            &UNEDF1         &3.941      &4.078     &4.161 \\
\hline
 $^{76}$Ge  &SkM$^*$        &4.016      &4.163     &4.245 \\
            &SLy6           &4.004      &4.138     &4.220 \\
            &SV-min         &3.978      &4.129     &4.212 \\
            &UNEDF1         &3.975      &4.140     &4.223 \\
\hline
 $^{127}$I  &SkM$^*$        &4.702      &4.836     &4.907 \\
            &SLy6           &4.698      &4.824     &4.895 \\
            &SV-min         &4.676      &4.814     &4.886 \\
            &UNEDF1         &4.676      &4.822     &4.894 \\
\hline
 $^{133}$Cs &SkM$^*$        &4.763      &4.898     &4.969 \\
            &SLy6           &4.759      &4.885     &4.955 \\
            &SV-min         &4.739      &4.878     &4.949 \\
            &UNEDF1         &4.740      &4.887     &4.958 \\
\hline \hline
\end{tabular}
\end{table}

\begin{table}[htbp]
\caption{Shell model predictions of the point-proton, point-neutron and weak radii, using the various shell model calculations for each of the isotopes considered. GCN5082* and SN100PN* refer to GCN5082 Expanded and SN100PN Expanded, respectively. \label{tab:SMradii}} 
\begin{tabular}{llccc}
\hline \hline 
Nucleus    &Interaction    &$R_p$ (fm) &$R_n$ (fm) & $R_{\text{w}}$ (fm) \\
\hline
 $^{40}$Ar  &SDPF-NR  &3.32852 &3.43528 &3.53293 \\
            &SDPF-U  &3.32852 &3.43528 &3.53282 \\
            &SDPF-MU  &3.32851 &3.43528 &3.53243 \\
            &EPQQM  &3.32851 &3.43528 &3.53348 \\
\hline
 $^{70}$Ge  &GCN2850  &3.95524&4.05401&4.13886\\
            &JUN45  &3.95843&4.05818&4.14331\\
            &jj44b  &3.95963&4.07061&4.15307\\
\hline
 $^{72}$Ge  &GCN2850  &3.97233&4.07882&4.16942\\
            &JUN45  &3.97412&4.11049&4.19586\\
            &jj44b  &3.97457&4.12156&4.20457\\
\hline
 $^{73}$Ge  &GCN2850  &3.97979&4.11144&4.20019\\
            &JUN45  &3.98047&4.13135&4.21644\\
            &jj44b  &3.97981&4.14185&4.22461\\
\hline
 $^{74}$Ge  &GCN2850  &3.98664&4.14437&4.23118\\
            &JUN45  &3.98951&4.16453&4.24903\\
            &jj44b  &3.98955&4.17113&4.25434\\
\hline
 $^{76}$Ge  &GCN2850  &4.00136&4.20041&4.28574\\
            &JUN45  &4.00456&4.21257&4.29678\\
            &jj44b  &4.00422&4.21636&4.29966\\
\hline
 $^{127}$I  &B\&D  &4.6651 &4.89861 &4.97959 \\
            &SN100PN  &4.66554 &4.92354 &4.99969 \\
            &Trun. SN100PN  &4.66559 &4.92575 &5.00184 \\
            &GCN5082  &4.66544 &4.91281 &4.98982 \\
            &GCN5082*  &4.66425 &4.91304 &4.98979 \\
\hline
 $^{133}$Cs &SN100PN  &4.72559 &4.99394 &5.06814 \\
            &SN100PN*  &4.7252 &4.99335 &5.06752 \\
            &GCN5082  &4.72545 &4.99188 &5.06646 \\
            &GCN5082*  &4.72493 &4.99011 &5.06455 \\
\hline \hline
\end{tabular}
\end{table}
\clearpage

\section{Details of the shell model calculations and comparison with experiment}\label{nuclear observable comparison}

Here we provide further details of the shell model calculations, including valence spaces and potential truncations, and nuclear interactions. We also compare experimental nuclear observables to the theoretical shell model values, namely the nuclear energy levels, as well as the magnetic dipole and electric quadrupole moments and transitions. For each of these observables, we provide the values for $^{40}$Ar, $^{70,72,74,76}$Ge and $^{133}$Cs below, whilst the $^{73}$Ge and $^{127}$I comparisons can be found in reference \cite{AbdelKhaleq:2023ipt}.  

Information on the ground state is most important for comparison with theoretical values, as it is used in the scattering calculations here. However, as the ground state of even-even nuclei is $0^+$, the magnetic dipole and electric quadrupole moment values are identically zero. Nevertheless, we employ the next non-zero state ($2_1^+$), which can give an indication of the overall accuracy of the calculation through information on the spectrum beyond the ground state. 

\subsection*{$^{40}$Ar}

$^{40}$Ar shell model calculations were performed in reference \cite{AbdelKhaleq:2023ipt} with the $sdpf$ valence space, which constitutes single particle levels $1d_{5/2}$, $2s_{1/2}$, $1d_{3/2}$, $1f_{7/2}$, $2p_{3/2}$, $1f_{5/2}$ and $2p_{1/2}$, and assuming  a $^{16}$O core. The valence space truncation employed consisted of an unrestricted $sd$ shell and an empty $pf$ shell for the protons, and a full $sd$ shell with an unrestricted $pf$ shell for the neutrons. Four shell model interactions were used: SDPF-NR \cite{Nummela2001SpectroscopyStates}, SDPF-U \cite{Nowacki2009NewSpace}, EPQQM \cite{Kaneko2011Shell-modelNuclei} and SDPF-MU \cite{Utsuno2012ShapeEffect}. 

\begin{figure}[htb]
\centering 
\includegraphics[width=1\linewidth]{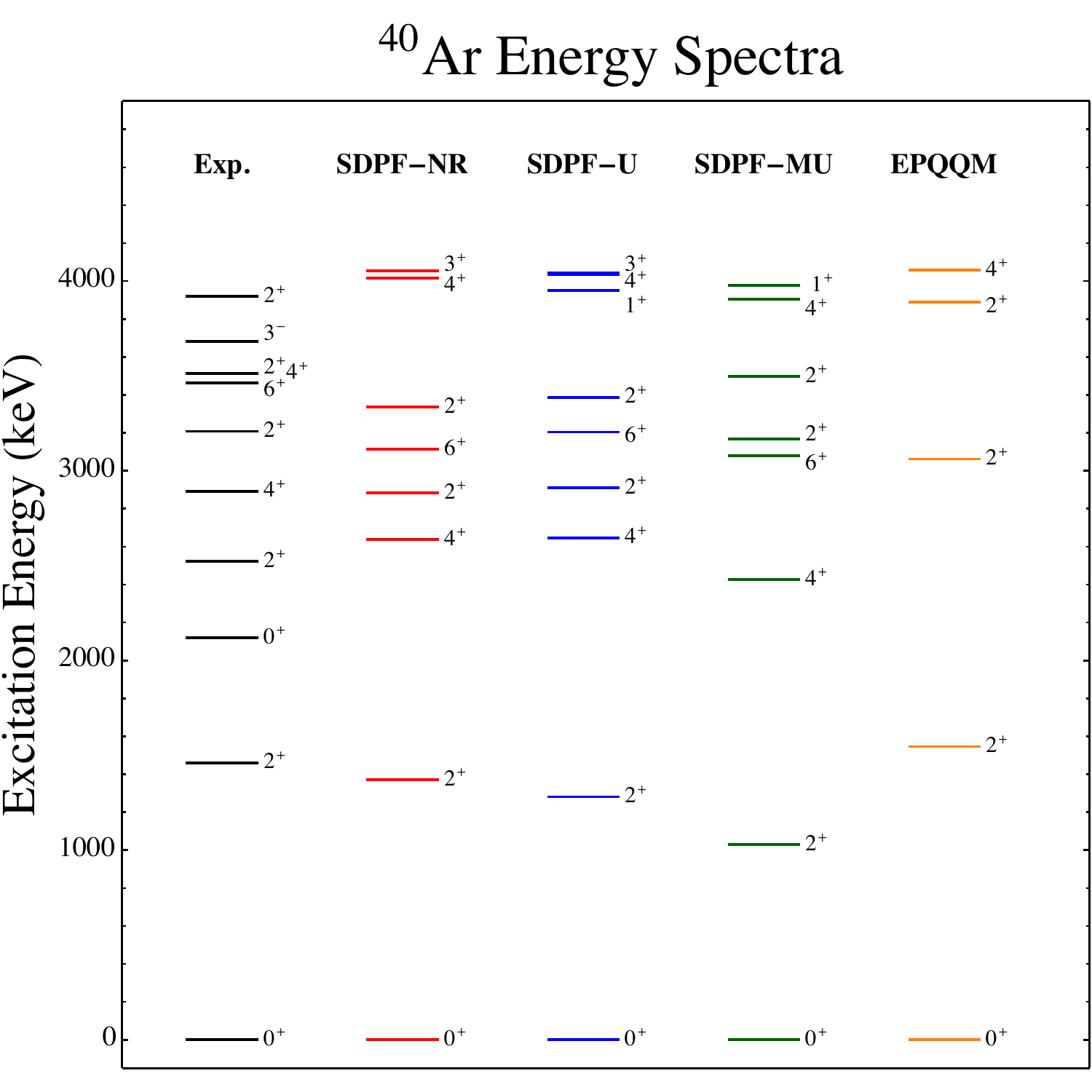}
\caption{Experimental energy levels (keV) for $^{40}$Ar are plotted up to 4000~keV and are compared with levels from the SDPF-NR, SDPF-U, SDPF-MU and EPQQM shell model calculations.\label{fig:40Arlevels}}
\centering
\end{figure}

The $^{40}$Ar theoretical energy levels are plotted against the experimental values in Fig.~\ref{fig:40Arlevels}. The first two theoretical energy levels for all interactions are consistent with experiment within $\sim200$ keV, except for SDPF-MU which is within $\sim400$ keV from experiment for the $2_1^+$ state. Overall, the energy spectrum of the EPQQM interaction is least consistent with the other shell model calculations and with experiment, with the spectrum being much more sparse and spread out in the level spacings. Note that all calculations miss the first-excited 0$^+$ state, which is due to a collective multiparticle-multihole excitation outside the valence space and will be remarked upon further in the following comparison with electromagnetic observables.

The magnetic dipole and electric quadrupole moments and transitions are presented in Table~\ref{tab:transitionsAr}, where the effective $g$ factors and charges employed take on values of $e_p = 1.5$, $e_n = 0.5$, $g_{sp} = 5.586$, $g_{sn} = -3.826$, $g_{lp} = 1$, and $g_{ln} = 0$. It is found that the shell model moments are not consistent with experimental values, and most experimental transitions are poorly reproduced as well.

The reason for the failure of the shell model for the case of $^{40}$Ar is understood: the assumption that $^{40}$Ca is an inert core is not justified. Its ground state is near spherical, as assumed in the shell model, but it has low-excitation deformed states formed by exciting pairs of protons and neutrons across the shell gaps at $N=20$ and $Z=20$. This leads to mixing between the spherical shell model states and the deformed states in nuclei like $^{40}$Ar. The case of $^{42}$Ca, an isotone of $^{40}$Ar, is best studied, as has recently been reviewed by Stuchbery and Wood \cite{physics4030048}. They infer that the first 2$^+$ state of $^{42}$Ca is about 50\% due to the spherical shell model configuration. The ground state is about 65\% of the shell model configuration. Similar mixing is present in $^{40}$Ar. Thus the shell model calculations performed here are limited by the presence of configuration mixing well beyond the usual valence space.


{\tiny 
\begin{table*}[htbp]
\caption{Magnetic dipole and electric quadrupole moments, and electric quadrupole [B(E2)] and magnetic dipole [B(M1)] transitions, between low-lying states of $^{40}$Ar. Experimental transition values are from \cite{Chen:2017ngq}, while moments are taken from \cite{INDC0816,indc.nds.0833}. \label{tab:transitionsAr}}
\begin{tabular}{lccccccccccc}
\hline \hline 
&   \multicolumn{5}{c}{Q [$e$fm$^2$]} & & \multicolumn{5}{c}{$\mu$ [n.m.]} \\
\cline{2-6}
\cline{8-12}
 State & SDPF-NR & SDPF-U & SDPF-MU & EPQQM & Exp. &  & SDPF-NR & SDPF-U & SDPF-MU & EPQQM & Exp. \\
\hline
 $2_1^+$ & 8.30 & 9.01 & 10.59 &  -0.27 & +1(4) &  & -0.401 &  -0.456  & -0.741 & 0.103 & -0.04(6) \\
\hline \hline
&   \multicolumn{5}{c}{B(E2) [$e^2$fm$^4$]} & & \multicolumn{5}{c}{B(M1) [n.m.$^2$]} \\
\hline
 $2_1^+ \rightarrow 0_{\text{gs}}^+$ & 50.31 &  48.86 & 46.33 & 59.58 & 73.13(325) &  & & &  &  &   \\
$2_2^+ \rightarrow 0_{\text{gs}}^+$ & 8.024 & 10.40 & 7.671 & 
7.294 & 9.670(1463) & & &  &  &  &    \\
$0_2^+ \rightarrow 2_1^+$ & 6.980 & 14.50 & 14.68 & 0.03158 & 43.07(650) & & &  &  &  &   \\
$2_2^+ \rightarrow 2_1^+$ & 23.76 & 17.83  & 28.13 & 60.55 & 146.26(4063) &  & 0.2710 & 0.3055 & 0.2952 & 0.1203 & 0.06623(1074)  \\
\hline \hline 
\end{tabular}
\end{table*}
}


\subsection*{$^{70,72,73,74,76}$Ge}

The germanium stable isotopes were studied in the unrestricted $f_5pg_9$ model space ($2p_{3/2}$,  $1f_{5/2}$, $2p_{1/2}$ and $1g_{9/2}$  single-particle levels) with the JUN45 \cite{Honma2009NewNuclei} and jj44b \cite{Mukhopadhyay2017NuclearCalculations} shell model interactions in Ref.~\cite{AbdelKhaleq:2023ipt}. In addition, the GCN2850 interaction \cite{Menendez2009DisassemblingDecay} was used in reference \cite{Fitzpatrick:2012ix,Anand:2013yka} with a valence space truncation that consisted of limiting the occupation number of the $1g_{9/2}$ level to no more than two nucleons above the minimum occupation for all isotopes.  

\begin{figure*}[htb]
\centering 
\includegraphics[width=7cm]{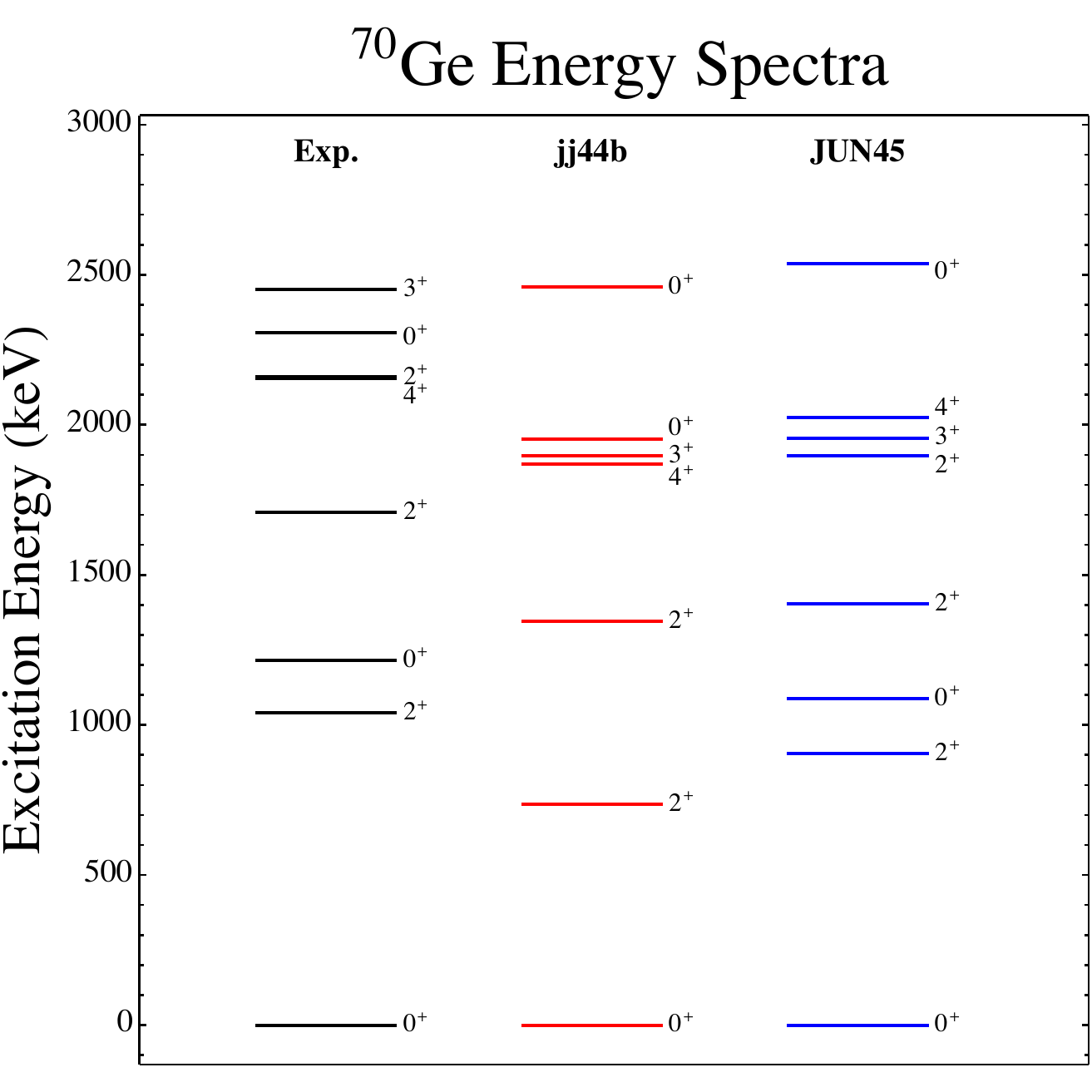}
\includegraphics[width=7cm]{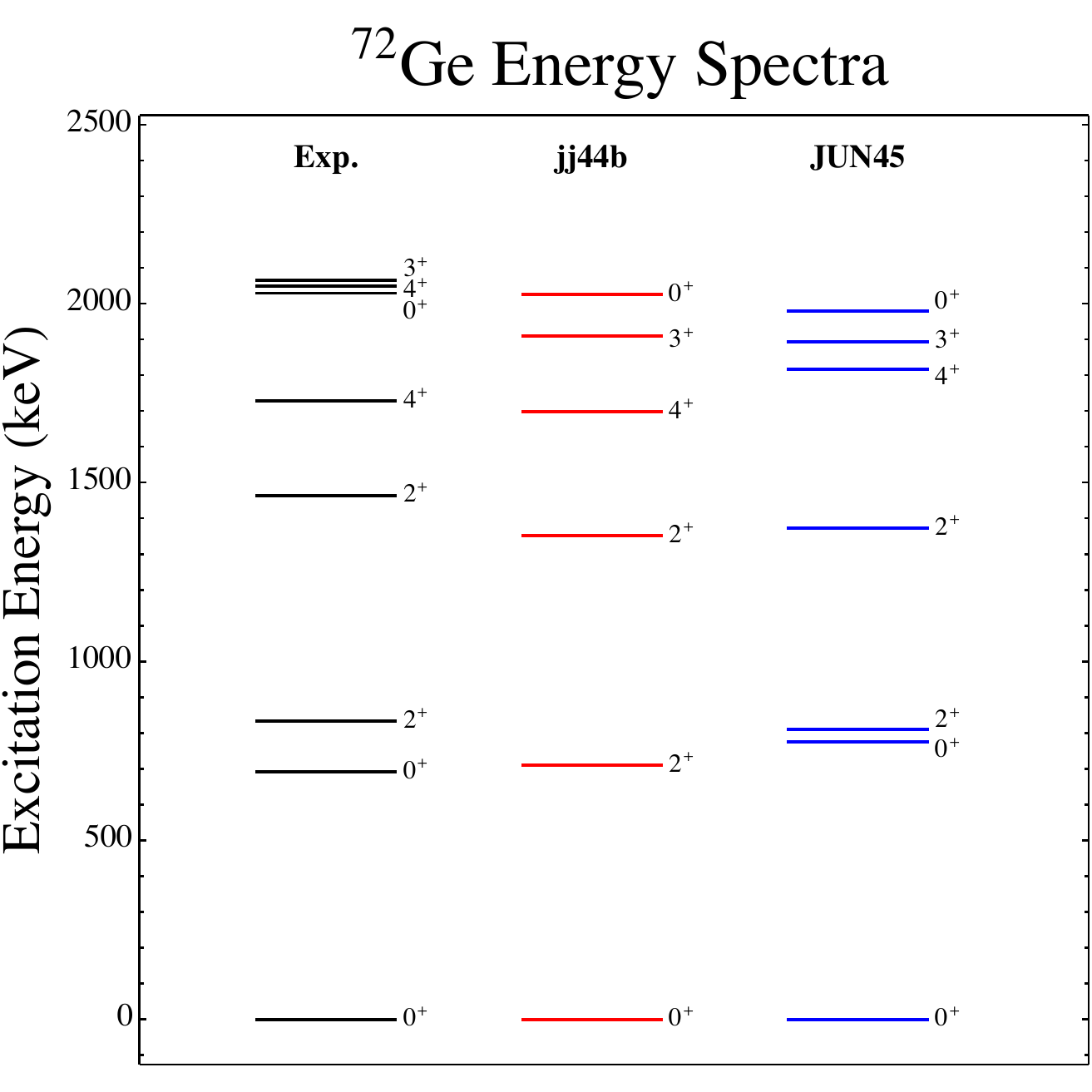}\\
\includegraphics[width=7cm]{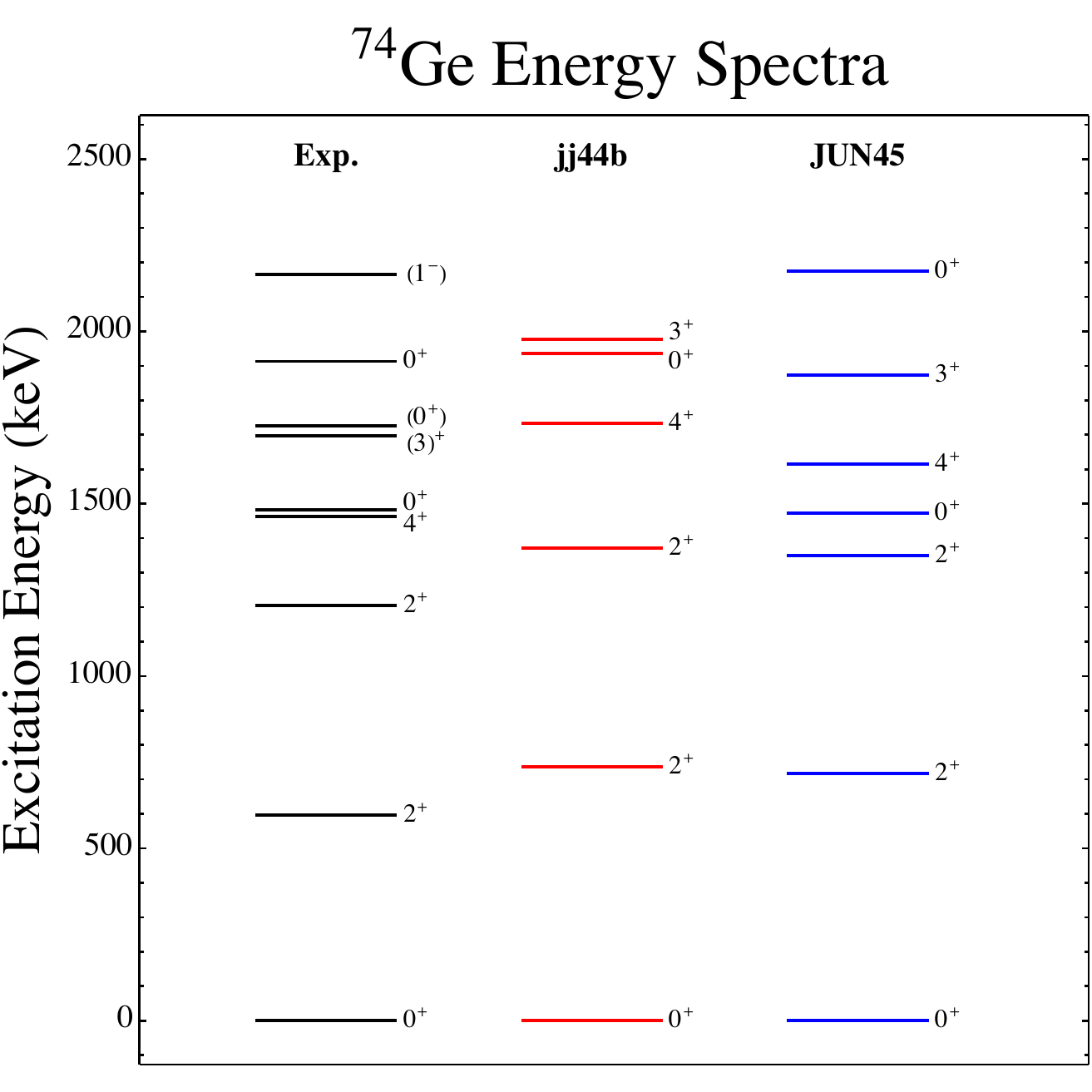}
\includegraphics[width=7cm]{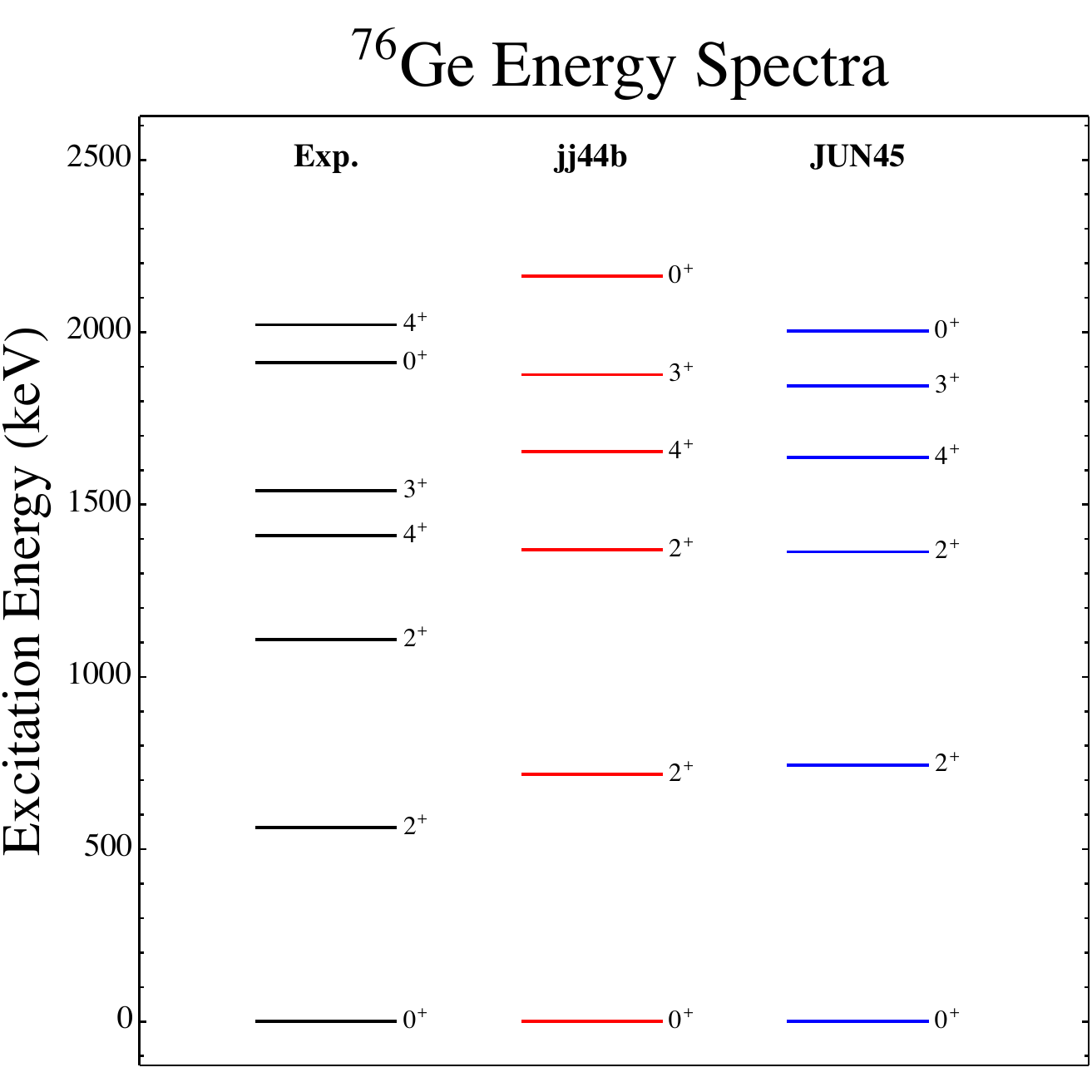}
\caption{Experimental energy levels (keV) for germanium isotopes are plotted up to 2500~keV ($^{70}$Ge), 2100~keV ($^{72}$Ge), 2165~keV ($^{74}$Ge) and 2200~keV ($^{76}$Ge), and compared with levels from shell model calculations using the JUN45 and jj44b interactions. \label{fig:Gelevels}}
\centering
\end{figure*}

The experimental energy spectra are plotted against JUN45 and jj44b results for the even isotopes in Fig.~\ref{fig:Gelevels}.  
In the cases of $^{70,72,74}$Ge, the low-lying experimental spectra are a little better reproduced by the JUN45 interaction, which, unlike jj44b, was fitted to the Ge isotopes in consideration. 
The shell model spectra for $^{76}$Ge are practically identical to one another up to $\sim1.9$ MeV. An agreement between theoretical and experimental levels of $\sim 200$ keV would be considered satisfactory for such shell model calculations using heavier nuclei, however this agreement exists only for some levels of the even isotopes. See also \cite{AbdelKhaleq:2023ipt} for $^{73}$Ge results, where the theoretical spectra were found to be inconsistent with another, and not particularly consistent with experiment.

The experimental and theoretical magnetic dipole and electric quadrupole moments and transitions are provided for the even isotopes in Tables~\ref{tab:momentsGe} and \ref{tab:transitionsGe}. 
For the magnetic dipole moments of the even isotopes, the experimental $2_1^+$ and $4_1^+$ states are well reproduced by both shell model interactions, whereas the jj44b value of the $2_2^+$ state is consistently closer to experiment compared to JUN45. Note that the compiler, Stone~\cite{INDC0816}, has increased the uncetainties on the magnetic moments of the $2_1^+$ states reported by McCormick {\it et al.}~\cite{McCormick2019} based on their statement that the absolute scale of the moments remains uncertain on the level of 20\%.
The $2_1^+$ electric quadrupole moment is quite inconsistent with the experimental values for $^{70,72,74}$Ge, where only the jj44b $^{76}$Ge value lies within the experimental range. 
Such discrepancies are not uncommon in medium-mass and heavy nuclei: the shell model is better equipped to describe single-nucleon features of the wavefunction than collective features like the shape of the charge distribution.



{\tiny 
\begin{table*}[htbp]
\caption{Magnetic dipole and electric quadrupole moments for low-lying states of $^{70,72,74,76}$Ge. Experimental moments are taken from \cite{INDC0816,indc.nds.0833,Gurdal:2013oma,McCormick2019}. \label{tab:momentsGe}}
\begin{tabular}{l|cccccccc|ccccccc}
\hline \hline 
&  \multicolumn{7}{c}{$^{70}$Ge} & &  \multicolumn{7}{c}{$^{72}$Ge} \\
\cline{2-8}
\cline{10-16}
&   \multicolumn{3}{c}{Q [$e$fm$^2$]} & & \multicolumn{3}{c}{$\mu$ [n.m.]}  & & \multicolumn{3}{c}{Q [$e$fm$^2$]} & & \multicolumn{3}{c}{$\mu$ [n.m.]}\\
\cline{2-4}
\cline{6-8}
\cline{10-12}
\cline{14-16}
State & JUN45 & jj44b & Exp. & & JUN45 & jj44b & Exp. & & JUN45  & jj44b & Exp. &  & JUN45 & jj44b & Exp. \\
\hline
$2_1^+$ & +14.32 & +21.35 & +3(6) & & +0.72 & +0.59  &  +0.64(10)  &  & +18.30 & +15.70 & -13(6) & & +0.62 & +0.52 & +0.56(8) \\
$2_2^+$ & -18.63 &  -21.70 &  &  & +1.3 & +1.1 & +1.0(4) &  & -19.09  & -16.14 & & & +1.4 & +1.0  & +0.5(3) \\
$4_1^+$ & +4.03 & +6.48 &  &  & +1.32 & +1.12  & +0.88(124)  &  &  +13.53 & +5.98  & & & +1.12 & +0.67 & +1.0(3)  \\
\hline \hline
&  \multicolumn{7}{c}{$^{74}$Ge} & &  \multicolumn{7}{c}{$^{76}$Ge} \\
\cline{2-8}
\cline{10-16}
\hline
$2_1^+$ & +16.99 & -7.22  & -19(2) &  & +0.59 & +0.61 & +0.56(8)  &  & +2.99 & -18.81 & -19(6) & & +0.737 & +0.596  & +0.53(8) \\
$2_2^+$ & -16.13 & +6.67  & -26(6) & & +1.25 & +0.96 & +0.75(15)  &  & -0.72 & +20.20 & & & +1.16 & +1.07  & +0.64(10) \\
$4_1^+$ & +16.89 &  -10.15 &  &  & +0.81 & +0.91 &  +1.3(4) &  & -0.76 & -18.06 &  &  & +1.20 & +0.87 & +0.8(6) \\
\hline \hline
\end{tabular}
\end{table*}
}

{\tiny 
\begin{table*}[htbp]
\caption{Electric quadrupole [B(E2)] transitions between low-lying states of $^{70,72,74,76}$Ge. Experimental transitions are taken from \cite{Gurdal:2016opi,Abriola:2010ome,Singh:2006spf,Singh:2024adz}. \label{tab:transitionsGe}}
\begin{tabular}{l|cccc|cccc|cccc|ccc}
\hline \hline 
&  \multicolumn{3}{c}{$^{70}$Ge} & &  \multicolumn{3}{c}{$^{72}$Ge} & & \multicolumn{3}{c}{$^{74}$Ge} & & \multicolumn{3}{c}{$^{76}$Ge}\\
\cline{2-16}
&   \multicolumn{3}{c}{B(E2) [$e^2$fm$^4$]} & & \multicolumn{3}{c}{B(E2) [$e^2$fm$^4$]} & & \multicolumn{3}{c}{B(E2) [$e^2$fm$^4$]} & & \multicolumn{3}{c}{B(E2) [$e^2$fm$^4$]}  \\
\cline{2-4}
\cline{6-8}
\cline{10-12}
\cline{14-16}
Transition & JUN45 & jj44b & Exp. & & JUN45  & jj44b & Exp. & & JUN45 & jj44b & Exp. & & JUN45  & jj44b & Exp. \\
\hline
$2_1^+ \rightarrow 0_{\text{gs}}^+$ & 437.3  & 602.4  & 356.4(69)  &  & 462.6 & 632.0  & 418.1(71) & & 542.8  &  648.3  & 609.0(74) & & 544.8  & 608.0 & 550.9(40) \\
$0_2^+ \rightarrow 2_1^+$ & 475.9 & 104.7  & 822.5(1199) &  &  &  & & & & & & & & &   \\
$2_2^+ \rightarrow 0_{\text{gs}}^+$ & 16.80 & 41.54 & 8.57(137) & & 30.22  &  35.65  & 2.31($^{+32}_{-43}$) & & 35.88  &  2.897  & 13.1(20) & &  11.53 & 0.4896 & 14.2($^{+15}_{-13}$)  \\
$2_1^+ \rightarrow 0_2^+$ &  &  & & &  116.0 & 25.3 & 316.7(53) & & & & & & & &  \\
$4_1^+ \rightarrow 2_1^+$ & & & & & & & & & 772.7 &   885.4  & 756.6(554) & & 738.3 & 809.7 & 698.0(153)  \\
\hline \hline
\end{tabular}
\end{table*}
}

\subsection*{$^{127}$I}\label{127I}

Shell model calculations for $^{127}$I were performed in reference \cite{AbdelKhaleq:2023ipt} in the model space which includes all proton and neutron orbits in the major shell between magic numbers 50 and 82, with single particle levels $1g_{7/2}$, $2d_{5/2}$, $2d_{3/2}$, $3s_{1/2}$ and $1h_{11/2}$, and with the SN100PN \cite{Brown2005Magnetic132Sn} and GCN5082 \cite{Menendez2009DisassemblingDecay} interactions.
The model space restrictions are described in reference \cite{AbdelKhaleq:2023ipt}, with the four different calculations given the names ``SN100PN" and ``GCN5082" (which share the same truncation), as well as ``Truncated SN100PN" (with a more truncated scheme), and ``GCN5082 Expanded" (with an alternative truncation to ``GCN5082"). 
The valence space restriction employed in reference \cite{Fitzpatrick:2012ix,Anand:2013yka} with the Baldridge and Dalton (B\&D) interaction \cite{Baldridge1978Shell-modelCases} involves fixing the occupation number of $1h_{11/2}$ to the minimum allowed nucleon number.
See Ref.~\cite{AbdelKhaleq:2023ipt} for a comparison between shell model predictions and experimental nuclear structure data.



\subsection*{$^{133}$Cs}

Here, we  consider the same valence space and shell model interactions (SN100PN and GCN5082) as $^{127}$I. We have performed four calculations for $^{133}$Cs, two of which employ the same truncations as the ``SN100PN" and ``GCN5082" calculations for $^{127}$I (and are given the same names here). In order to investigate the sensitivity of the shell model predictions to the truncation scheme, we consider two additional calculations, one each for the SN100PN and GCN5082 interactions, where the neutron $2d_{5/2}$ level now has a minimum of 4 neutrons (as opposed to previously being completely filled), with the remainder of the proton and neutrons levels retaining the previous truncations. This is to explore the effect of a less truncated valence space on the final results. These two additional calculations are termed ``SN100PN Expanded" and ``GCN5082 Expanded". 

\begin{figure}[htbp]
\centering 
\includegraphics[width=1\linewidth]{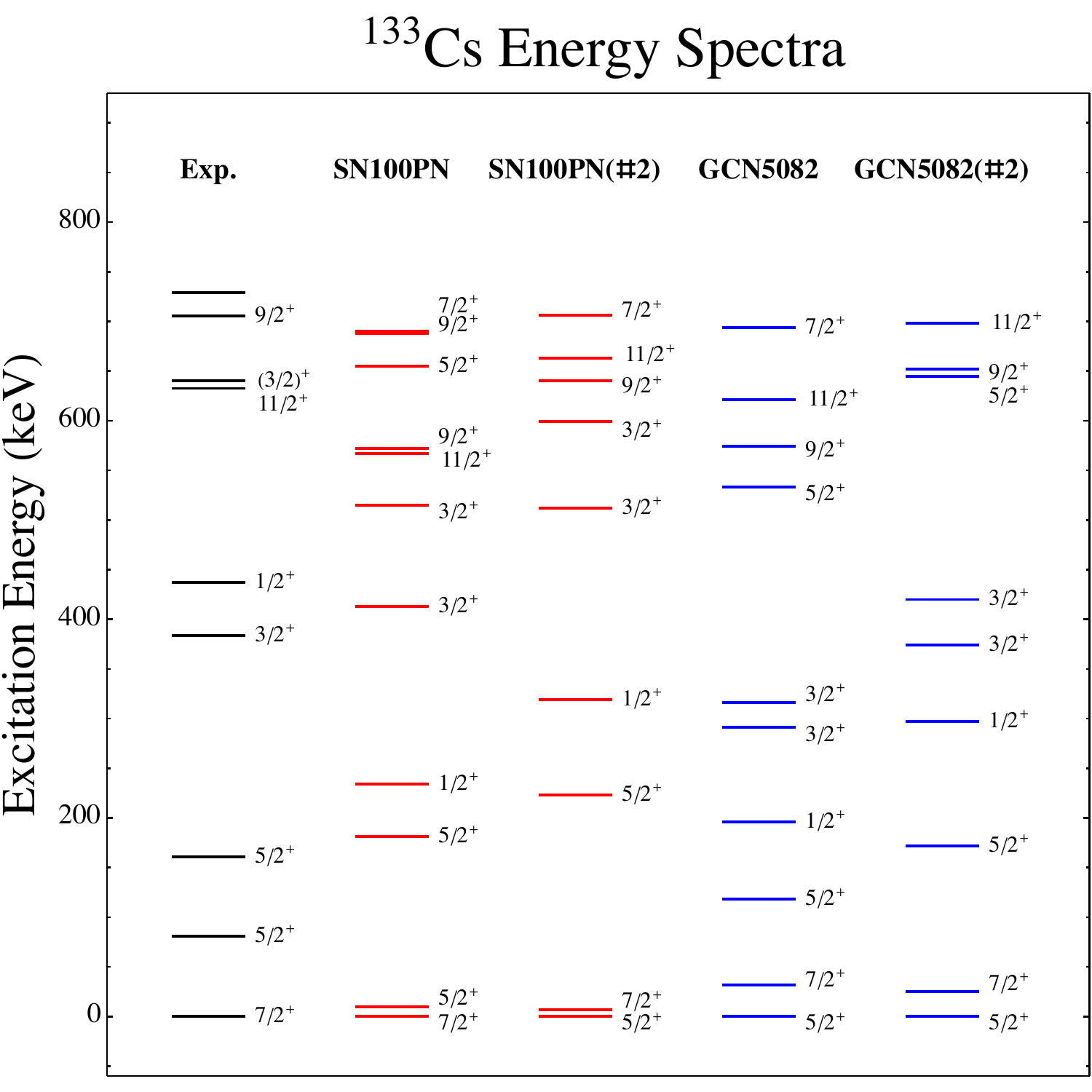}
\caption{Experimental energy levels (keV) are plotted up to 750~keV for $^{133}$Cs, compared with levels from SN100PN and GCN5082 shell model calculations for two different valence space truncations (\#2 indicates ``Expanded'' truncations).\label{fig:133CslevelsAll}}
\centering
\end{figure}

The experimental energy spectra are plotted against the theoretical values for all the above calculations in Fig. \ref{fig:133CslevelsAll}. Only the SN100PN result reproduces the correct ground state spin and parity of $7/2^+$, however all four calculations reproduce the correct spins and parities of the three lowest-lying states within $200$ keV of experiment. For both shell model interactions, using a less truncated valence space causes the spacing of the levels to widen compared to the original calculations, generally pushing the spectrum further up. This is likely due to the opening up of the $2d_{5/2}$ level in these calculations, which increases pairing correlations in the ground state of the even-even configuration space that remains with the odd nucleon removed. 

The experimental and theoretical magnetic dipole and electric quadrupole moments and transitions are provided in Tables \ref{tab:momentsCs} and \ref{tab:transitionsCs}. The proton and neutron effective charges and $g$ factors employed here are the same as those used in reference \cite{AbdelKhaleq:2023ipt} for $^{127}$I. The experimental magnetic dipole moments are much better reproduced by the shell model calculations compared to the electric quadrupole ones. The experimental electric quadrupole transitions are only reproduced within the uncertainty margins by the GCN5082 calculation, specifically for transitions from the $5/2_1^+$ and $3/2_1^+$ states to the ground state. On the other hand, the theoretical magnetic dipole transition values are not consistent with experiment. However, the important test of the shell model wavefunction is the magnetic dipole moment, which is sensitive to the wavefunction of the single state, and to the single nucleon structure of the state, rather than to emerging collective properties, which affect the E2 transition rates and the quadrupole moments. The magnetic dipole transitions are suppressed among the dominant low-excitation configurations, as is evident from their small values, and so become very sensitive to small components in the wavefunction that are not of interest here. Despite the lack of agreement for some observables, the present shell model calculations can be trusted to include the important single-particle configurations in the ground-state wavefunction. 



{\tiny 
\begin{table*}[htbp]
\caption{Magnetic dipole and electric quadrupole moments for low-lying states of $^{133}$Cs. Experimental moments are taken from \cite{NSR2019STZV,INDC0816,indc.nds.0833,Campbell1977}. ``\#2'' refers to the expanded truncation schemes.  \label{tab:momentsCs}}
\centering
\begin{tabular}{lccccccccccc}
\hline \hline 
&   \multicolumn{5}{c}{Q [$e$fm$^2$]} & & \multicolumn{5}{c}{$\mu$ [n.m.]}  \\
\cline{2-6}
\cline{8-12}
 State & SN100PN & \#2 & GCN5082 & \#2 & Exp. & & SN100PN & \#2 & GCN5082 & \#2 & Exp.  \\
\hline
 $7/2_{\text{gs}}^+ $ & -14.11 &  -13.28 & -18.71 & -17.13 & -0.343(10) & & 2.6649  & 2.6880 & 2.6715 & 2.7222 &+2.5778(14)  \\
 $5/2_1^+$ & -66.40 & -67.47 &  -63.36 & -60.36 & $-$22(5)\tablenotemark[1] & & 3.54 & 3.62 & 3.36  & 3.50 & +3.45(2)  \\
 $5/2_2^+$ & & & & & & & 2.22 & 2.19 & 2.38 & 2.33 & +2.25(15)   \\
 \hline \hline
\end{tabular}
\tablenotetext[1]{From \cite{Campbell1977}. A different value, without sign, is given in Ref.~\cite{indc.nds.0833}.}
\end{table*}
}

{\tiny 
\begin{table*}[htbp]
\caption{Electric quadrupole [B(E2)] and magnetic dipole [B(M1)] transitions between low-lying states of $^{133}$Cs. Experimental transition values are taken from \cite{Khazov:2011sug}. ``\#2'' refers to the expanded truncation schemes. \label{tab:transitionsCs}}
\centering
\begin{tabular}{lccccc}
\hline \hline 
&   \multicolumn{5}{c}{B(E2) [$e^2$fm$^4$]}  \\
\cline{2-6}
 Transition & SN100PN & \#2 & GCN5082 & \#2 & Exp. \\
\hline
 $5/2_1^+ \rightarrow 7/2_{\text{gs}}^+ $ & 178.7 & 155.2  & 242.8 & 177.2  & 233.9(161) \\
 $5/2_2^+ \rightarrow 7/2_{\text{gs}}^+ $ & 1772 & 1863  & 1715  & 1752  & 1153.3(726) \\
$3/2_1^+ \rightarrow 7/2_{\text{gs}}^+ $ & 1005 & 733.5  & 525.5  &  217.6 &  483.9(1210) \\
\hline \hline
 &   \multicolumn{5}{c}{B(M1) [n.m.$^2$]}  \\
\hline
 $5/2_1^+ \rightarrow 7/2_{\text{gs}}^+ $ &  0.000006111  & 0.00008033  & 0.0003863 & 0.0001393 & 0.00426(4) \\
 $5/2_2^+ \rightarrow 7/2_{\text{gs}}^+ $ & 0.01032 & 0.004174  & 0.01096  & 0.004590  & 0.00226(14) \\
\hline \hline 
\end{tabular}
\end{table*}
}

\section{EFT/Nuclear Operators}\label{operator appendix}

The nuclear operators employed in Eq.~(\ref{FF eqn}) have the form

\begin{widetext}
\begin{equation}
\begin{split}
    M_{JM} (q\vec{x}) & \equiv j_J(qx) Y_{JM} (\Omega_x), \\
    \Sigma'_{JM} (q \vec{x}) & \equiv -i \left[ \frac{\vec{\nabla}}{q} \times \vec{M}^M_{J J} (q \vec{x}) \right] \cdot \vec{\sigma}_N = [J]^{-1} \left[-\sqrt{J} \ \vec{M}^M_{J J+1} (q \vec{x}) +\sqrt{J+1} \ \vec{M}^M_{J J-1} (q \vec{x}) \right] \cdot \vec{\sigma}_N ,\\
    \Phi''_{JM} (q \vec{x}) & \equiv i \left(\frac{\vec{\nabla}}{q} M_{JM} (q\vec{x}) \right)\cdot \left( \vec{\sigma}_N \times \frac{1}{q} \vec{\nabla} \right), 
\end{split}
\end{equation} 
\end{widetext}
where $\hspace{1mm}$ $\vec{M}_{JL}^{M} \equiv j_L(qx) \vec{Y}_{JLM}$, $[J]=\sqrt{2J+1}$, and $\vec{\sigma}_N$ is the nucleon spin operator. Here, $j_J(qx)$ is the spherical Bessel function, $Y_{JM}$ is the spherical harmonic, and $ \vec{Y}_{JLM}$ is the vector spherical harmonic.\\

The Wigner-Eckart theorem has been used to express matrix elements in reduced form, using the convention 

\begin{equation}
\begin{split}
    & \langle j' m' | T_{JM}  | j m \rangle =  (-1)^{j'-m'} \begin{pmatrix}
         \vspace{2mm} j'  &  J  &  j \\
         \vspace{2mm} -m' &  M &  m  \hspace{2mm}
         \end{pmatrix} \langle j' || T_J || j \rangle.
\end{split}
\end{equation}

The one-nucleon matrix elements in Eq.~(\ref{reduced nuclear matrix element}) have analytic expressions of the form \cite{PhysRevD.102.074018}:

\begin{widetext}
\begin{align}
\Big\langle n'l'\frac{1}{2} j'\Big|\Big|M_{J}\Big|\Big|nl \frac{1}{2}j\Big\rangle
&=\langle n'l' | j_{J}(qr_i)| nl \rangle (-1)^{j+1/2+J}
\sqrt{\frac{1}{4\pi}} \, \bigl[(2j'+1)(2j+1)\bigr]^{\frac{1}{2}}
\bigl[(2J+1)(2l+1)(2l'+1)\bigr]^{\frac{1}{2}} \nonumber \\
&\quad\times \left(
\begin{array}{ccc}
l' & J & l \\
0 & 0 & 0
\end{array}\right)			
\left\lbrace
\begin{matrix}
l' & j' & \nicefrac{1}{2} \\
j & l & J
\end{matrix}
\right\rbrace,
\end{align}

\begin{align}
\Big\langle n'l' \frac{1}{2} j' \Big|\Big|j_{J'}(pr_i) \bigl[Y_{J'}(\hat{\mathbf{r}}_i) \bm{\sigma}_i\bigr]^J\Big|\Big|nl\frac{1}{2}j\Big\rangle
&=\langle n'l' | j_{J'}(qr_i) | nl \rangle (-1)^{l'}
\sqrt{\frac{6}{4\pi}} \, \bigl[(2l'+1)(2l+1)(2j'+1)(2j+1)\bigr]^{\frac{1}{2}} \nonumber \\
& \times \bigl[(2J'+1)(2J+1)\bigr]^{\frac{1}{2}} 
\left( \begin{matrix}
l' & J' & l \\
0 & 0 & 0
\end{matrix} \right)			
\left\lbrace
\begin{matrix}
l' & l & J' \\
\nicefrac{1}{2} & \nicefrac{1}{2} & 1 \\
j' & j & J
\end{matrix}\right\rbrace,
\end{align}

\begin{align}
\langle n' l' j'||\Phi''_J||n l j\rangle&=
(-1)^{l'}\frac{6}{\sqrt{4\pi}}\sqrt{(2j'+1)(2j+1)(2l'+1)}  \\
&\times\left\lbrace
\sqrt{(J+1)(2J+3)}
\sum_{L=J}^{J+1}(-1)^{J+L}(2L+1)
\Bigg\lbrace
\begin{matrix}                                                
J+1 & 1 & L \\
1 & J & 1
\end{matrix}
\right\rbrace
\left\lbrace
\begin{matrix}                                                
l' & l & L \\
\nicefrac{1}{2} & \nicefrac{1}{2} & 1 \\
j'  & j   & J
\end{matrix}
\right\rbrace \nonumber \\
&\quad\times\bigg[
\sqrt{(l+1)(2l+3)} 
\left\lbrace
\begin{matrix}                                                
J+1 & 1 & L \\
l & l' & l+1
\end{matrix}
\right\rbrace
\left(
\begin{matrix}                                                
l' & J+ 1 & l+1 \\
0 & 0 & 0
\end{matrix}
\right)
\Big\langle n' l'\Big|j_{J+1}(qr_i)
\left(\frac{\partial}{\partial(qr_i)}-\frac{l}{qr_i}\right) \Big|n l\Big\rangle 
\nonumber \\
&\quad\quad-
\sqrt{l(2l-1)} 
\left\lbrace
\begin{matrix}                                                
J+1 & 1 & L \\
l & l' & l-1
\end{matrix}
\right\rbrace
\left(
\begin{matrix}                                                
l' & J+ 1 & l-1 \\
0 & 0 & 0
\end{matrix}
\right)
\Big\langle n' l'\Big|j_{J+1}(qr_i)
\left(\frac{\partial}{\partial(qr_i)}+\frac{l+1}{qr_i}\right) \Big|n l\Big\rangle 
\bigg]  \nonumber \\
&+
\sqrt{J(2J-1)}
\sum_{L=J-1}^{J}(-1)^{J+L}(2L+1)
\left\lbrace
\begin{matrix}                                                
J-1 & 1 & L \\
1 & J & 1
\end{matrix}
\right\rbrace
\left\lbrace
\begin{matrix}                                                
l' & l & L \\
\nicefrac{1}{2} & \nicefrac{1}{2} & 1 \\
j'  & j   & J
\end{matrix}
\right\rbrace \nonumber \\
&\quad\times\bigg[
\sqrt{(l+1)(2l+3)} 
\left\lbrace
\begin{matrix}                                                
J-1 & 1 & L \\
l & l' & l+1
\end{matrix}
\right\rbrace
\left(
\begin{matrix}                                                
l' & J-1 & l+1 \\
0 & 0 & 0
\end{matrix}
\right)
\Big\langle n' l'\Big|j_{J-1}(qr_i)
\left(\frac{\partial}{\partial(qr_i)}-\frac{l}{qr_i}\right) \Big|n l\Big\rangle 
\nonumber \\
&\qquad\quad- 
\sqrt{l(2l-1)} 
\left\lbrace
\begin{matrix}                                                
J-1 & 1 & L \\
l & l' & l-1
\end{matrix}
\right\rbrace
\left(
\begin{matrix}                                                
l' & J-1 & l-1 \\
0 & 0 & 0
\end{matrix}
\right)
\Big\langle n' l'\Big|j_{J-1}(qr_i)
\left(\frac{\partial}{\partial(qr_i)}+\frac{l+1}{qr_i}\right) \Big|n l\Big\rangle 
\bigg]
\Bigg\rbrace.\notag
\end{align}
\end{widetext}

\section{Form factors}\label{form factor appendix}

In this section we plot the form factors for the individual responses $M$, $\Phi''$ and $\Sigma'$ using the shell model results, where the $\Sigma'$ response is only plotted for isotopes with non-zero nuclear ground state spin. This is presented for $^{40}$Ar, $^{70,72,73,74,76}$Ge, $^{127}$I, and $^{133}$Cs, highlighting the corresponding COHERENT range of interest for each of these nuclei. The form factor expressions employed to plot these are presented in reference \cite{AbdelKhaleq:2023ipt}, except for $^{133}$Cs which is provided in Appendix~\ref{Cs FF expressions}. The KN and SHF form factors were plotted using $(Z F \left(q,r_{\text{rms}}^p) + N F(q,r_{\text{rms}}^n)\right)^2$, while the shell model counterparts were plotted using $\sum_{N,N'=p,n} F^{(N,N')}_{X,X} (q^2)$, where $F^{(N,N')}_{X,X} (q^2)$ is defined in Eq.~(\ref{Fitzpatrick FF}).

\subsection{Argon-40}





\begin{figure}[H]
\centering 
\includegraphics[width=1\linewidth]{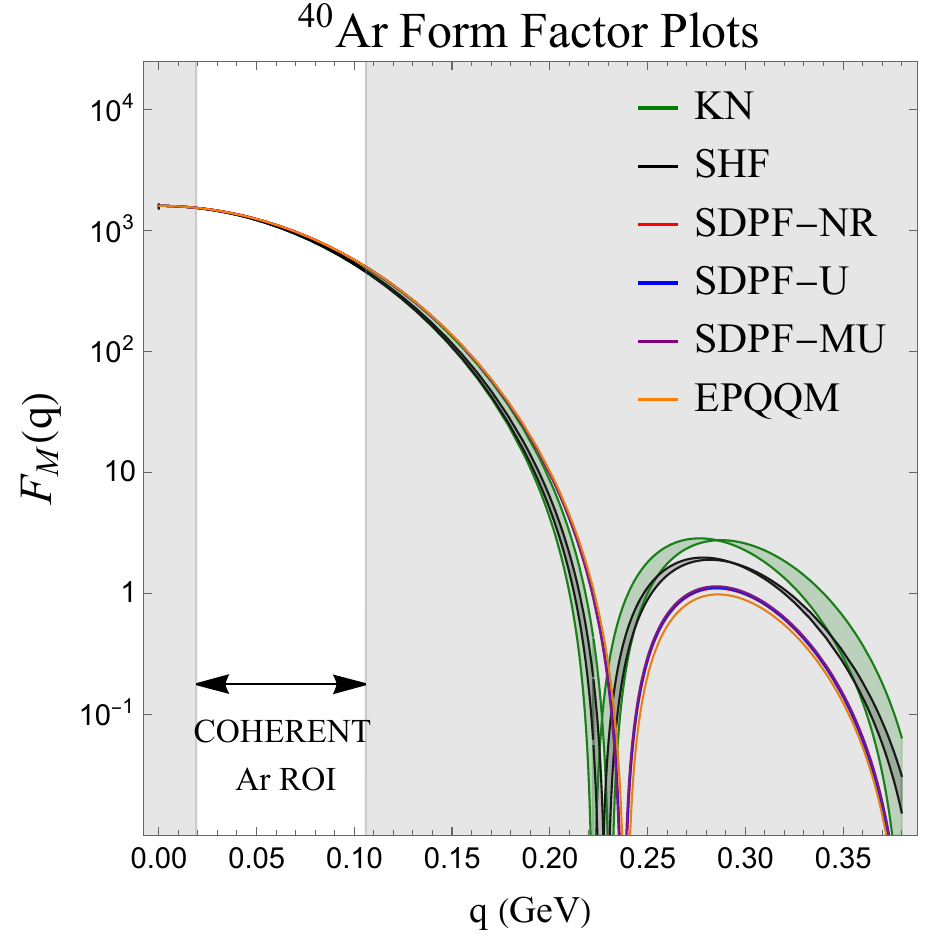}
\caption{SI ($M$) form factors employing the different shell model interactions used for this isotope. This response is compared against the KN and SHF distributions, where the uncertainty bands stem from the $r^n_{\text{rms}}={r^p_{\text{rms}}} ^{+0.2}_{-0}$. The COHERENT region-of-interest is also displayed. \label{fig:40ArFFM}}
\centering
\end{figure}



\begin{figure}[H]
\centering 
\includegraphics[width=1\linewidth]{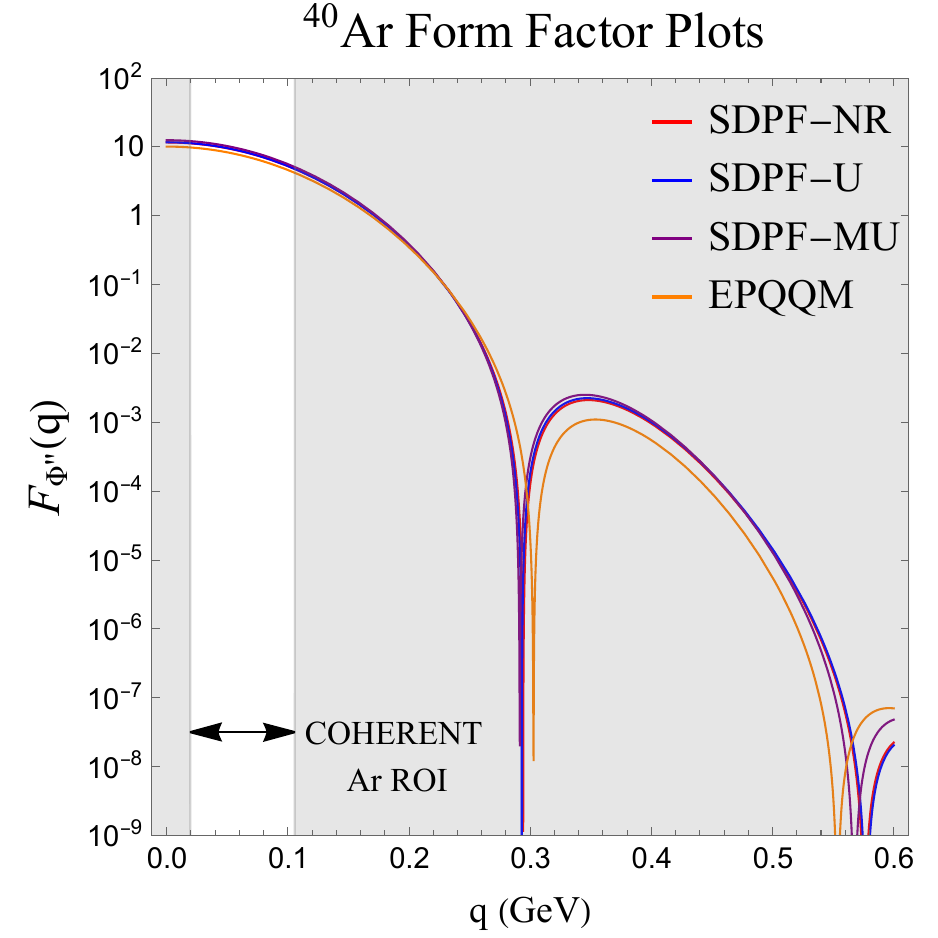}
\caption{$\Phi''$ form factors employing the different shell model interactions used for this isotope, plotted against the COHERENT range-of-interest. \label{fig:40ArFFPhi}}
\centering
\end{figure}

The maximum difference between the form factors for the $M$ response at the minimum $q$ value is $\sim0.3\%$, and $\sim11\%$ at the maximum $q$ value. For $\Phi''$, these values are $\sim24\%$ and $\sim21\%$, respectively, where the largest differences occur between the SDPF-MU and EPQQM interactions.
 
\subsection{Germanium isotopes}

The $M$ response is consistent amongst all calculations, with a maximum difference of $\sim 0.2\%$ at the lower end of the COHERENT $q$ region-of-interest (ROI) for the even isotopes, and $\sim3\%$ at the higher end for $^{70,72}$Ge. For $^{73}$Ge these take on the values of $\sim0.3\%$, and $\sim20\%$ at the higher end for $^{73,74,76}$Ge. 
For the $\Phi''$ channel, the maximum difference is observed between the GCN2850 and jj44b shell model calculations, which is $\sim80\%$ for $^{70}$Ge at both ends of the $q$ spectrum, factor of $\sim2.4$ for $^{72}$Ge, factor of $\sim2.3$ for $^{73}$Ge, factor of $\sim2.2$ for $^{74}$Ge, and $\sim60\%$ for $^{76}$Ge. The difference in the SD $\Sigma'$ channel for $^{73}$Ge is $\sim30\%$ at both ends of the $q$ ROI.\\




\begin{figure}[H]
\centering 
\includegraphics[width=1\linewidth]{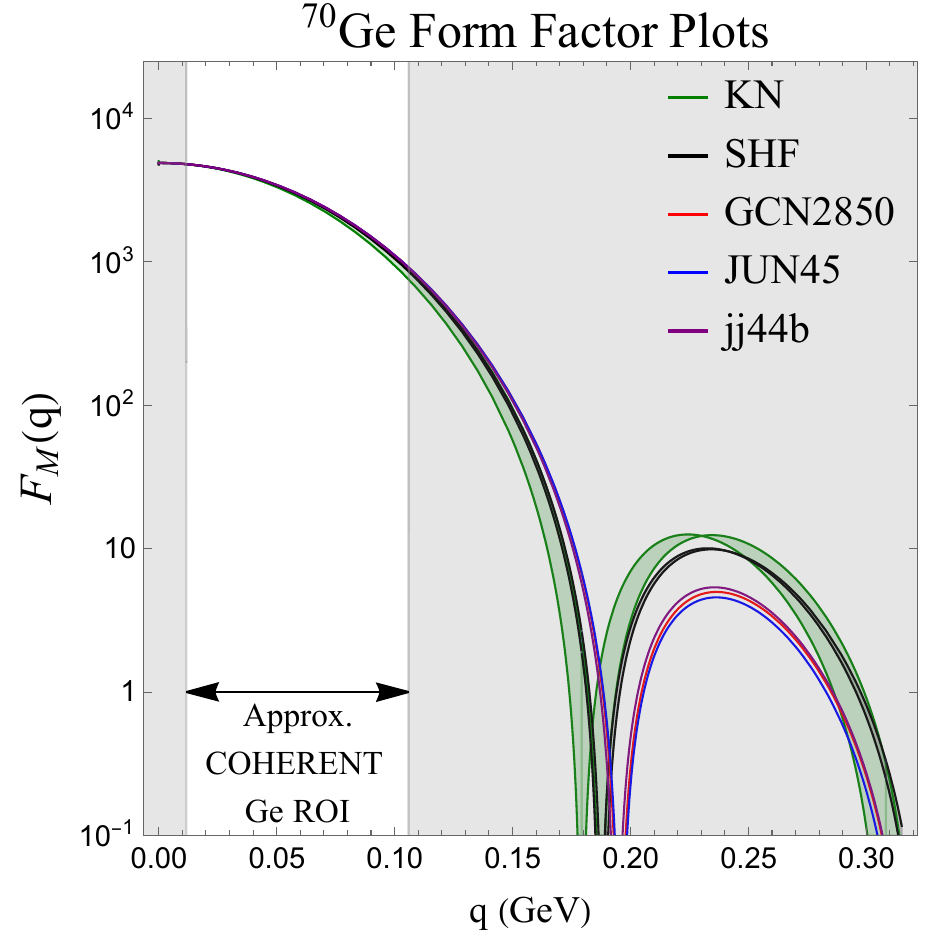}
\caption{SI ($M$) form factors employing the different shell model interactions used for this isotope. This response is compared against the KN and SHF distributions, where the uncertainty bands stem from the $r^n_{\text{rms}}$ uncertainty. The COHERENT range-of-interest is also displayed. \label{fig:70GeFFM}}
\centering
\end{figure}


\begin{figure}[H]
\centering 
\includegraphics[width=1\linewidth]{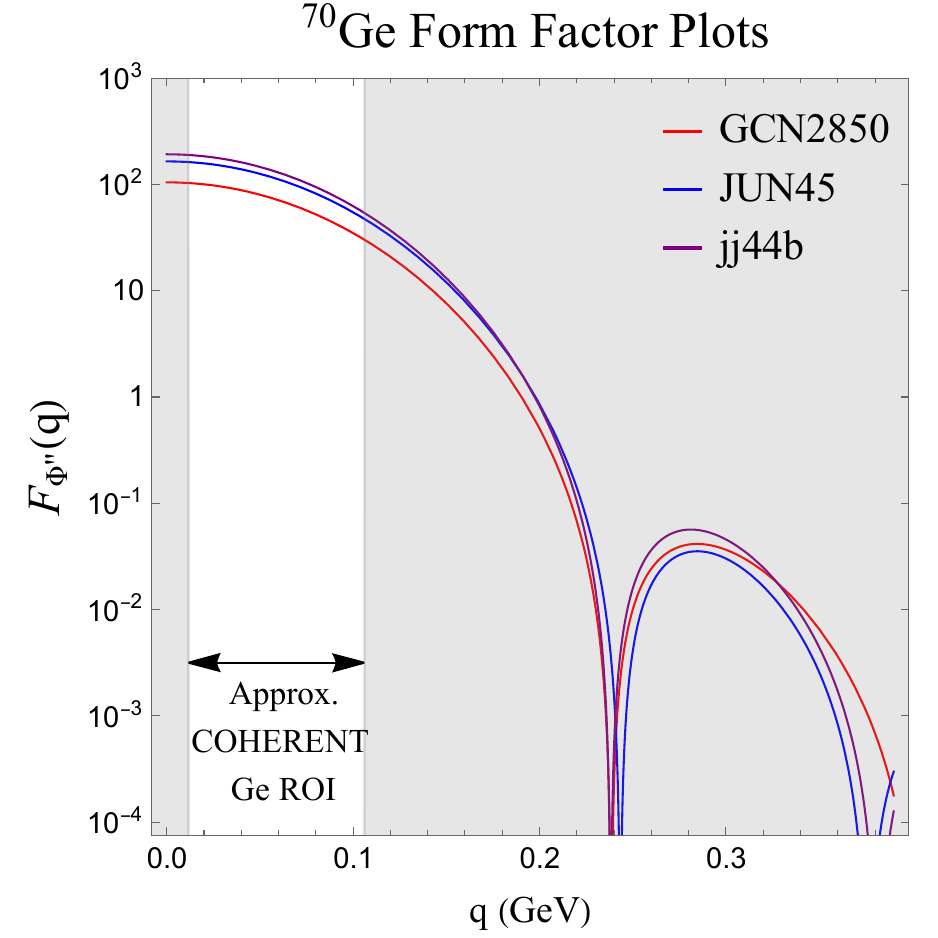}
\caption{$\Phi''$ form factors employing the different shell model interactions used for this isotope, plotted against the COHERENT range-of-interest. \label{fig:70GeFFPhi}}
\centering
\end{figure}




\begin{figure}[H]
\centering 
\includegraphics[width=1\linewidth]{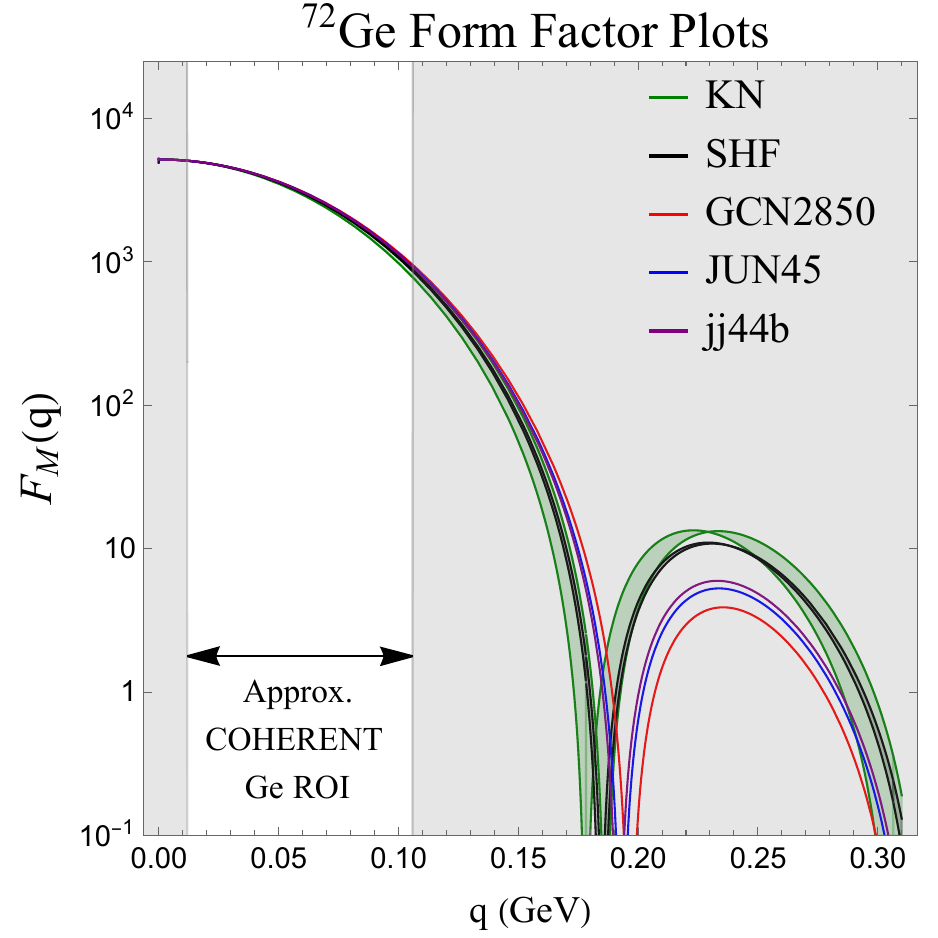}
\caption{SI ($M$) form factors employing the different shell model interactions used for this isotope. This response is compared against the KN and SHF distributions, where the uncertainty bands stem from the $r^n_{\text{rms}}$ uncertainty. The COHERENT range-of-interest is also displayed. \label{fig:72GeFFM}}
\centering
\end{figure}


\begin{figure}[H]
\centering 
\includegraphics[width=1\linewidth]{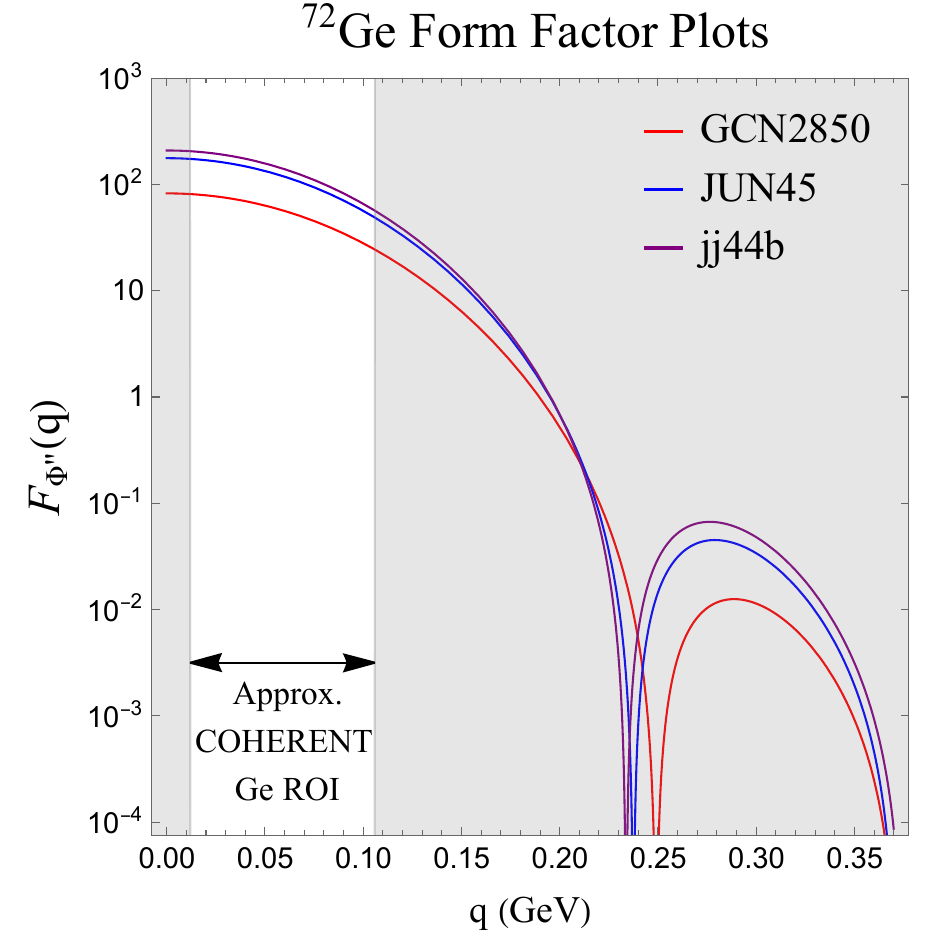}
\caption{$\Phi''$ form factors employing the different shell model interactions used for this isotope, plotted against the COHERENT range-of-interest. \label{fig:72GeFFPhi}}
\centering
\end{figure}




\begin{figure}[H]
\centering 
\includegraphics[width=1\linewidth]{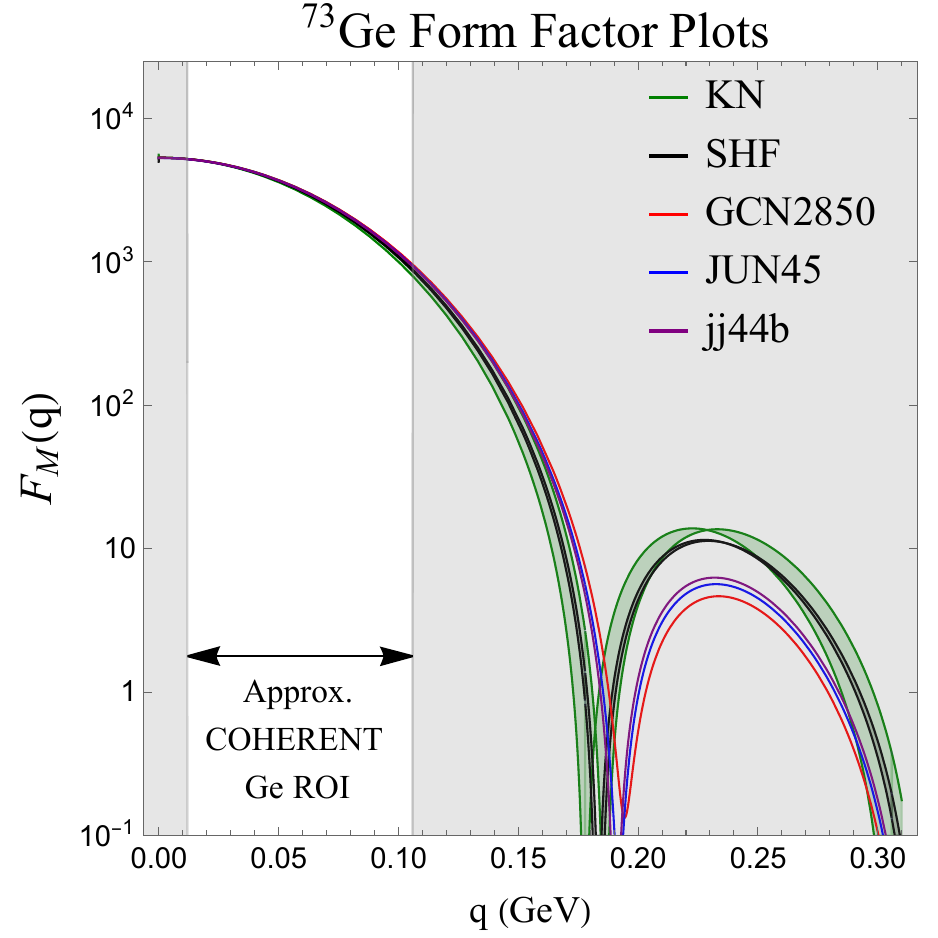}
\caption{SI ($M$) form factors employing the different shell model interactions used for this isotope. This response is compared against the KN and SHF distributions, where the uncertainty bands stem from the $r^n_{\text{rms}}$ uncertainty. The COHERENT range-of-interest is also displayed. \label{fig:73GeFFM}}
\centering
\end{figure}


\begin{figure}[H]
\centering 
\includegraphics[width=1\linewidth]{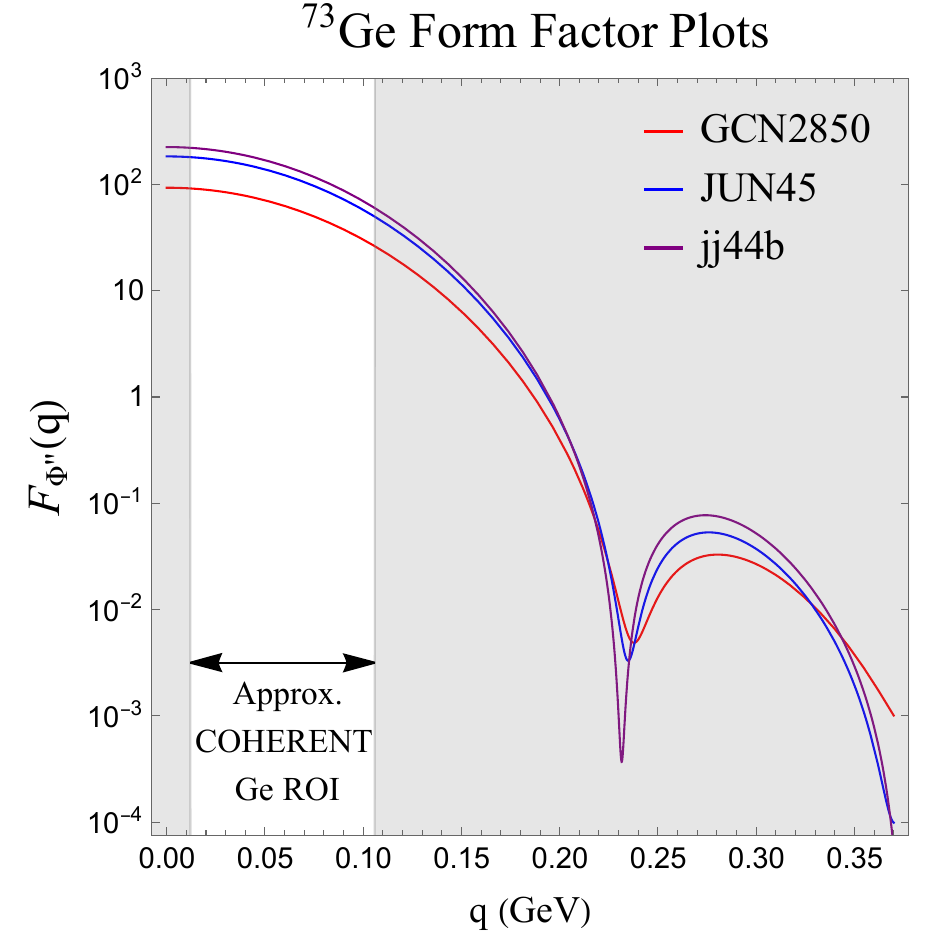}
\caption{$\Phi''$ form factors employing the different shell model interactions used for this isotope, plotted against the COHERENT range-of-interest. \label{fig:73GeFFPhi}}
\centering
\end{figure}


\begin{figure}[H]
\centering 
\includegraphics[width=1\linewidth]{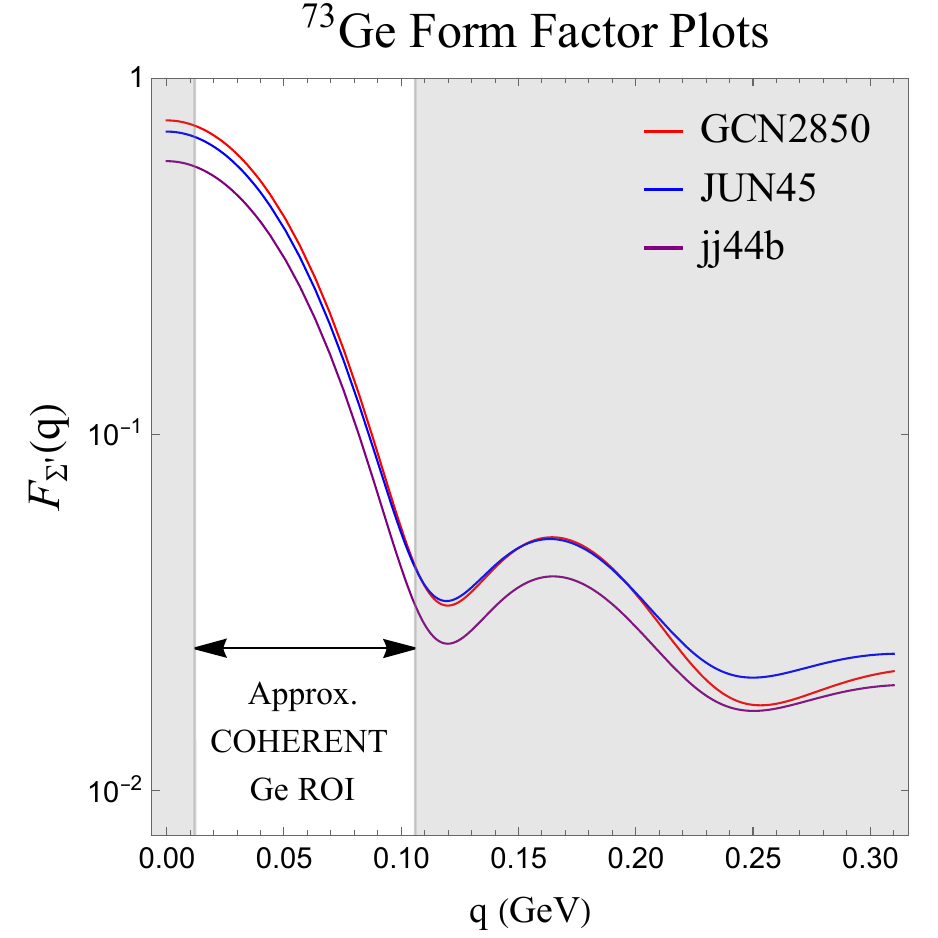}
\caption{$\Sigma'$ form factors employing the different shell model interactions used for this isotope, plotted against the COHERENT range-of-interest. \label{fig:73GeFFSigma}}
\centering
\end{figure}




\begin{figure}[H]
\centering 
\includegraphics[width=1\linewidth]{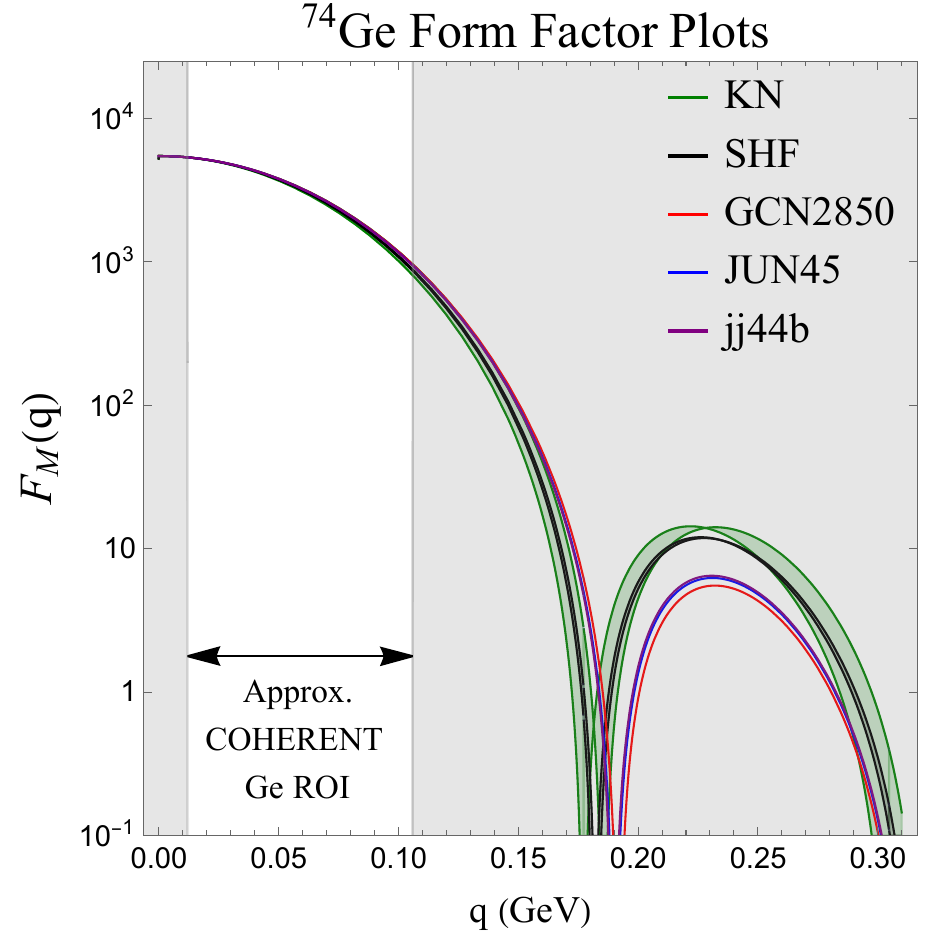}
\caption{SI ($M$) form factors employing the different shell model interactions used for this isotope. This response is compared against the KN and SHF distributions, where the uncertainty bands stem from the $r^n_{\text{rms}}$ uncertainty. The COHERENT range-of-interest is also displayed. \label{fig:74GeFFM}}
\centering
\end{figure}


\begin{figure}[H]
\centering 
\includegraphics[width=1\linewidth]{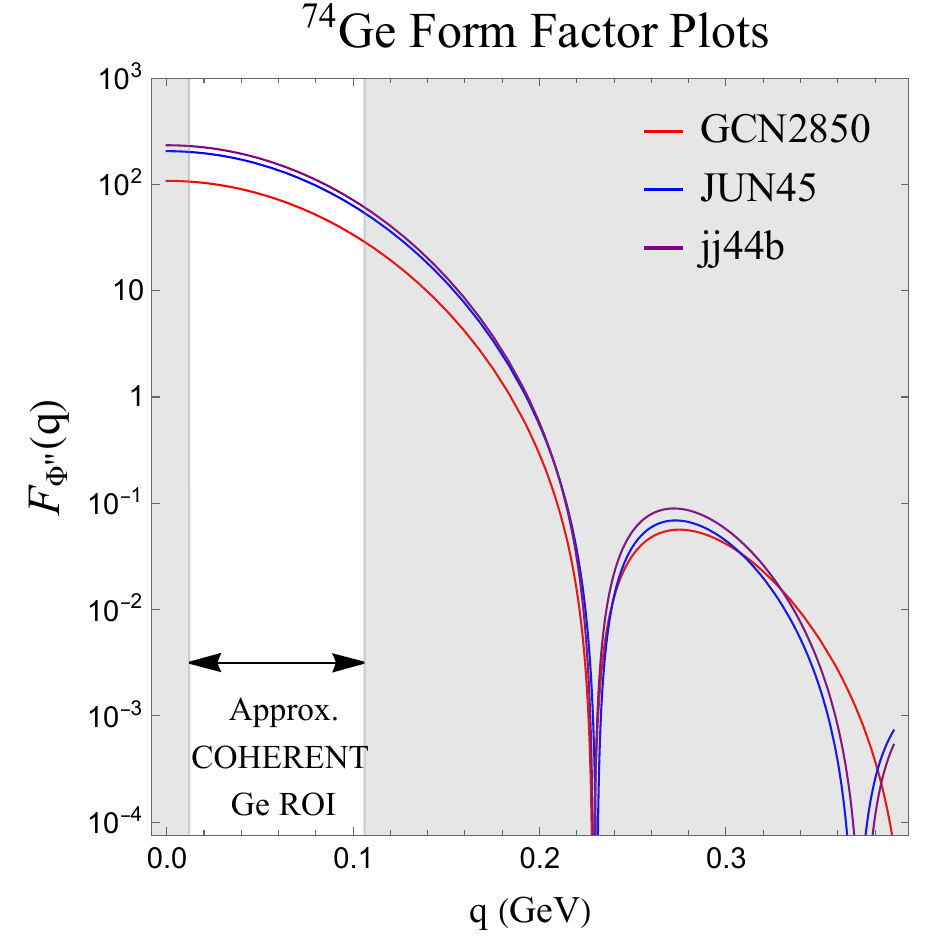}
\caption{$\Phi''$ form factors employing the different shell model interactions used for this isotope, plotted against the COHERENT range-of-interest. \label{fig:74GeFFPhi}}
\centering
\end{figure}




\begin{figure}[H]
\centering 
\includegraphics[width=1\linewidth]{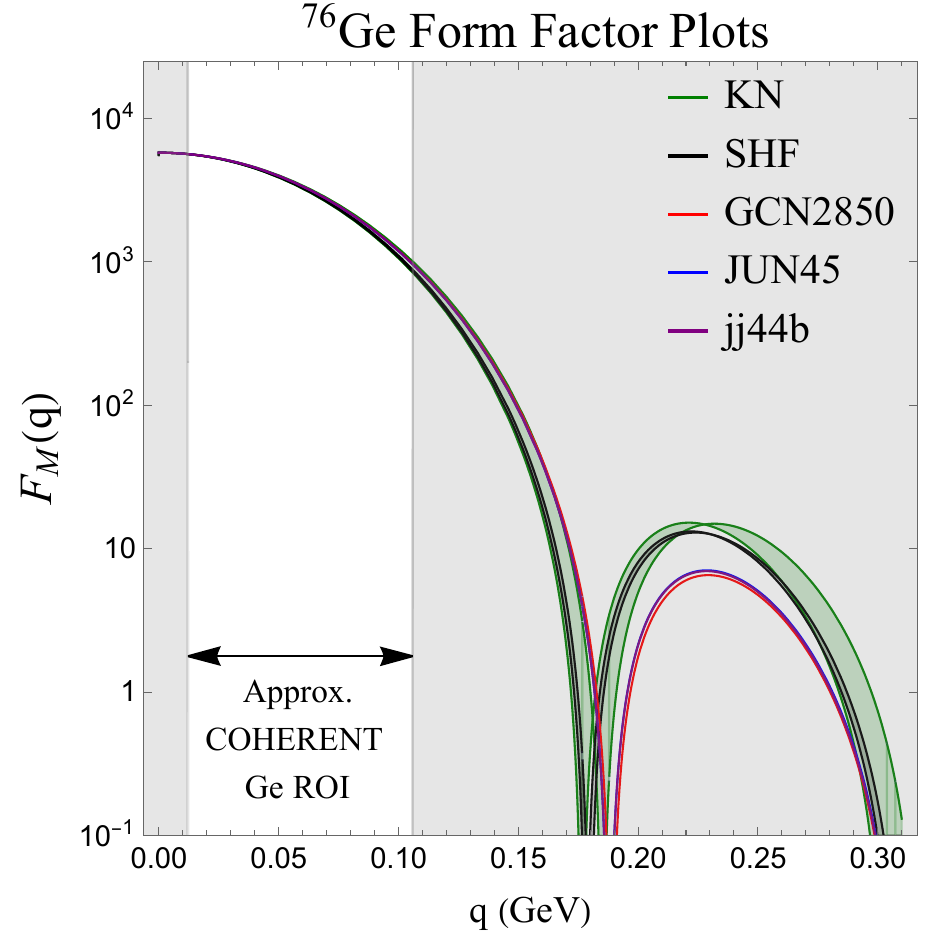}
\caption{SI ($M$) form factors employing the different shell model interactions used for this isotope. This response is compared against the KN and SHF distributions, where the uncertainty bands stem from the $r^n_{\text{rms}}$ uncertainty. The COHERENT range-of-interest is also displayed. \label{fig:76GeFFM}}
\centering
\end{figure}


\begin{figure}[H]
\centering 
\includegraphics[width=1\linewidth]{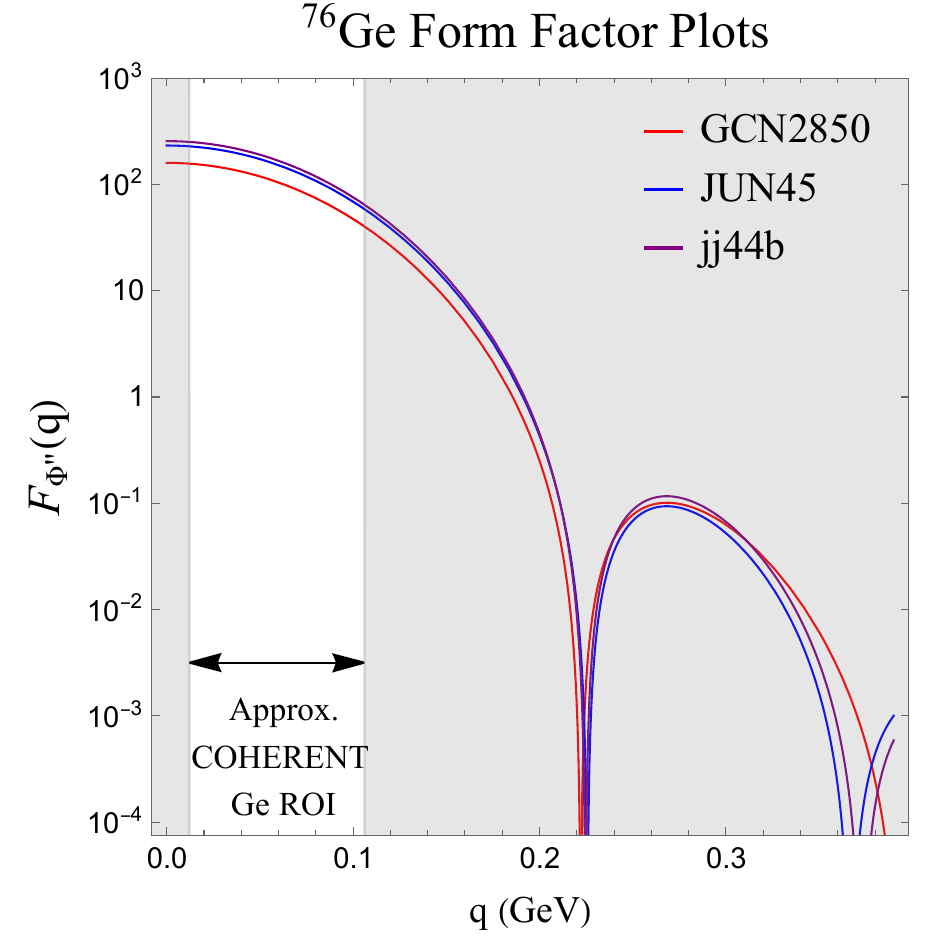}
\caption{$\Phi''$ form factors employing the different shell model interactions used for this isotope, plotted against the COHERENT range-of-interest. \label{fig:75GeFFPhi}}
\centering
\end{figure}

\subsection{Iodine-127}

The maximum difference between the form factors for the $M$ channel at the minimum $q$ value is $\sim1\%$, and $\sim30\%$ at the maximum $q$ value. For $\Phi''$, these values are $\sim55\%$ and $\sim50\%$, respectively. For $\Sigma'$, the factor differences are $\sim44$ and $\sim81$, due to the large difference between the Truncated SN100PN calculation and the remaining calculations. This is consistent with the integrated form factor (IFF) results in reference \cite{AbdelKhaleq:2023ipt}, which showed a more pronounced difference between the SD Truncated SN100PN IFF results and other calculations. This indicates that a more severe valence space truncation for a shell model interaction has a larger effect compared to a second more comparable truncation.




\begin{figure}[H]
\centering 
\includegraphics[width=1\linewidth]{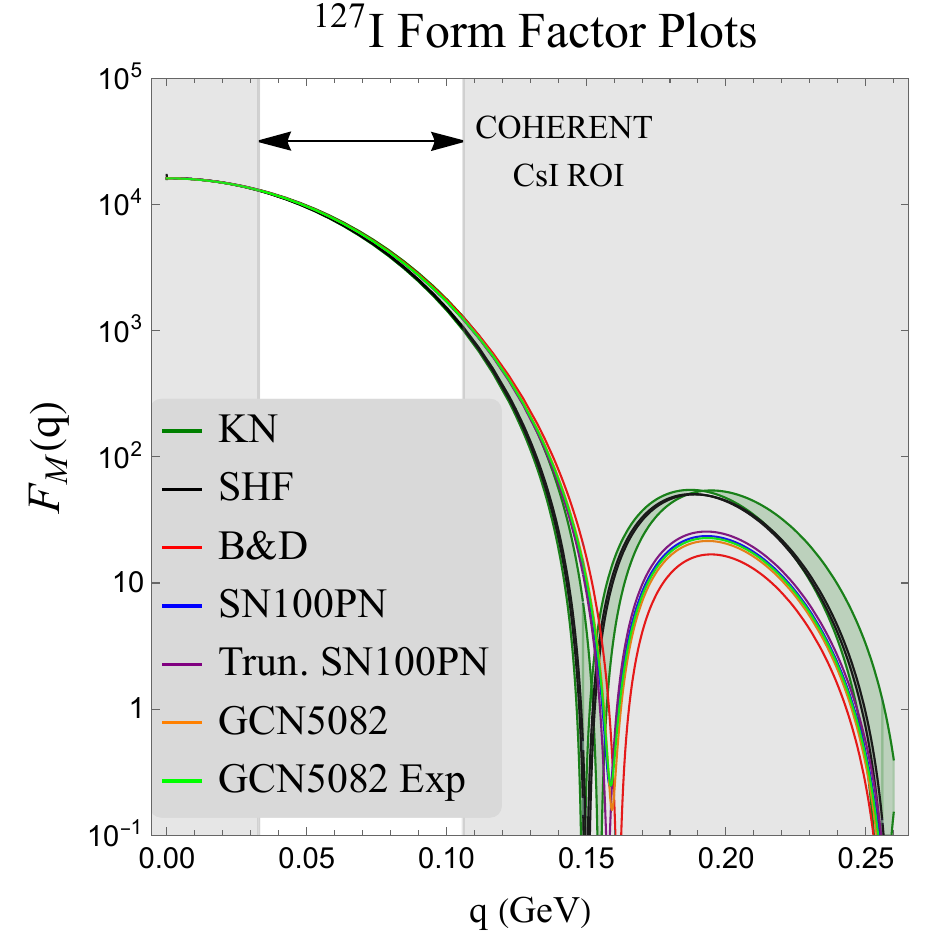}
\caption{SI ($M$) form factors employing the different shell model interactions used for this isotope. This response is compared against the KN and SHF distributions, where the uncertainty bands stem from the $r^n_{\text{rms}}$ uncertainty. The COHERENT range-of-interest is also displayed. \label{fig:127IFFM}}
\centering
\end{figure}



\begin{figure}[H]
\centering 
\includegraphics[width=1\linewidth]{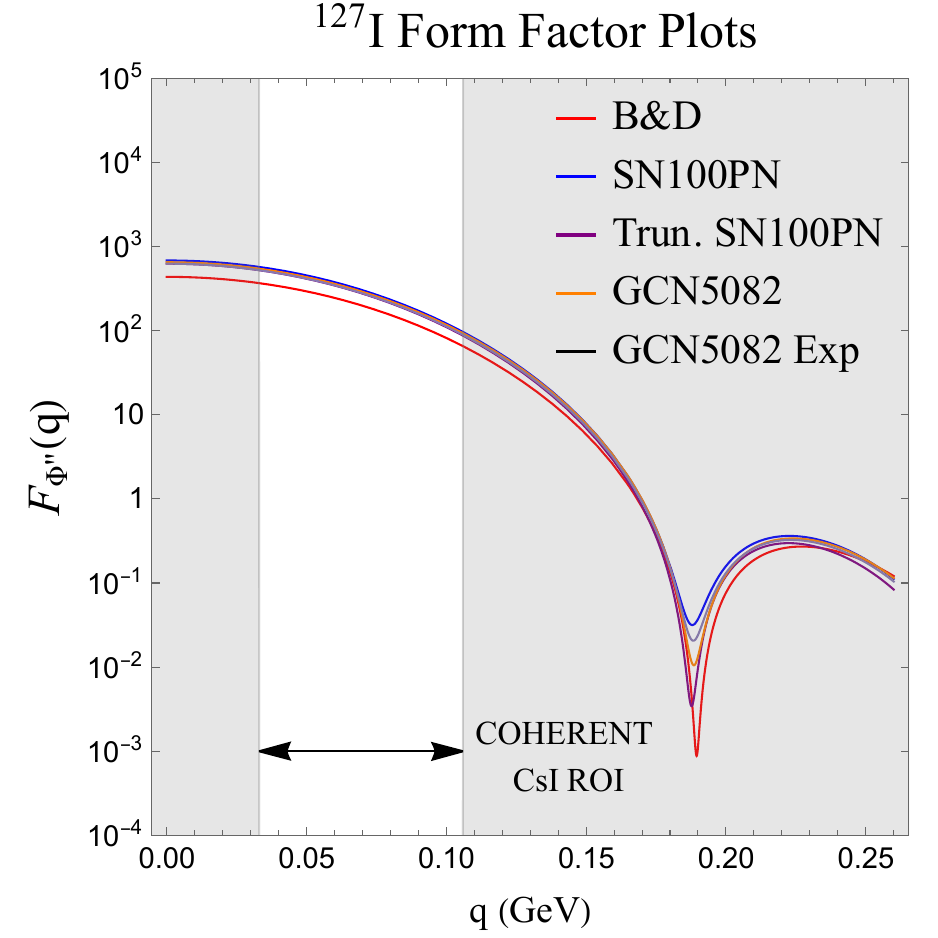}
\caption{$\Phi''$ form factors employing the different shell model interactions used for this isotope, plotted against the COHERENT range-of-interest. \label{fig:127IFFPhi}}
\centering
\end{figure}



\begin{figure}[H]
\centering 
\includegraphics[width=1\linewidth]{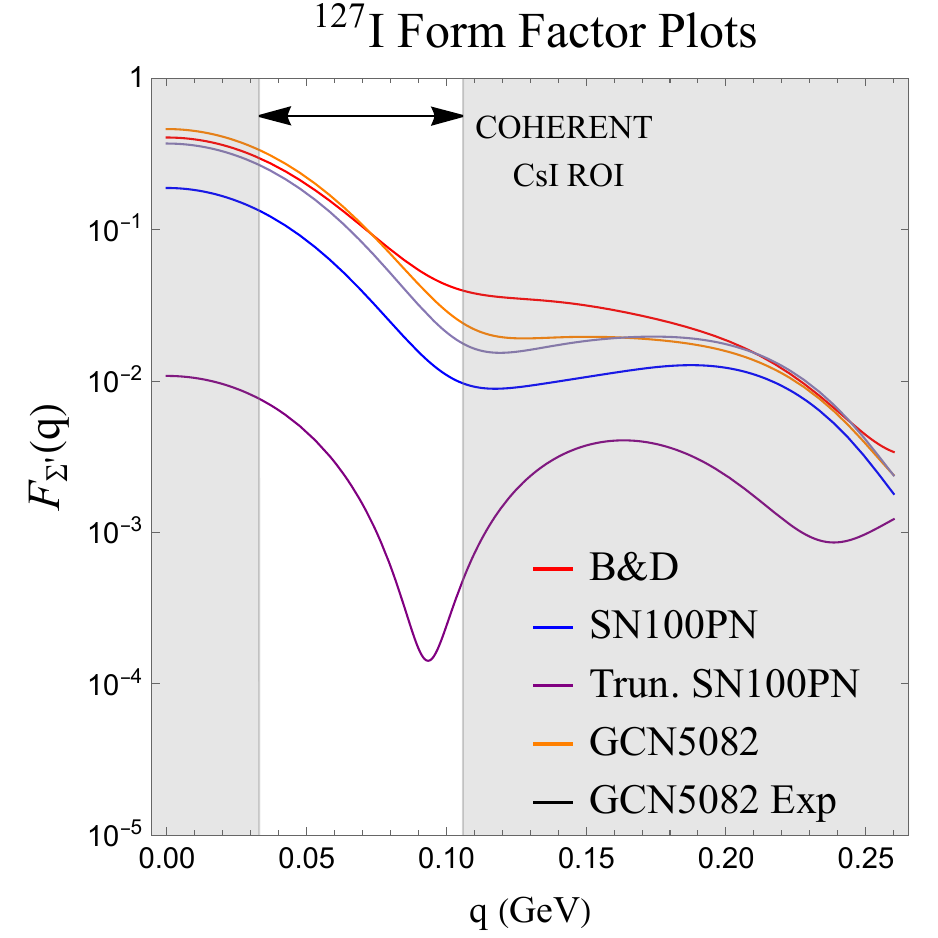}
\caption{$\Sigma'$ form factors employing the different shell model interactions used for this isotope, plotted against the COHERENT range-of-interest. \label{fig:127IFFSigma}}
\centering
\end{figure}

\subsection{Cesium-133}

The maximum difference between the form factors for the $M$ channel at the minimum $q$ value is $\sim1\%$, and $\sim20\%$ at the maximum $q$ value. For $\Phi''$, these values are $\sim6\%$ and $\sim7\%$, respectively. For $\Sigma'$, these differences are $\sim13\%$ and $\sim18\%$.\\



\begin{figure}[H]
\centering 
\includegraphics[width=1\linewidth]{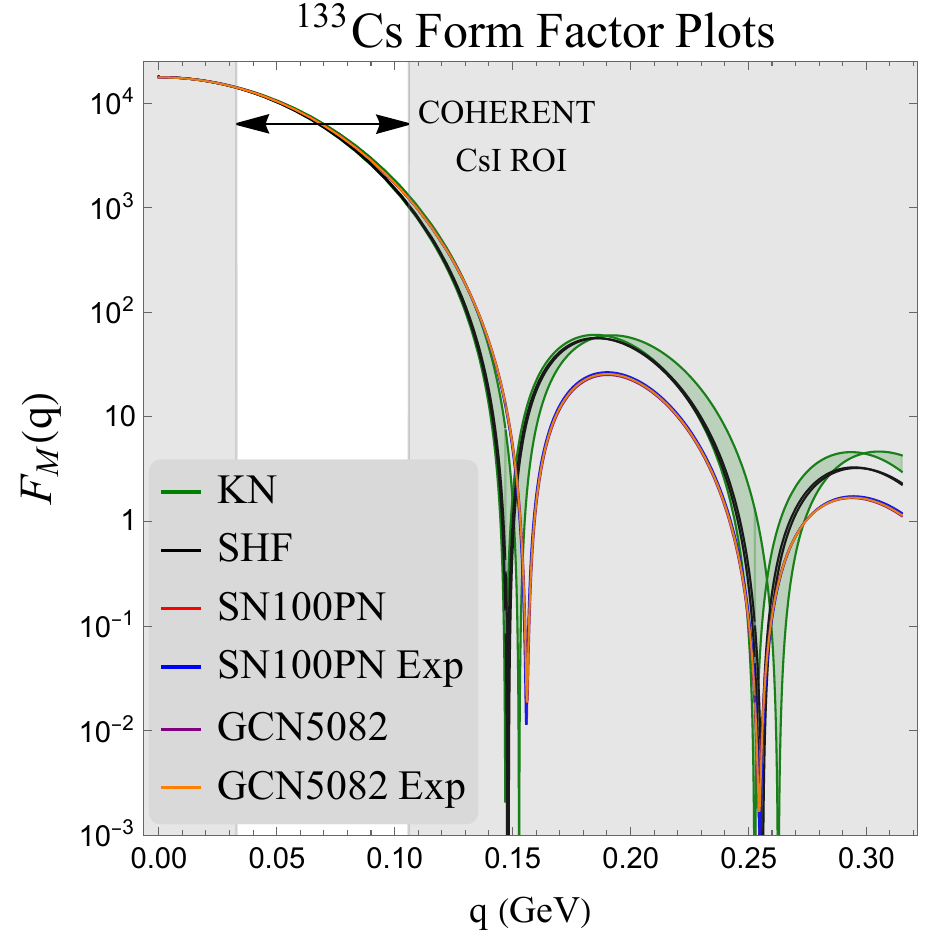}
\caption{SI ($M$) form factors employing the different shell model interactions used for this isotope. This response is compared against the KN and SHF distributions, where the uncertainty bands stem from the $r^n_{\text{rms}}$ uncertainty. The COHERENT range-of-interest is also displayed. \label{fig:133CsFFM}}
\centering
\end{figure}


\begin{figure}[H]
\centering 
\includegraphics[width=1\linewidth]{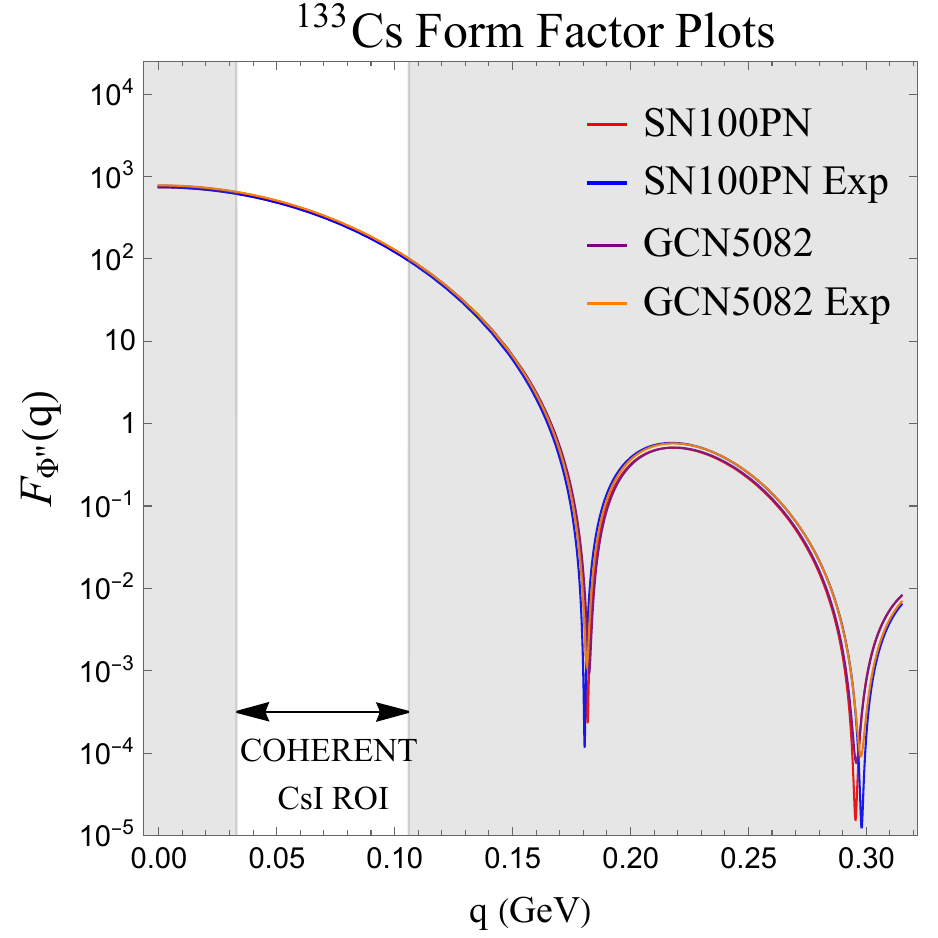}
\caption{$\Phi''$ form factors employing the different shell model interactions used for this isotope, plotted against the COHERENT range-of-interest. \label{fig:133CsFFPhi}}
\centering
\end{figure}


\begin{figure}[H]
\centering 
\includegraphics[width=1\linewidth]{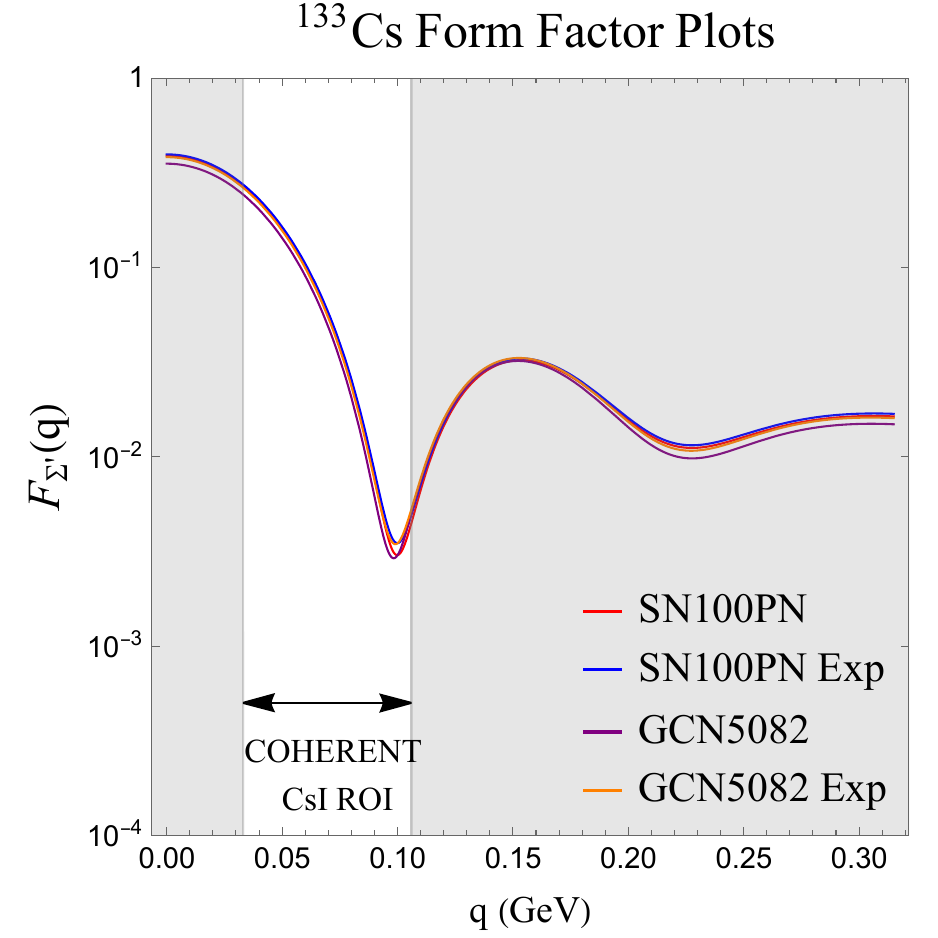}
\caption{$\Sigma'$ form factors employing the different shell model interactions used for this isotope, plotted against the COHERENT range-of-interest. \label{fig:133CsFFSigma}}
\centering
\end{figure}

\newpage

\onecolumngrid

\subsection*{$^{133}$Cs Form Factor Expressions}\label{Cs FF expressions}





The explicit nuclear form factors for $^{133}$Cs are given below using the convention of Fitzpatrick et al.~\cite{Fitzpatrick:2012ix}:

\begin{eqnarray}\label{Fitzpatrick FF}
     F^{(N,N')}_{X,Y} (q^2) \equiv  \frac{4\pi}{2J_i+1}  \sum_{J=0}^{2J_i} \langle J_i || X_J^{(N)} || J_i \rangle \langle J_i || Y^{(N')}_J || J_i \rangle, \\\nonumber
\end{eqnarray}

\noindent and can be used to obtain the $\mathcal{F}_N^{\Sigma'}(q^2)$ responses in Eq.~(\ref{FF eqn}) through the relation:

\begin{equation}
    |\mathcal{F}^{\Sigma'}_{N}|^2 (q^2)  = \frac{1}{4} \frac{2J_i+1}{4\pi} F_{\Sigma'}^{N,N}
\end{equation}

This can be used to show that, in the isospin basis, the spin structure functions are:

\begin{eqnarray}
    S^{\mathcal{T}}_{00} &=&\frac{2J_i+1}{4\pi} F^{\Sigma'}_{00}(\mathbf{q}^2)\nonumber\\
    S^{\mathcal{T}}_{11} &=& (1+\delta'(\mathbf{q}^2))^2 \left(\frac{2J_i+1}{4\pi} F^{\Sigma'}_{11}(\mathbf{q}^2)\right)\nonumber\\
    S^{\mathcal{T}}_{01} &=& 2(1+\delta'(\mathbf{q}^2))\left(\frac{2J_i+1}{4\pi}F^{\Sigma'}_{01}(q^2)\right)\nonumber\\ 
\end{eqnarray}

While describing CEvNS only requires the operators $M, \Phi''$ and $\Sigma'$, below we provide the form factors for the full list of operators in~\cite{Fitzpatrick:2012ix}. 

\subsubsection*{SN100PN}
 
$F_{M}^{(p,p)}= e^{-2 y} (3020.-11000. y+15500. y^2-10800. y^3+4130. y^4-881. y^5+103. y^6-6. y^7+0.145 y^8-0.000591 y^9+ 6.62\times10^{-7} y^{10})$

$F_{\Sigma''}^{(p,p)}=  e^{-2 y} (0.204-0.18 y+0.0841 y^2+0.168 y^3-0.0456 y^4-0.00246 y^5+0.014 y^6-0.00278 y^7+0.000806 y^8+1.9\times10^{-8} y^9+1.19\times10^{-12} y^{10})$

$F_{\Sigma'}^{(p,p)}= e^{-2 y} (0.408-3.09 y+8.18 y^2-9.16 y^3+5.23 y^4-1.62 y^5+0.287 y^6-0.0273 y^7+0.00167 y^8+1.8\times10^{-7} y^9+2.77\times10^{-11} y^{10}) $

$F_{\Phi''}^{(p,p)}=e^{-2 y} (76.-185. y+171. y^2-75.9 y^3+17.4 y^4-2.01 y^5+0.101 y^6-0.0012 y^7+4.14\times10^{-6} y^8) $

$F_{\tilde{\Phi}'}^{(p,p)}= e^{-2 y} (0.00327+0.00151 y-0.00238 y^2-0.000124 y^3+0.00104 y^4-0.00025 y^5+0.000026 y^6-4.22\times10^{-9} y^7+2.11\times10^{-9} y^8)  $

$F_{\Delta}^{(p,p)}= e^{-2 y} (2.73-6.56 y+6.44 y^2-2.97 y^3+0.704 y^4-0.0813 y^5+0.00378 y^6-1.37\times10^{-6} y^7+1.91\times10^{-10} y^8) $

$F_{M \Phi''}^{(p,p)}= e^{-2 y} (-479.+1460. y-1680. y^2+948. y^3-284. y^4+45.8 y^5-3.72 y^6+0.127 y^7-0.000979 y^8+1.66\times10^{-6} y^9)$

$F_{\Sigma' \Delta}^{(p,p)}= e^{-2 y} (1.06-5.26 y+8.26 y^2-6.02 y^3+2.24 y^4-0.451 y^5+0.0461 y^6-0.00218 y^7+1.81\times10^{-8} y^8+7.25\times10^{-11} y^9) $

\vspace{5mm}

$F_{M}^{(p,n)}= e^{-2 y} (4280.-17000. y+26500. y^2-20900. y^3+9210. y^4-2350. y^5+342. y^6-27. y^7+1. y^8-0.0128 y^9+0.000025 y^{10})  $

$F_{\Sigma''}^{(p,n)}= e^{-2 y} (-0.0048+0.0153 y-0.0323 y^2+0.0186 y^3+0.000919 y^4-0.00533 y^5+0.00219 y^6-0.000397 y^7+0.0000337 y^8-9.97\times10^{-7} y^9+1.92\times10^{-11} y^{10}) $

$F_{\Sigma'}^{(p,n)}= e^{-2 y} (-0.00961+0.0804 y-0.264 y^2+0.382 y^3-0.279 y^4+0.112 y^5-0.0259 y^6+0.0034 y^7-0.000232 y^8+6.38\times10^{-6} y^9+8.41\times10^{-10} y^{10}) $

$F_{\Phi''}^{(p,n)}= e^{-2 y} (161.-449. y+478. y^2-252. y^3+70.9 y^4-10.8 y^5+0.831 y^6-0.0265 y^7+0.000156 y^8) $

$F_{\tilde{\Phi}'}^{(p,n)}= e^{-2 y} (-0.0171+0.0166 y+0.00245 y^2-0.0105 y^3+0.00504 y^4-0.00096 y^5+0.000072 y^6-1.28\times10^{-6} y^7-2.6\times10^{-8} y^8)$

$F_{\Delta}^{(p,n)}=  e^{-2 y} (0.199-0.526 y+0.584 y^2-0.324 y^3+0.0954 y^4-0.0148 y^5+0.0011 y^6-0.0000299 y^7+6.02\times10^{-9} y^8) $

$F_{M \Phi''}^{(p,n)}= e^{-2 y} (-1020.+3440. y-4510. y^2+2950. y^3-1060. y^4+213. y^5-23.6 y^6+1.31 y^7-0.0295 y^8+0.0000625 y^9)$

$F_{\Sigma' \Delta}^{(p,n)}= e^{-2 y} (0.0768-0.401 y+0.694 y^2-0.581 y^3+0.26 y^4-0.0643 y^5+0.00865 y^6-0.000584 y^7+0.0000155 y^8+2.25\times10^{-9} y^9) $

\vspace{5mm}

$F_{M \Phi''}^{(n,p)}= e^{-2 y} (-679.+2290. y-2960. y^2+1920. y^3-680. y^4+135. y^5-14.5 y^6+0.749 y^7-0.0132 y^8+0.0000625 y^9) $

$F_{\Sigma' \Delta}^{(n,p)}= e^{-2 y} (-0.0249+0.144 y-0.304 y^2+0.29 y^3-0.141 y^4+0.0379 y^5-0.00559 y^6+0.000415 y^7-0.0000118 y^8+2.25\times10^{-9} y^9) $

\vspace{5mm}

$F_{M}^{(n,n)}=e^{-2 y} (6070.-26100. y+44600. y^2-39200. y^3+19700. y^4-5870. y^5+1040. y^6-106. y^7+5.58 y^8-0.124 y^9+0.000948 y^{10}) $

$F_{\Sigma''}^{(n,n)}= e^{-2 y} (0.000113-0.00062 y+0.00352 y^2-0.00581 y^3+0.00463 y^4-0.00205 y^5+0.00053 y^6-0.0000795 y^7+6.33\times10^{-6} y^8-2.1\times10^{-7} y^9+2.77\times10^{-9} y^{10}) $

$F_{\Sigma'}^{(n,n)}= e^{-2 y} (0.000226-0.00207 y+0.00999 y^2-0.018 y^3+0.0157 y^4-0.00743 y^5+0.00209 y^6-0.000355 y^7+0.0000357 y^8-1.84\times10^{-6} y^9+3.92\times10^{-8} y^{10}) $

$F_{\Phi''}^{(n,n)}= e^{-2 y} (343.-1070. y+1300. y^2-797. y^3+269. y^4-51.4 y^5+5.41 y^6-0.287 y^7+0.00593 y^8) $

$F_{\tilde{\Phi}'}^{(n,n)}= e^{-2 y} (0.0891-0.215 y+0.223 y^2-0.122 y^3+0.038 y^4-0.00687 y^5+0.000686 y^6-0.0000332 y^7+6.04\times10^{-7} y^8)$

$F_{\Delta}^{(n,n)}= e^{-2 y} (0.0144-0.0417 y+0.0517 y^2-0.0334 y^3+0.012 y^4-0.00238 y^5+0.000252 y^6-0.0000128 y^7+2.54\times10^{-7} y^8) $

$F_{M \Phi''}^{(n,n)}=e^{-2 y} (-1440.+5360. y-7780. y^2+5760. y^3-2400. y^4+578. y^5-80.2 y^6+6.04 y^7-0.213 y^8+0.00237 y^9) $

$F_{\Sigma' \Delta}^{(n,n)}= e^{-2 y} (-0.00181+0.0109 y-0.0246 y^2+0.0263 y^3-0.015 y^4+0.00482 y^5-0.000901 y^6+0.0000936 y^7-4.82\times10^{-6} y^8+9.94\times10^{-8} y^9) $

\vspace{5mm}

\subsubsection*{SN100PN Expanded}
 
$F_{M}^{(p,p)}= e^{-2 y} (3020.-11000. y+15500. y^2-10800. y^3+4120. y^4-876. y^5+102. y^6-5.91 y^7+0.143 y^8-0.000608 y^9+7.17\times10^{-7} y^{10})$

$F_{\Sigma''}^{(p,p)}=  e^{-2 y} (0.206-0.19 y+0.1 y^2+0.166 y^3-0.0494 y^4+0.000185 y^5+0.0135 y^6-0.00271 y^7+0.000818 y^8+2.12\times10^{-8} y^9+1.62\times10^{-12} y^{10})$

$F_{\Sigma'}^{(p,p)}= e^{-2 y} (0.412-3.11 y+8.21 y^2-9.17 y^3+5.21 y^4-1.61 y^5+0.283 y^6-0.0269 y^7+0.00167 y^8+2.15\times10^{-7} y^9+3.99\times10^{-11} y^{10}) $

$F_{\Phi''}^{(p,p)}=e^{-2 y} (75.3-184. y+170. y^2-75.5 y^3+17.3 y^4-2.01 y^5+0.102 y^6-0.00125 y^7+4.48\times10^{-6} y^8) $

$F_{\tilde{\Phi}'}^{(p,p)}= e^{-2 y} (0.0028+0.00224 y-0.00221 y^2-0.000663 y^3+0.00134 y^4-0.000307 y^5+0.0000285 y^6+1.9\times10^{-8} y^7+2.31\times10^{-9} y^8)  $

$F_{\Delta}^{(p,p)}= e^{-2 y} (2.77-6.65 y+6.53 y^2-3.01 y^3+0.713 y^4-0.0822 y^5+0.00382 y^6-1.66\times10^{-6} y^7+2.77\times10^{-10} y^8) $

$F_{M \Phi''}^{(p,p)}= e^{-2 y} (-477.+1450. y-1670. y^2+944. y^3-283. y^4+45.6 y^5-3.71 y^6+0.126 y^7-0.00101 y^8+1.79\times10^{-6} y^9)$

$F_{\Sigma' \Delta}^{(p,p)}= e^{-2 y} (1.07-5.31 y+8.33 y^2-6.06 y^3+2.25 y^4-0.452 y^5+0.046 y^6-0.00217 y^7+1.65\times10^{-8} y^8+1.05\times10^{-10} y^9) $

\vspace{5mm}

$F_{M}^{(p,n)}= e^{-2 y} (4280.-17000. y+26500. y^2-20800. y^3+9180. y^4-2330. y^5+339. y^6-26.7 y^7+0.998 y^8-0.013 y^9+0.0000266 y^{10})  $

$F_{\Sigma''}^{(p,n)}= e^{-2 y} (-0.0042+0.0114 y-0.0194 y^2+0.00958 y^3+0.00118 y^4-0.00307 y^5+0.00117 y^6-0.000213 y^7+0.00002 y^8-7.23\times10^{-7} y^9+1.64\times10^{-11} y^{10}) $

$F_{\Sigma'}^{(p,n)}= e^{-2 y} (-0.00841+0.0695 y-0.22 y^2+0.309 y^3-0.219 y^4+0.0848 y^5-0.0189 y^6+0.00241 y^7-0.000163 y^8+4.55\times10^{-6} y^9+7.28\times10^{-10} y^{10}) $

$F_{\Phi''}^{(p,n)}= e^{-2 y} (161.-449. y+479. y^2-252. y^3+70.9 y^4-10.8 y^5+0.84 y^6-0.0273 y^7+0.000166 y^8) $

$F_{\tilde{\Phi}'}^{(p,n)}= e^{-2 y} (-0.0115+0.00924 y+0.00526 y^2-0.00957 y^3+0.00411 y^4-0.00075 y^5+0.0000598 y^6-1.43\times10^{-6} y^7-2.15\times10^{-8} y^8)$

$F_{\Delta}^{(p,n)}=  e^{-2 y} (0.18-0.468 y+0.513 y^2-0.285 y^3+0.0835 y^4-0.0127 y^5+0.000896 y^6-0.0000218 y^7+5.22\times10^{-9} y^8) $

$F_{M \Phi''}^{(p,n)}= e^{-2 y} (-1020.+3460. y-4530. y^2+2960. y^3-1060. y^4+213. y^5-23.6 y^6+1.32 y^7-0.0299 y^8+0.0000666 y^9)$

$F_{\Sigma' \Delta}^{(p,n)}= e^{-2 y} (0.0695-0.359 y+0.608 y^2-0.506 y^3+0.226 y^4-0.0552 y^5+0.00718 y^6-0.000457 y^7+0.000011 y^8+1.95\times10^{-9} y^9) $

\vspace{5mm}

$F_{M \Phi''}^{(n,p)}= e^{-2 y} (-675.+2280. y-2950. y^2+1910. y^3-677. y^4+134. y^5-14.5 y^6+0.752 y^7-0.0136 y^8+0.0000666 y^9) $

$F_{\Sigma' \Delta}^{(n,p)}= e^{-2 y} (-0.0218+0.124 y-0.253 y^2+0.234 y^3-0.11 y^4+0.0286 y^5-0.00409 y^6+0.000299 y^7-8.57\times10^{-6} y^8+1.95\times10^{-9} y^9) $

\vspace{5mm}

$F_{M}^{(n,n)}=e^{-2 y} (6060.-26100. y+44600. y^2-39200. y^3+19600. y^4-5840. y^5+1030. y^6-105. y^7+5.58 y^8-0.127 y^9+0.000994 y^{10}) $

$F_{\Sigma''}^{(n,n)}= e^{-2 y} (0.0000858-0.000388 y+0.00137 y^2-0.00192 y^3+0.00139 y^4-0.000568 y^5+0.000138 y^6-0.0000202 y^7+1.72\times10^{-6} y^8-7.53\times10^{-8} y^9+1.43\times10^{-9} y^{10}) $

$F_{\Sigma'}^{(n,n)}= e^{-2 y} (0.000172-0.00154 y+0.0061 y^2-0.0102 y^3+0.0086 y^4-0.00401 y^5+0.00112 y^6-0.000189 y^7+0.0000185 y^8-9.03\times10^{-7} y^9+2.03\times10^{-8} y^{10}) $

$F_{\Phi''}^{(n,n)}= e^{-2 y} (344.-1080. y+1310. y^2-803. y^3+270. y^4-51.5 y^5+5.45 y^6-0.293 y^7+0.00621 y^8) $

$F_{\tilde{\Phi}'}^{(n,n)}= e^{-2 y} (0.0471-0.113 y+0.115 y^2-0.0603 y^3+0.0179 y^4-0.00312 y^5+0.000316 y^6-0.0000171 y^7+3.82\times10^{-7} y^8)$

$F_{\Delta}^{(n,n)}= e^{-2 y} (0.0117-0.0328 y+0.0397 y^2-0.0256 y^3+0.00922 y^4-0.0018 y^5+0.000179 y^6-7.73\times10^{-6} y^7+1.32\times10^{-7} y^8) $

$F_{M \Phi''}^{(n,n)}=e^{-2 y} (-1440.+5380. y-7820. y^2+5780. y^3-2400. y^4+577. y^5-80.2 y^6+6.07 y^7-0.217 y^8+0.00248 y^9) $

$F_{\Sigma' \Delta}^{(n,n)}= e^{-2 y} (-0.00142+0.00836 y-0.0178 y^2+0.0182 y^3-0.0101 y^4+0.00319 y^5-0.000576 y^6+0.0000566 y^7-2.65\times10^{-6} y^8+5.16\times10^{-8} y^9) $

\vspace{5mm}

\subsubsection*{GCN5082}
 
$F_{M}^{(p,p)}= e^{-2 y} (3020.-11000. y+15500. y^2-10900. y^3+4220. y^4-914. y^5+109. y^6-6.58 y^7+0.165 y^8-0.000663 y^9+7.35\times10^{-7} y^{10})$

$F_{\Sigma''}^{(p,p)}=  e^{-2 y} (0.185-0.0951 y-0.0302 y^2+0.167 y^3+0.0107 y^4-0.0318 y^5+0.0186 y^6-0.00298 y^7+0.000747 y^8+1.84\times10^{-8} y^9+2.34\times10^{-12} y^{10})$

$F_{\Sigma'}^{(p,p)}= e^{-2 y} (0.37-2.86 y+7.74 y^2-8.75 y^3+5.02 y^4-1.57 y^5+0.281 y^6-0.0275 y^7+0.00167 y^8+3.2\times10^{-7} y^9+7.21\times10^{-11} y^{10}) $

$F_{\Phi''}^{(p,p)}=e^{-2 y} (92.1-225. y+208. y^2-93. y^3+21.5 y^4-2.52 y^5+0.128 y^6-0.00143 y^7+4.59\times10^{-6} y^8) $

$F_{\tilde{\Phi}'}^{(p,p)}= e^{-2 y} (0.0147-0.00888 y+0.000482 y^2+0.00251 y^3-(3.38*10^(-6)) y^4-0.0000492 y^5+0.0000216 y^6+4.51\times10^{-7} y^7+3.47\times10^{-9} y^8)  $

$F_{\Delta}^{(p,p)}= e^{-2 y} (2.65-6.37 y+6.21 y^2-2.86 y^3+0.678 y^4-0.0785 y^5+0.00367 y^6-2.32\times10^{-6} y^7+5.1\times10^{-10} y^8) $

$F_{M \Phi''}^{(p,p)}= e^{-2 y} (-528.+1600. y-1850. y^2+1050. y^3-320. y^4+52.3 y^5-4.34 y^6+0.15 y^7-0.00112 y^8+1.84\times10^{-6} y^9)$

$F_{\Sigma' \Delta}^{(p,p)}= e^{-2 y} (0.991-5.02 y+7.93 y^2-5.78 y^3+2.16 y^4-0.436 y^5+0.045 y^6-0.00214 y^7+3.71\times10^{-8} y^8+1.91\times10^{-10} y^9) $

\vspace{5mm}

$F_{M}^{(p,n)}= e^{-2 y} (4280.-17000. y+26500. y^2-21000. y^3+9290. y^4-2380. y^5+351. y^6-28.1 y^7+1.06 y^8-0.0135 y^9+0.0000258 y^{10})  $

$F_{\Sigma''}^{(p,n)}= e^{-2 y} (-0.00421+0.0148 y-0.029 y^2+0.0133 y^3+0.00321 y^4-0.00556 y^5+0.0021 y^6-0.000387 y^7+0.0000362 y^8-1.17\times10^{-6} y^9+2.84\times10^{-11} y^{10}) $

$F_{\Sigma'}^{(p,n)}= e^{-2 y} (-0.00841+0.0722 y-0.243 y^2+0.357 y^3-0.265 y^4+0.108 y^5-0.0256 y^6+0.00346 y^7-0.000246 y^8+7.12\times10^{-6} y^9+1.42\times10^{-9} y^{10}) $

$F_{\Phi''}^{(p,n)}= e^{-2 y} (175.-486. y+519. y^2-274. y^3+77.4 y^4-11.8 y^5+0.916 y^6-0.0292 y^7+0.000162 y^8) $

$F_{\tilde{\Phi}'}^{(p,n)}= e^{-2 y} (-0.0389+0.0585 y-0.0328 y^2+0.00374 y^3+0.00233 y^4-0.000795 y^5+0.0000795 y^6-1.68\times10^{-6} y^7-4.12\times10^{-8} y^8)$

$F_{\Delta}^{(p,n)}=  e^{-2 y} (0.223-0.588 y+0.643 y^2-0.351 y^3+0.102 y^4-0.0157 y^5+0.00117 y^6-0.0000316 y^7+1.03\times10^{-8} y^8) $

$F_{M \Phi''}^{(p,n)}= e^{-2 y} (-1000.+3390. y-4440. y^2+2920. y^3-1050. y^4+214. y^5-23.9 y^6+1.35 y^7-0.031 y^8+0.0000646 y^9)$

$F_{\Sigma' \Delta}^{(p,n)}= e^{-2 y} (0.0832-0.441 y+0.762 y^2-0.632 y^3+0.28 y^4-0.0688 y^5+0.00921 y^6-0.000627 y^7+0.000017 y^8+3.83\times10^{-9} y^9) $

\vspace{5mm}

$F_{M \Phi''}^{(n,p)}= e^{-2 y} (-747.+2520. y-3260. y^2+2120. y^3-753. y^4+150. y^5-16.3 y^6+0.839 y^7-0.0144 y^8+0.0000646 y^9) $

$F_{\Sigma' \Delta}^{(n,p)}= e^{-2 y} (-0.0225+0.133 y-0.283 y^2+0.272 y^3-0.134 y^4+0.0366 y^5-0.00553 y^6+0.000422 y^7-0.0000125 y^8+3.83\times10^{-9} y^9) $

\vspace{5mm}

$F_{M}^{(n,n)}=e^{-2 y} (6070.-26100. y+44500. y^2-39200. y^3+19700. y^4-5880. y^5+1040. y^6-106. y^7+5.58 y^8-0.122 y^9+0.000919 y^{10}) $

$F_{\Sigma''}^{(n,n)}= e^{-2 y} (0.0000957-0.000626 y+0.0036 y^2-0.00574 y^3+0.00447 y^4-0.00195 y^5+0.000504 y^6-0.0000768 y^7+6.4\times10^{-6} y^8-2.28\times10^{-7} y^9+3.28\times10^{-9} y^{10}) $

$F_{\Sigma'}^{(n,n)}= e^{-2 y} (0.000191-0.0018 y+0.00906 y^2-0.0167 y^3+0.0148 y^4-0.00718 y^5+0.00205 y^6-0.000357 y^7+0.0000371 y^8-1.98\times10^{-6} y^9+4.34\times10^{-8} y^{10}) $

$F_{\Phi''}^{(n,n)}= e^{-2 y} (333.-1040. y+1260. y^2-773. y^3+261. y^4-49.7 y^5+5.23 y^6-0.277 y^7+0.00574 y^8) $

$F_{\tilde{\Phi}'}^{(n,n)}= e^{-2 y} (0.103-0.247 y+0.254 y^2-0.137 y^3+0.0425 y^4-0.00762 y^5+0.000756 y^6-0.0000363 y^7+6.56\times10^{-7} y^8)$

$F_{\Delta}^{(n,n)}= e^{-2 y} (0.0187-0.0538 y+0.0652 y^2-0.0412 y^3+0.0145 y^4-0.00283 y^5+0.000293 y^6-0.0000145 y^7+2.79\times10^{-7} y^8) $

$F_{M \Phi''}^{(n,n)}=e^{-2 y} (-1420.+5270. y-7660. y^2+5670. y^3-2360. y^4+569. y^5-79. y^6+5.94 y^7-0.209 y^8+0.0023 y^9) $

$F_{\Sigma' \Delta}^{(n,n)}= e^{-2 y} (-0.00189+0.0116 y-0.026 y^2+0.0276 y^3-0.0157 y^4+0.00509 y^5-0.000965 y^6+0.000102 y^7-5.31\times10^{-6} y^8+1.1\times10^{-7} y^9) $

\vspace{5mm}

\subsubsection*{GCN5082 Expanded}
 
$F_{M}^{(p,p)}= e^{-2 y} (3020.-11000. y+15500. y^2-10900. y^3+4200. y^4-907. y^5+108. y^6-6.46 y^7+0.162 y^8-0.000701 y^9+8.42\times10^{-7} y^{10})$

$F_{\Sigma''}^{(p,p)}=  e^{-2 y} (0.193-0.121 y+0.00697 y^2+0.175 y^3-0.00996 y^4-0.0217 y^5+0.0172 y^6-0.0029 y^7+0.000786 y^8+3.71\times10^{-8} y^9+3.42\times10^{-12} y^{10})$

$F_{\Sigma'}^{(p,p)}= e^{-2 y} (0.386-2.97 y+7.97 y^2-8.97 y^3+5.13 y^4-1.6 y^5+0.284 y^6-0.0276 y^7+0.0017 y^8+3.15\times10^{-7} y^9+8.5\times10^{-11} y^{10}) $

$F_{\Phi''}^{(p,p)}=e^{-2 y} (91.3-223. y+207. y^2-92.9 y^3+21.6 y^4-2.55 y^5+0.131 y^6-0.00154 y^7+5.26\times10^{-6} y^8) $

$F_{\tilde{\Phi}'}^{(p,p)}= e^{-2 y} (0.015-0.00813 y+0.000808 y^2+0.00263 y^3-8.21\times10^{-6} y^4-0.0000477 y^5+0.0000242 y^6+5.25\times10^{-7} y^7+4.03\times10^{-9} y^8)  $

$F_{\Delta}^{(p,p)}= e^{-2 y} (2.74-6.58 y+6.44 y^2-2.97 y^3+0.703 y^4-0.0812 y^5+0.00379 y^6-2.48\times10^{-6} y^7+5.93\times10^{-10} y^8) $

$F_{M \Phi''}^{(p,p)}= e^{-2 y} (-525.+1600. y-1850. y^2+1050. y^3-319. y^4+52.3 y^5-4.34 y^6+0.151 y^7-0.00118 y^8+2.1\times10^{-6} y^9)$

$F_{\Sigma' \Delta}^{(p,p)}= e^{-2 y} (1.03-5.18 y+8.17 y^2-5.96 y^3+2.22 y^4-0.447 y^5+0.046 y^6-0.00218 y^7+2.02\times10^{-8} y^8+2.24\times10^{-10} y^9) $

\vspace{5mm}

$F_{M}^{(p,n)}= e^{-2 y} (4280.-17000. y+26400. y^2-20900. y^3+9250. y^4-2370. y^5+348. y^6-27.8 y^7+1.05 y^8-0.0137 y^9+0.0000284 y^{10})  $

$F_{\Sigma''}^{(p,n)}= e^{-2 y} (-0.000933+0.00115 y-0.00851 y^2+0.00811 y^3-0.00352 y^4+0.0000644 y^5+0.00036 y^6-0.000128 y^7+0.0000185 y^8-8.1\times10^{-7} y^9+1.54\times10^{-11} y^{10}) $

$F_{\Sigma'}^{(p,n)}= e^{-2 y} (-0.00187+0.0274 y-0.124 y^2+0.206 y^3-0.16 y^4+0.0672 y^5-0.0161 y^6+0.00221 y^7-0.000159 y^8+4.77\times10^{-6} y^9+1.01\times10^{-9} y^{10}) $

$F_{\Phi''}^{(p,n)}= e^{-2 y} (176.-490. y+524. y^2-276. y^3+78.1 y^4-12. y^5+0.935 y^6-0.0303 y^7+0.000177 y^8) $

$F_{\tilde{\Phi}'}^{(p,n)}= e^{-2 y} (-0.0278+0.0407 y-0.0221 y^2+0.00217 y^3+0.00162 y^4-0.000525 y^5+0.0000548 y^6-1.43\times10^{-6} y^7-3.34\times10^{-8} y^8)$

$F_{\Delta}^{(p,n)}=  e^{-2 y} (0.188-0.488 y+0.526 y^2-0.287 y^3+0.0828 y^4-0.0124 y^5+0.000881 y^6-0.0000216 y^7+7.34\times10^{-9} y^8) $

$F_{M \Phi''}^{(p,n)}= e^{-2 y} (-1010.+3430. y-4490. y^2+2940. y^3-1060. y^4+215. y^5-24. y^6+1.36 y^7-0.0314 y^8+0.0000709 y^9)$

$F_{\Sigma' \Delta}^{(p,n)}= e^{-2 y} (0.0706-0.37 y+0.625 y^2-0.512 y^3+0.225 y^4-0.0545 y^5+0.00706 y^6-0.000454 y^7+0.0000113 y^8+2.73\times10^{-9} y^9) $

\vspace{5mm}

$F_{M \Phi''}^{(n,p)}= e^{-2 y} (-743.+2510. y-3250. y^2+2110. y^3-751. y^4+150. y^5-16.3 y^6+0.848 y^7-0.015 y^8+0.0000709 y^9) $

$F_{\Sigma' \Delta}^{(n,p)}= e^{-2 y} (-0.00497+0.0599 y-0.158 y^2+0.164 y^3-0.0838 y^4+0.0234 y^5-0.00359 y^6+0.00028 y^7-8.51\times10^{-6} y^8+2.73\times10^{-9} y^9 )$

\vspace{5mm}

$F_{M}^{(n,n)}=e^{-2 y} (6050.-26100. y+44400. y^2-39100. y^3+19600. y^4-5850. y^5+1040. y^6-105. y^7+5.58 y^8-0.125 y^9+0.000963 y^{10}) $

$F_{\Sigma''}^{(n,n)}= e^{-2 y} (4.51\times10^{-6}-8.31\times10^{-6} y+0.000803 y^2-0.00169 y^3+0.0016 y^4-0.000798 y^5+0.000224 y^6-0.0000347 y^7+2.67\times10^{-6} y^8-7.76\times10^{-8} y^9+1.46\times10^{-9} y^{10}) $

$F_{\Sigma'}^{(n,n)}= e^{-2 y} (9.02\times10^{-6}-0.000196 y+0.00242 y^2-0.00533 y^3+0.00506 y^4-0.00254 y^5+0.000758 y^6-0.000139 y^7+0.0000153 y^8- 8.33\times10^{-7} y^9+1.95\times10^{-8} y^{10}) $

$F_{\Phi''}^{(n,n)}= e^{-2 y} (339.-1060. y+1290. y^2-788. y^3+265. y^4-50.4 y^5+5.32 y^6-0.286 y^7+0.00602 y^8) $

$F_{\tilde{\Phi}'}^{(n,n)}= e^{-2 y} (0.0514-0.123 y+0.123 y^2-0.0637 y^3+0.0189 y^4-0.00332 y^5+0.000335 y^6-0.0000177 y^7+3.75\times10^{-7} y^8)$

$F_{\Delta}^{(n,n)}= e^{-2 y} (0.0129-0.036 y+0.0427 y^2-0.0268 y^3+0.00939 y^4-0.0018 y^5+0.000175 y^6-7.44\times10^{-6} y^7+1.25\times10^{-7} y^8) $

$F_{M \Phi''}^{(n,n)}=e^{-2 y} (-1430.+5330. y-7740. y^2+5720. y^3-2370. y^4+572. y^5-79.4 y^6+6. y^7-0.213 y^8+0.00241 y^9) $

$F_{\Sigma' \Delta}^{(n,n)}= e^{-2 y} (-0.000341+0.00418 y-0.0111 y^2+0.0125 y^3-0.00739 y^4+0.00247 y^5-0.000477 y^6+0.0000503 y^7-2.49\times10^{-6} y^8+4.92\times10^{-8} y^9) $

\vspace{5mm}

\subsection*{$^{133}$Cs Occupation Number Table}\label{Cs Occupation Table}

The occupation numbers $n^N_{nlj}$ of each valence single-nucleon orbital are given in order of increasing orbit, with $N=\{p,n\}$ denoting protons and neutrons respectively, and $n$ begins at $1$.  ``Max Occ" refers to the maximum occupation number allowed for each orbital. ``\#2'' refers the ``Expanded'' truncation schemes.

\begin{table}[H]
\centering
\begin{tabular}{lccc|ccc|ccc|ccc|cc}
\hline \hline 
 &  &  & & \multicolumn{2}{c}{SN100PN} & & \multicolumn{2}{c}{SN100PN(\#2)} & & \multicolumn{2}{c}{GCN5082} & & \multicolumn{2}{c}{GCN5082(\#2)}  \\
\cline{5-6}
\cline{8-9}
\cline{11-12}
\cline{14-15}
$n$  & $l$ & $j$ & {\footnotesize Max Occ} & $n^p_{nlj}$ & $n^n_{nlj}$ & & $n^p_{nlj}$ &  $n^n_{nlj}$ & &$n^p_{nlj}$ & $n^n_{nlj}$ & &  $n^p_{nlj}$ & $n^n_{nlj}$\\
\hline 
1   & 4  &  7/2 & 8 & 3.32  & 7.73 & & 3.36 & 7.75 &  & 2.90 & 7.71 &  & 2.96 & 7.70 \\
2  &  2 &  5/2 &  6 & 1.04 & 6.00 & & 1.03 & 5.67 & & 1.38 & 6.00 & & 1.35 & 5.73 \\
2   & 2 &  3/2 &  4 &  0.23 & 2.63 & & 0.22 & 2.77 & & 0.30 &  2.68 & & 0.27 & 2.77 \\
3  &  0 &  1/2 & 2 & 0.14 & 1.61 &  & 0.12 & 1.51 & & 0.14 &  1.73 & & 0.12 & 1.64 \\
1  &  5 &  11/2 &  12 & 0.26 & 10.02 & & 0.28 & 10.28 & & 0.28  & 9.87  & & 0.30 & 10.14 \\
\hline \hline
\end{tabular}
\end{table}

\newpage

\twocolumngrid

\bibliography{main}

\begin{thebibliography}{70}%
\makeatletter
\providecommand \@ifxundefined [1]{%
 \@ifx{#1\undefined}
}%
\providecommand \@ifnum [1]{%
 \ifnum #1\expandafter \@firstoftwo
 \else \expandafter \@secondoftwo
 \fi
}%
\providecommand \@ifx [1]{%
 \ifx #1\expandafter \@firstoftwo
 \else \expandafter \@secondoftwo
 \fi
}%
\providecommand \natexlab [1]{#1}%
\providecommand \enquote  [1]{``#1''}%
\providecommand \bibnamefont  [1]{#1}%
\providecommand \bibfnamefont [1]{#1}%
\providecommand \citenamefont [1]{#1}%
\providecommand \href@noop [0]{\@secondoftwo}%
\providecommand \href [0]{\begingroup \@sanitize@url \@href}%
\providecommand \@href[1]{\@@startlink{#1}\@@href}%
\providecommand \@@href[1]{\endgroup#1\@@endlink}%
\providecommand \@sanitize@url [0]{\catcode `\\12\catcode `\$12\catcode `\&12\catcode `\#12\catcode `\^12\catcode `\_12\catcode `\%12\relax}%
\providecommand \@@startlink[1]{}%
\providecommand \@@endlink[0]{}%
\providecommand \url  [0]{\begingroup\@sanitize@url \@url }%
\providecommand \@url [1]{\endgroup\@href {#1}{\urlprefix }}%
\providecommand \urlprefix  [0]{URL }%
\providecommand \Eprint [0]{\href }%
\providecommand \doibase [0]{https://doi.org/}%
\providecommand \selectlanguage [0]{\@gobble}%
\providecommand \bibinfo  [0]{\@secondoftwo}%
\providecommand \bibfield  [0]{\@secondoftwo}%
\providecommand \translation [1]{[#1]}%
\providecommand \BibitemOpen [0]{}%
\providecommand \bibitemStop [0]{}%
\providecommand \bibitemNoStop [0]{.\EOS\space}%
\providecommand \EOS [0]{\spacefactor3000\relax}%
\providecommand \BibitemShut  [1]{\csname bibitem#1\endcsname}%
\let\auto@bib@innerbib\@empty
\bibitem [{\citenamefont {Akimov}\ \emph {et~al.}(2017)\citenamefont {Akimov} \emph {et~al.}}]{COHERENT:2017ipa}%
  \BibitemOpen
  \bibfield  {author} {\bibinfo {author} {\bibfnamefont {D.}~\bibnamefont {Akimov}} \emph {et~al.} (\bibinfo {collaboration} {COHERENT}),\ }\bibfield  {title} {\bibinfo {title} {{Observation of Coherent Elastic Neutrino-Nucleus Scattering}},\ }\href {https://doi.org/10.1126/science.aao0990} {\bibfield  {journal} {\bibinfo  {journal} {Science}\ }\textbf {\bibinfo {volume} {357}},\ \bibinfo {pages} {1123} (\bibinfo {year} {2017})},\ \Eprint {https://arxiv.org/abs/1708.01294} {arXiv:1708.01294 [nucl-ex]} \BibitemShut {NoStop}%
\bibitem [{\citenamefont {Akimov}\ \emph {et~al.}(2021{\natexlab{a}})\citenamefont {Akimov} \emph {et~al.}}]{COHERENT:2020iec}%
  \BibitemOpen
  \bibfield  {author} {\bibinfo {author} {\bibfnamefont {D.}~\bibnamefont {Akimov}} \emph {et~al.} (\bibinfo {collaboration} {COHERENT}),\ }\bibfield  {title} {\bibinfo {title} {{First Measurement of Coherent Elastic Neutrino-Nucleus Scattering on Argon}},\ }\href {https://doi.org/10.1103/PhysRevLett.126.012002} {\bibfield  {journal} {\bibinfo  {journal} {Phys. Rev. Lett.}\ }\textbf {\bibinfo {volume} {126}},\ \bibinfo {pages} {012002} (\bibinfo {year} {2021}{\natexlab{a}})},\ \Eprint {https://arxiv.org/abs/2003.10630} {arXiv:2003.10630 [nucl-ex]} \BibitemShut {NoStop}%
\bibitem [{\citenamefont {Adamski}\ \emph {et~al.}(2024)\citenamefont {Adamski} \emph {et~al.}}]{Adamski:2024yqt}%
  \BibitemOpen
  \bibfield  {author} {\bibinfo {author} {\bibfnamefont {S.}~\bibnamefont {Adamski}} \emph {et~al.},\ }\href@noop {} {\bibinfo {title} {{First detection of coherent elastic neutrino-nucleus scattering on germanium}}} (\bibinfo {year} {2024}),\ \Eprint {https://arxiv.org/abs/2406.13806} {arXiv:2406.13806 [hep-ex]} \BibitemShut {NoStop}%
\bibitem [{\citenamefont {Bowen}\ and\ \citenamefont {Huber}(2020)}]{Bowen:2020unj}%
  \BibitemOpen
  \bibfield  {author} {\bibinfo {author} {\bibfnamefont {M.}~\bibnamefont {Bowen}}\ and\ \bibinfo {author} {\bibfnamefont {P.}~\bibnamefont {Huber}},\ }\bibfield  {title} {\bibinfo {title} {{Reactor neutrino applications and coherent elastic neutrino nucleus scattering}},\ }\href {https://doi.org/10.1103/PhysRevD.102.053008} {\bibfield  {journal} {\bibinfo  {journal} {Phys. Rev. D}\ }\textbf {\bibinfo {volume} {102}},\ \bibinfo {pages} {053008} (\bibinfo {year} {2020})},\ \Eprint {https://arxiv.org/abs/2005.10907} {arXiv:2005.10907 [physics.ins-det]} \BibitemShut {NoStop}%
\bibitem [{\citenamefont {Cadeddu}\ \emph {et~al.}(2018)\citenamefont {Cadeddu}, \citenamefont {Giunti}, \citenamefont {Li},\ and\ \citenamefont {Zhang}}]{Cadeddu:2017etk}%
  \BibitemOpen
  \bibfield  {author} {\bibinfo {author} {\bibfnamefont {M.}~\bibnamefont {Cadeddu}}, \bibinfo {author} {\bibfnamefont {C.}~\bibnamefont {Giunti}}, \bibinfo {author} {\bibfnamefont {Y.~F.}\ \bibnamefont {Li}},\ and\ \bibinfo {author} {\bibfnamefont {Y.~Y.}\ \bibnamefont {Zhang}},\ }\bibfield  {title} {\bibinfo {title} {{Average CsI neutron density distribution from COHERENT data}},\ }\href {https://doi.org/10.1103/PhysRevLett.120.072501} {\bibfield  {journal} {\bibinfo  {journal} {Phys. Rev. Lett.}\ }\textbf {\bibinfo {volume} {120}},\ \bibinfo {pages} {072501} (\bibinfo {year} {2018})},\ \Eprint {https://arxiv.org/abs/1710.02730} {arXiv:1710.02730 [hep-ph]} \BibitemShut {NoStop}%
\bibitem [{\citenamefont {Papoulias}\ \emph {et~al.}(2020)\citenamefont {Papoulias}, \citenamefont {Kosmas}, \citenamefont {Sahu}, \citenamefont {Kota},\ and\ \citenamefont {Hota}}]{Papoulias:2019lfi}%
  \BibitemOpen
  \bibfield  {author} {\bibinfo {author} {\bibfnamefont {D.~K.}\ \bibnamefont {Papoulias}}, \bibinfo {author} {\bibfnamefont {T.~S.}\ \bibnamefont {Kosmas}}, \bibinfo {author} {\bibfnamefont {R.}~\bibnamefont {Sahu}}, \bibinfo {author} {\bibfnamefont {V.~K.~B.}\ \bibnamefont {Kota}},\ and\ \bibinfo {author} {\bibfnamefont {M.}~\bibnamefont {Hota}},\ }\bibfield  {title} {\bibinfo {title} {{Constraining nuclear physics parameters with current and future COHERENT data}},\ }\href {https://doi.org/10.1016/j.physletb.2019.135133} {\bibfield  {journal} {\bibinfo  {journal} {Phys. Lett. B}\ }\textbf {\bibinfo {volume} {800}},\ \bibinfo {pages} {135133} (\bibinfo {year} {2020})},\ \Eprint {https://arxiv.org/abs/1903.03722} {arXiv:1903.03722 [hep-ph]} \BibitemShut {NoStop}%
\bibitem [{\citenamefont {Coloma}\ \emph {et~al.}(2020)\citenamefont {Coloma}, \citenamefont {Esteban}, \citenamefont {Gonzalez-Garcia},\ and\ \citenamefont {Menendez}}]{Coloma:2020nhf}%
  \BibitemOpen
  \bibfield  {author} {\bibinfo {author} {\bibfnamefont {P.}~\bibnamefont {Coloma}}, \bibinfo {author} {\bibfnamefont {I.}~\bibnamefont {Esteban}}, \bibinfo {author} {\bibfnamefont {M.~C.}\ \bibnamefont {Gonzalez-Garcia}},\ and\ \bibinfo {author} {\bibfnamefont {J.}~\bibnamefont {Menendez}},\ }\bibfield  {title} {\bibinfo {title} {{Determining the nuclear neutron distribution from Coherent Elastic neutrino-Nucleus Scattering: current results and future prospects}},\ }\href {https://doi.org/10.1007/JHEP08(2020)030} {\bibfield  {journal} {\bibinfo  {journal} {JHEP}\ }\textbf {\bibinfo {volume} {08}}\bibfield  {number} {\bibinfo  {number} { (08)},\ \bibinfo {pages} {030}},\ }\Eprint {https://arxiv.org/abs/2006.08624} {arXiv:2006.08624 [hep-ph]} \BibitemShut {NoStop}%
\bibitem [{\citenamefont {Bisset}\ \emph {et~al.}(2023)\citenamefont {Bisset}, \citenamefont {Dutta}, \citenamefont {Huang},\ and\ \citenamefont {Strigari}}]{Bisset:2023oxt}%
  \BibitemOpen
  \bibfield  {author} {\bibinfo {author} {\bibfnamefont {I.~A.}\ \bibnamefont {Bisset}}, \bibinfo {author} {\bibfnamefont {B.}~\bibnamefont {Dutta}}, \bibinfo {author} {\bibfnamefont {W.-C.}\ \bibnamefont {Huang}},\ and\ \bibinfo {author} {\bibfnamefont {L.~E.}\ \bibnamefont {Strigari}},\ }\bibfield  {title} {\bibinfo {title} {{Short Baseline Neutrino Anomalies at Stopped Pion Experiments}},\ }\href@noop {} {\  (\bibinfo {year} {2023})},\ \Eprint {https://arxiv.org/abs/2310.13194} {arXiv:2310.13194 [hep-ph]} \BibitemShut {NoStop}%
\bibitem [{\citenamefont {Denton}\ \emph {et~al.}(2018)\citenamefont {Denton}, \citenamefont {Farzan},\ and\ \citenamefont {Shoemaker}}]{Denton:2018xmq}%
  \BibitemOpen
  \bibfield  {author} {\bibinfo {author} {\bibfnamefont {P.~B.}\ \bibnamefont {Denton}}, \bibinfo {author} {\bibfnamefont {Y.}~\bibnamefont {Farzan}},\ and\ \bibinfo {author} {\bibfnamefont {I.~M.}\ \bibnamefont {Shoemaker}},\ }\bibfield  {title} {\bibinfo {title} {{Testing large non-standard neutrino interactions with arbitrary mediator mass after COHERENT data}},\ }\href {https://doi.org/10.1007/JHEP07(2018)037} {\bibfield  {journal} {\bibinfo  {journal} {JHEP}\ }\textbf {\bibinfo {volume} {07}},\ \bibinfo {pages} {037}},\ \Eprint {https://arxiv.org/abs/1804.03660} {arXiv:1804.03660 [hep-ph]} \BibitemShut {NoStop}%
\bibitem [{\citenamefont {Coloma}\ \emph {et~al.}(2017)\citenamefont {Coloma}, \citenamefont {Gonzalez-Garcia}, \citenamefont {Maltoni},\ and\ \citenamefont {Schwetz}}]{Coloma:2017ncl}%
  \BibitemOpen
  \bibfield  {author} {\bibinfo {author} {\bibfnamefont {P.}~\bibnamefont {Coloma}}, \bibinfo {author} {\bibfnamefont {M.~C.}\ \bibnamefont {Gonzalez-Garcia}}, \bibinfo {author} {\bibfnamefont {M.}~\bibnamefont {Maltoni}},\ and\ \bibinfo {author} {\bibfnamefont {T.}~\bibnamefont {Schwetz}},\ }\bibfield  {title} {\bibinfo {title} {{COHERENT Enlightenment of the Neutrino Dark Side}},\ }\href {https://doi.org/10.1103/PhysRevD.96.115007} {\bibfield  {journal} {\bibinfo  {journal} {Phys. Rev. D}\ }\textbf {\bibinfo {volume} {96}},\ \bibinfo {pages} {115007} (\bibinfo {year} {2017})},\ \Eprint {https://arxiv.org/abs/1708.02899} {arXiv:1708.02899 [hep-ph]} \BibitemShut {NoStop}%
\bibitem [{\citenamefont {Dent}\ \emph {et~al.}(2018)\citenamefont {Dent}, \citenamefont {Dutta}, \citenamefont {Liao}, \citenamefont {Newstead}, \citenamefont {Strigari},\ and\ \citenamefont {Walker}}]{Dent:2017mpr}%
  \BibitemOpen
  \bibfield  {author} {\bibinfo {author} {\bibfnamefont {J.~B.}\ \bibnamefont {Dent}}, \bibinfo {author} {\bibfnamefont {B.}~\bibnamefont {Dutta}}, \bibinfo {author} {\bibfnamefont {S.}~\bibnamefont {Liao}}, \bibinfo {author} {\bibfnamefont {J.~L.}\ \bibnamefont {Newstead}}, \bibinfo {author} {\bibfnamefont {L.~E.}\ \bibnamefont {Strigari}},\ and\ \bibinfo {author} {\bibfnamefont {J.~W.}\ \bibnamefont {Walker}},\ }\bibfield  {title} {\bibinfo {title} {{Accelerator and reactor complementarity in coherent neutrino-nucleus scattering}},\ }\href {https://doi.org/10.1103/PhysRevD.97.035009} {\bibfield  {journal} {\bibinfo  {journal} {Phys. Rev. D}\ }\textbf {\bibinfo {volume} {97}},\ \bibinfo {pages} {035009} (\bibinfo {year} {2018})},\ \Eprint {https://arxiv.org/abs/1711.03521} {arXiv:1711.03521 [hep-ph]} \BibitemShut {NoStop}%
\bibitem [{\citenamefont {Kosmas}\ \emph {et~al.}(2017)\citenamefont {Kosmas}, \citenamefont {Papoulias}, \citenamefont {T\'ortola},\ and\ \citenamefont {Valle}}]{PhysRevD.96.063013}%
  \BibitemOpen
  \bibfield  {author} {\bibinfo {author} {\bibfnamefont {T.~S.}\ \bibnamefont {Kosmas}}, \bibinfo {author} {\bibfnamefont {D.~K.}\ \bibnamefont {Papoulias}}, \bibinfo {author} {\bibfnamefont {M.}~\bibnamefont {T\'ortola}},\ and\ \bibinfo {author} {\bibfnamefont {J.~W.~F.}\ \bibnamefont {Valle}},\ }\bibfield  {title} {\bibinfo {title} {Probing light sterile neutrino signatures at reactor and spallation neutron source neutrino experiments},\ }\href {https://doi.org/10.1103/PhysRevD.96.063013} {\bibfield  {journal} {\bibinfo  {journal} {Phys. Rev. D}\ }\textbf {\bibinfo {volume} {96}},\ \bibinfo {pages} {063013} (\bibinfo {year} {2017})}\BibitemShut {NoStop}%
\bibitem [{\citenamefont {Billard}\ \emph {et~al.}(2014)\citenamefont {Billard}, \citenamefont {Strigari},\ and\ \citenamefont {Figueroa-Feliciano}}]{Billard:2013qya}%
  \BibitemOpen
  \bibfield  {author} {\bibinfo {author} {\bibfnamefont {J.}~\bibnamefont {Billard}}, \bibinfo {author} {\bibfnamefont {L.}~\bibnamefont {Strigari}},\ and\ \bibinfo {author} {\bibfnamefont {E.}~\bibnamefont {Figueroa-Feliciano}},\ }\bibfield  {title} {\bibinfo {title} {{Implication of neutrino backgrounds on the reach of next generation dark matter direct detection experiments}},\ }\href {https://doi.org/10.1103/PhysRevD.89.023524} {\bibfield  {journal} {\bibinfo  {journal} {Phys. Rev. D}\ }\textbf {\bibinfo {volume} {89}},\ \bibinfo {pages} {023524} (\bibinfo {year} {2014})},\ \Eprint {https://arxiv.org/abs/1307.5458} {arXiv:1307.5458 [hep-ph]} \BibitemShut {NoStop}%
\bibitem [{\citenamefont {O'Hare}(2021)}]{OHare:2021utq}%
  \BibitemOpen
  \bibfield  {author} {\bibinfo {author} {\bibfnamefont {C.~A.~J.}\ \bibnamefont {O'Hare}},\ }\bibfield  {title} {\bibinfo {title} {{New Definition of the Neutrino Floor for Direct Dark Matter Searches}},\ }\href {https://doi.org/10.1103/PhysRevLett.127.251802} {\bibfield  {journal} {\bibinfo  {journal} {Phys. Rev. Lett.}\ }\textbf {\bibinfo {volume} {127}},\ \bibinfo {pages} {251802} (\bibinfo {year} {2021})},\ \Eprint {https://arxiv.org/abs/2109.03116} {arXiv:2109.03116 [hep-ph]} \BibitemShut {NoStop}%
\bibitem [{\citenamefont {Aprile}\ \emph {et~al.}(2024)\citenamefont {Aprile} \emph {et~al.}}]{XENON:2024ijk}%
  \BibitemOpen
  \bibfield  {author} {\bibinfo {author} {\bibfnamefont {E.}~\bibnamefont {Aprile}} \emph {et~al.} (\bibinfo {collaboration} {XENON}),\ }\href@noop {} {\bibinfo {title} {{First Measurement of Solar $^8$B Neutrinos via Coherent Elastic Neutrino-Nucleus Scattering with XENONnT}}} (\bibinfo {year} {2024}),\ \Eprint {https://arxiv.org/abs/2408.02877} {arXiv:2408.02877 [nucl-ex]} \BibitemShut {NoStop}%
\bibitem [{\citenamefont {Bo}\ \emph {et~al.}(2024)\citenamefont {Bo} \emph {et~al.}}]{PandaX:2024muv}%
  \BibitemOpen
  \bibfield  {author} {\bibinfo {author} {\bibfnamefont {Z.}~\bibnamefont {Bo}} \emph {et~al.} (\bibinfo {collaboration} {PandaX}),\ }\href@noop {} {\bibinfo {title} {{First Indication of Solar $^8$B Neutrino Flux through Coherent Elastic Neutrino-Nucleus Scattering in PandaX-4T}}} (\bibinfo {year} {2024}),\ \Eprint {https://arxiv.org/abs/2407.10892} {arXiv:2407.10892 [hep-ex]} \BibitemShut {NoStop}%
\bibitem [{\citenamefont {Akimov}\ \emph {et~al.}(2021{\natexlab{b}})\citenamefont {Akimov} \emph {et~al.}}]{COHERENT:2021xhx}%
  \BibitemOpen
  \bibfield  {author} {\bibinfo {author} {\bibfnamefont {D.}~\bibnamefont {Akimov}} \emph {et~al.} (\bibinfo {collaboration} {COHERENT}),\ }\bibfield  {title} {\bibinfo {title} {{A D$_2$O detector for flux normalization of a pion decay-at-rest neutrino source}},\ }\href {https://doi.org/10.1088/1748-0221/16/08/P08048} {\bibfield  {journal} {\bibinfo  {journal} {JINST}\ }\textbf {\bibinfo {volume} {16}}\bibfield  {number} {\bibinfo  {number} { (08)},\ \bibinfo {pages} {P08048}},\ }\Eprint {https://arxiv.org/abs/2104.09605} {arXiv:2104.09605 [physics.ins-det]} \BibitemShut {NoStop}%
\bibitem [{\citenamefont {Hoferichter}\ \emph {et~al.}(2020)\citenamefont {Hoferichter}, \citenamefont {Men\'endez},\ and\ \citenamefont {Schwenk}}]{PhysRevD.102.074018}%
  \BibitemOpen
  \bibfield  {author} {\bibinfo {author} {\bibfnamefont {M.}~\bibnamefont {Hoferichter}}, \bibinfo {author} {\bibfnamefont {J.}~\bibnamefont {Men\'endez}},\ and\ \bibinfo {author} {\bibfnamefont {A.}~\bibnamefont {Schwenk}},\ }\bibfield  {title} {\bibinfo {title} {Coherent elastic neutrino-nucleus scattering: Eft analysis and nuclear responses},\ }\href {https://doi.org/10.1103/PhysRevD.102.074018} {\bibfield  {journal} {\bibinfo  {journal} {Phys. Rev. D}\ }\textbf {\bibinfo {volume} {102}},\ \bibinfo {pages} {074018} (\bibinfo {year} {2020})}\BibitemShut {NoStop}%
\bibitem [{\citenamefont {Payne}\ \emph {et~al.}(2019)\citenamefont {Payne}, \citenamefont {Bacca}, \citenamefont {Hagen}, \citenamefont {Jiang},\ and\ \citenamefont {Papenbrock}}]{Payne:2019wvy}%
  \BibitemOpen
  \bibfield  {author} {\bibinfo {author} {\bibfnamefont {C.~G.}\ \bibnamefont {Payne}}, \bibinfo {author} {\bibfnamefont {S.}~\bibnamefont {Bacca}}, \bibinfo {author} {\bibfnamefont {G.}~\bibnamefont {Hagen}}, \bibinfo {author} {\bibfnamefont {W.}~\bibnamefont {Jiang}},\ and\ \bibinfo {author} {\bibfnamefont {T.}~\bibnamefont {Papenbrock}},\ }\bibfield  {title} {\bibinfo {title} {{Coherent elastic neutrino-nucleus scattering on $^{40}$Ar from first principles}},\ }\href {https://doi.org/10.1103/PhysRevC.100.061304} {\bibfield  {journal} {\bibinfo  {journal} {Phys. Rev. C}\ }\textbf {\bibinfo {volume} {100}},\ \bibinfo {pages} {061304} (\bibinfo {year} {2019})},\ \Eprint {https://arxiv.org/abs/1908.09739} {arXiv:1908.09739 [nucl-th]} \BibitemShut {NoStop}%
\bibitem [{\citenamefont {Van~Dessel}\ \emph {et~al.}(2023)\citenamefont {Van~Dessel}, \citenamefont {Pandey}, \citenamefont {Ray},\ and\ \citenamefont {Jachowicz}}]{VanDessel:2020epd}%
  \BibitemOpen
  \bibfield  {author} {\bibinfo {author} {\bibfnamefont {N.}~\bibnamefont {Van~Dessel}}, \bibinfo {author} {\bibfnamefont {V.}~\bibnamefont {Pandey}}, \bibinfo {author} {\bibfnamefont {H.}~\bibnamefont {Ray}},\ and\ \bibinfo {author} {\bibfnamefont {N.}~\bibnamefont {Jachowicz}},\ }\bibfield  {title} {\bibinfo {title} {{Cross Sections for Coherent Elastic and Inelastic Neutrino-Nucleus Scattering}},\ }\href {https://doi.org/10.3390/universe9050207} {\bibfield  {journal} {\bibinfo  {journal} {Universe}\ }\textbf {\bibinfo {volume} {9}},\ \bibinfo {pages} {207} (\bibinfo {year} {2023})},\ \Eprint {https://arxiv.org/abs/2007.03658} {arXiv:2007.03658 [nucl-th]} \BibitemShut {NoStop}%
\bibitem [{\citenamefont {Freedman}(1974)}]{PhysRevD.9.1389}%
  \BibitemOpen
  \bibfield  {author} {\bibinfo {author} {\bibfnamefont {D.~Z.}\ \bibnamefont {Freedman}},\ }\bibfield  {title} {\bibinfo {title} {Coherent effects of a weak neutral current},\ }\href {https://doi.org/10.1103/PhysRevD.9.1389} {\bibfield  {journal} {\bibinfo  {journal} {Phys. Rev. D}\ }\textbf {\bibinfo {volume} {9}},\ \bibinfo {pages} {1389} (\bibinfo {year} {1974})}\BibitemShut {NoStop}%
\bibitem [{\citenamefont {Lewin}\ and\ \citenamefont {Smith}(1996)}]{Lewin:1995rx}%
  \BibitemOpen
  \bibfield  {author} {\bibinfo {author} {\bibfnamefont {J.~D.}\ \bibnamefont {Lewin}}\ and\ \bibinfo {author} {\bibfnamefont {P.~F.}\ \bibnamefont {Smith}},\ }\bibfield  {title} {\bibinfo {title} {{Review of mathematics, numerical factors, and corrections for dark matter experiments based on elastic nuclear recoil}},\ }\href {https://doi.org/10.1016/S0927-6505(96)00047-3} {\bibfield  {journal} {\bibinfo  {journal} {Astropart. Phys.}\ }\textbf {\bibinfo {volume} {6}},\ \bibinfo {pages} {87} (\bibinfo {year} {1996})}\BibitemShut {NoStop}%
\bibitem [{\citenamefont {Klein}\ and\ \citenamefont {Nystrand}(1999)}]{PhysRevC.60.014903}%
  \BibitemOpen
  \bibfield  {author} {\bibinfo {author} {\bibfnamefont {S.~R.}\ \bibnamefont {Klein}}\ and\ \bibinfo {author} {\bibfnamefont {J.}~\bibnamefont {Nystrand}},\ }\bibfield  {title} {\bibinfo {title} {Exclusive vector meson production in relativistic heavy ion collisions},\ }\href {https://doi.org/10.1103/PhysRevC.60.014903} {\bibfield  {journal} {\bibinfo  {journal} {Phys. Rev. C}\ }\textbf {\bibinfo {volume} {60}},\ \bibinfo {pages} {014903} (\bibinfo {year} {1999})}\BibitemShut {NoStop}%
\bibitem [{\citenamefont {Aristizabal~Sierra}\ \emph {et~al.}(2019)\citenamefont {Aristizabal~Sierra}, \citenamefont {Liao},\ and\ \citenamefont {Marfatia}}]{AristizabalSierra:2019zmy}%
  \BibitemOpen
  \bibfield  {author} {\bibinfo {author} {\bibfnamefont {D.}~\bibnamefont {Aristizabal~Sierra}}, \bibinfo {author} {\bibfnamefont {J.}~\bibnamefont {Liao}},\ and\ \bibinfo {author} {\bibfnamefont {D.}~\bibnamefont {Marfatia}},\ }\bibfield  {title} {\bibinfo {title} {{Impact of form factor uncertainties on interpretations of coherent elastic neutrino-nucleus scattering data}},\ }\href {https://doi.org/10.1007/JHEP06(2019)141} {\bibfield  {journal} {\bibinfo  {journal} {JHEP}\ }\textbf {\bibinfo {volume} {06}},\ \bibinfo {pages} {141}},\ \Eprint {https://arxiv.org/abs/1902.07398} {arXiv:1902.07398 [hep-ph]} \BibitemShut {NoStop}%
\bibitem [{\citenamefont {Tomalak}\ \emph {et~al.}(2021)\citenamefont {Tomalak}, \citenamefont {Machado}, \citenamefont {Pandey},\ and\ \citenamefont {Plestid}}]{Tomalak:2020zfh}%
  \BibitemOpen
  \bibfield  {author} {\bibinfo {author} {\bibfnamefont {O.}~\bibnamefont {Tomalak}}, \bibinfo {author} {\bibfnamefont {P.}~\bibnamefont {Machado}}, \bibinfo {author} {\bibfnamefont {V.}~\bibnamefont {Pandey}},\ and\ \bibinfo {author} {\bibfnamefont {R.}~\bibnamefont {Plestid}},\ }\bibfield  {title} {\bibinfo {title} {{Flavor-dependent radiative corrections in coherent elastic neutrino-nucleus scattering}},\ }\href {https://doi.org/10.1007/JHEP02(2021)097} {\bibfield  {journal} {\bibinfo  {journal} {JHEP}\ }\textbf {\bibinfo {volume} {02}},\ \bibinfo {pages} {097}},\ \Eprint {https://arxiv.org/abs/2011.05960} {arXiv:2011.05960 [hep-ph]} \BibitemShut {NoStop}%
\bibitem [{\citenamefont {Barranco}\ \emph {et~al.}(2005)\citenamefont {Barranco}, \citenamefont {Miranda},\ and\ \citenamefont {Rashba}}]{Barranco:2005yy}%
  \BibitemOpen
  \bibfield  {author} {\bibinfo {author} {\bibfnamefont {J.}~\bibnamefont {Barranco}}, \bibinfo {author} {\bibfnamefont {O.~G.}\ \bibnamefont {Miranda}},\ and\ \bibinfo {author} {\bibfnamefont {T.~I.}\ \bibnamefont {Rashba}},\ }\bibfield  {title} {\bibinfo {title} {{Probing new physics with coherent neutrino scattering off nuclei}},\ }\href {https://doi.org/10.1088/1126-6708/2005/12/021} {\bibfield  {journal} {\bibinfo  {journal} {JHEP}\ }\textbf {\bibinfo {volume} {12}},\ \bibinfo {pages} {021}},\ \Eprint {https://arxiv.org/abs/hep-ph/0508299} {arXiv:hep-ph/0508299} \BibitemShut {NoStop}%
\bibitem [{\citenamefont {Dent}\ \emph {et~al.}(2017)\citenamefont {Dent}, \citenamefont {Dutta}, \citenamefont {Liao}, \citenamefont {Newstead}, \citenamefont {Strigari},\ and\ \citenamefont {Walker}}]{Dent:2016wcr}%
  \BibitemOpen
  \bibfield  {author} {\bibinfo {author} {\bibfnamefont {J.~B.}\ \bibnamefont {Dent}}, \bibinfo {author} {\bibfnamefont {B.}~\bibnamefont {Dutta}}, \bibinfo {author} {\bibfnamefont {S.}~\bibnamefont {Liao}}, \bibinfo {author} {\bibfnamefont {J.~L.}\ \bibnamefont {Newstead}}, \bibinfo {author} {\bibfnamefont {L.~E.}\ \bibnamefont {Strigari}},\ and\ \bibinfo {author} {\bibfnamefont {J.~W.}\ \bibnamefont {Walker}},\ }\bibfield  {title} {\bibinfo {title} {{Probing light mediators at ultralow threshold energies with coherent elastic neutrino-nucleus scattering}},\ }\href {https://doi.org/10.1103/PhysRevD.96.095007} {\bibfield  {journal} {\bibinfo  {journal} {Phys. Rev. D}\ }\textbf {\bibinfo {volume} {96}},\ \bibinfo {pages} {095007} (\bibinfo {year} {2017})},\ \Eprint {https://arxiv.org/abs/1612.06350} {arXiv:1612.06350 [hep-ph]} \BibitemShut {NoStop}%
\bibitem [{\citenamefont {Angeli}\ and\ \citenamefont {Marinova}(2013)}]{Angeli:2013epw}%
  \BibitemOpen
  \bibfield  {author} {\bibinfo {author} {\bibfnamefont {I.}~\bibnamefont {Angeli}}\ and\ \bibinfo {author} {\bibfnamefont {K.~P.}\ \bibnamefont {Marinova}},\ }\bibfield  {title} {\bibinfo {title} {{Table of experimental nuclear ground state charge radii: An update}},\ }\href {https://doi.org/10.1016/j.adt.2011.12.006} {\bibfield  {journal} {\bibinfo  {journal} {Atom. Data Nucl. Data Tabl.}\ }\textbf {\bibinfo {volume} {99}},\ \bibinfo {pages} {69} (\bibinfo {year} {2013})}\BibitemShut {NoStop}%
\bibitem [{\citenamefont {Adhikari}\ \emph {et~al.}(2021)\citenamefont {Adhikari} \emph {et~al.}}]{PREX:2021umo}%
  \BibitemOpen
  \bibfield  {author} {\bibinfo {author} {\bibfnamefont {D.}~\bibnamefont {Adhikari}} \emph {et~al.} (\bibinfo {collaboration} {PREX}),\ }\bibfield  {title} {\bibinfo {title} {{Accurate Determination of the Neutron Skin Thickness of $^{208}$Pb through Parity-Violation in Electron Scattering}},\ }\href {https://doi.org/10.1103/PhysRevLett.126.172502} {\bibfield  {journal} {\bibinfo  {journal} {Phys. Rev. Lett.}\ }\textbf {\bibinfo {volume} {126}},\ \bibinfo {pages} {172502} (\bibinfo {year} {2021})},\ \Eprint {https://arxiv.org/abs/2102.10767} {arXiv:2102.10767 [nucl-ex]} \BibitemShut {NoStop}%
\bibitem [{\citenamefont {Kumar}(2020)}]{Kumar:2020ejz}%
  \BibitemOpen
  \bibfield  {author} {\bibinfo {author} {\bibfnamefont {K.~S.}\ \bibnamefont {Kumar}} (\bibinfo {collaboration} {PREX, CREX}),\ }\bibfield  {title} {\bibinfo {title} {{Electroweak probe of neutron skins of nuclei}},\ }\href {https://doi.org/10.1016/j.aop.2019.168012} {\bibfield  {journal} {\bibinfo  {journal} {Annals Phys.}\ }\textbf {\bibinfo {volume} {412}},\ \bibinfo {pages} {168012} (\bibinfo {year} {2020})}\BibitemShut {NoStop}%
\bibitem [{\citenamefont {Reed}\ \emph {et~al.}(2020)\citenamefont {Reed}, \citenamefont {Jaffe}, \citenamefont {Horowitz},\ and\ \citenamefont {Sfienti}}]{Reed:2020fdf}%
  \BibitemOpen
  \bibfield  {author} {\bibinfo {author} {\bibfnamefont {B.}~\bibnamefont {Reed}}, \bibinfo {author} {\bibfnamefont {Z.}~\bibnamefont {Jaffe}}, \bibinfo {author} {\bibfnamefont {C.~J.}\ \bibnamefont {Horowitz}},\ and\ \bibinfo {author} {\bibfnamefont {C.}~\bibnamefont {Sfienti}},\ }\bibfield  {title} {\bibinfo {title} {{Measuring the surface thickness of the weak charge density of nuclei}},\ }\href {https://doi.org/10.1103/PhysRevC.102.064308} {\bibfield  {journal} {\bibinfo  {journal} {Phys. Rev. C}\ }\textbf {\bibinfo {volume} {102}},\ \bibinfo {pages} {064308} (\bibinfo {year} {2020})},\ \Eprint {https://arxiv.org/abs/2009.06664} {arXiv:2009.06664 [nucl-th]} \BibitemShut {NoStop}%
\bibitem [{\citenamefont {Trzci\ifmmode~\acute{n}\else \'{n}\fi{}ska}\ \emph {et~al.}(2001)\citenamefont {Trzci\ifmmode~\acute{n}\else \'{n}\fi{}ska}, \citenamefont {Jastrz\ifmmode~\mbox{\c{e}}\else \c{e}\fi{}bski}, \citenamefont {Lubi\ifmmode~\acute{n}\else \'{n}\fi{}ski}, \citenamefont {Hartmann}, \citenamefont {Schmidt}, \citenamefont {von Egidy},\ and\ \citenamefont {K\l{}os}}]{PhysRevLett.87.082501}%
  \BibitemOpen
  \bibfield  {author} {\bibinfo {author} {\bibfnamefont {A.}~\bibnamefont {Trzci\ifmmode~\acute{n}\else \'{n}\fi{}ska}}, \bibinfo {author} {\bibfnamefont {J.}~\bibnamefont {Jastrz\ifmmode~\mbox{\c{e}}\else \c{e}\fi{}bski}}, \bibinfo {author} {\bibfnamefont {P.}~\bibnamefont {Lubi\ifmmode~\acute{n}\else \'{n}\fi{}ski}}, \bibinfo {author} {\bibfnamefont {F.~J.}\ \bibnamefont {Hartmann}}, \bibinfo {author} {\bibfnamefont {R.}~\bibnamefont {Schmidt}}, \bibinfo {author} {\bibfnamefont {T.}~\bibnamefont {von Egidy}},\ and\ \bibinfo {author} {\bibfnamefont {B.}~\bibnamefont {K\l{}os}},\ }\bibfield  {title} {\bibinfo {title} {Neutron density distributions deduced from antiprotonic atoms},\ }\href {https://doi.org/10.1103/PhysRevLett.87.082501} {\bibfield  {journal} {\bibinfo  {journal} {Phys. Rev. Lett.}\ }\textbf {\bibinfo {volume} {87}},\ \bibinfo {pages} {082501} (\bibinfo {year} {2001})}\BibitemShut {NoStop}%
\bibitem [{\citenamefont {Orrigo}\ \emph {et~al.}(2016)\citenamefont {Orrigo}, \citenamefont {Alvarez-Ruso},\ and\ \citenamefont {Peña-Garay}}]{ORRIGO2016414}%
  \BibitemOpen
  \bibfield  {author} {\bibinfo {author} {\bibfnamefont {S.}~\bibnamefont {Orrigo}}, \bibinfo {author} {\bibfnamefont {L.}~\bibnamefont {Alvarez-Ruso}},\ and\ \bibinfo {author} {\bibfnamefont {C.}~\bibnamefont {Peña-Garay}},\ }\bibfield  {title} {\bibinfo {title} {A new approach to nuclear form factors for direct dark matter searches},\ }\href {https://doi.org/https://doi.org/10.1016/j.nuclphysbps.2015.09.060} {\bibfield  {journal} {\bibinfo  {journal} {Nuclear and Particle Physics Proceedings}\ }\textbf {\bibinfo {volume} {273-275}},\ \bibinfo {pages} {414} (\bibinfo {year} {2016})},\ \bibinfo {note} {37th International Conference on High Energy Physics (ICHEP)}\BibitemShut {NoStop}%
\bibitem [{\citenamefont {Reinhard}\ \emph {et~al.}(2021)\citenamefont {Reinhard}, \citenamefont {Schuetrumpf},\ and\ \citenamefont {Maruhn}}]{reinhard2021}%
  \BibitemOpen
  \bibfield  {author} {\bibinfo {author} {\bibfnamefont {P.-G.}\ \bibnamefont {Reinhard}}, \bibinfo {author} {\bibfnamefont {B.}~\bibnamefont {Schuetrumpf}},\ and\ \bibinfo {author} {\bibfnamefont {J.~A.}\ \bibnamefont {Maruhn}},\ }\bibfield  {title} {\bibinfo {title} {The {A}xial {H}artree-{F}ock + {BCS} {C}ode {S}ky{A}x},\ }\href {https://doi.org/10.1016/j.cpc.2020.107603} {\bibfield  {journal} {\bibinfo  {journal} {Comput. Phys. Commun.}\ }\textbf {\bibinfo {volume} {258}},\ \bibinfo {pages} {107603} (\bibinfo {year} {2021})}\BibitemShut {NoStop}%
\bibitem [{\citenamefont {Sprung}\ and\ \citenamefont {Martorell}(1997)}]{DWLSprung_1997}%
  \BibitemOpen
  \bibfield  {author} {\bibinfo {author} {\bibfnamefont {D.~W.~L.}\ \bibnamefont {Sprung}}\ and\ \bibinfo {author} {\bibfnamefont {J.}~\bibnamefont {Martorell}},\ }\bibfield  {title} {\bibinfo {title} {The symmetrized fermi function and its transforms},\ }\href {https://doi.org/10.1088/0305-4470/30/18/026} {\bibfield  {journal} {\bibinfo  {journal} {Journal of Physics A: Mathematical and General}\ }\textbf {\bibinfo {volume} {30}},\ \bibinfo {pages} {6525} (\bibinfo {year} {1997})}\BibitemShut {NoStop}%
\bibitem [{\citenamefont {O'Connell}\ \emph {et~al.}(1972)\citenamefont {O'Connell}, \citenamefont {Donnelly},\ and\ \citenamefont {Walecka}}]{PhysRevC.6.719}%
  \BibitemOpen
  \bibfield  {author} {\bibinfo {author} {\bibfnamefont {J.~S.}\ \bibnamefont {O'Connell}}, \bibinfo {author} {\bibfnamefont {T.~W.}\ \bibnamefont {Donnelly}},\ and\ \bibinfo {author} {\bibfnamefont {J.~D.}\ \bibnamefont {Walecka}},\ }\bibfield  {title} {\bibinfo {title} {Semileptonic weak interactions with ${\mathrm{c}}^{12}$},\ }\href {https://doi.org/10.1103/PhysRevC.6.719} {\bibfield  {journal} {\bibinfo  {journal} {Phys. Rev. C}\ }\textbf {\bibinfo {volume} {6}},\ \bibinfo {pages} {719} (\bibinfo {year} {1972})}\BibitemShut {NoStop}%
\bibitem [{\citenamefont {de~Forest~Jr.}\ and\ \citenamefont {Walecka}(1966)}]{doi:10.1080/00018736600101254}%
  \BibitemOpen
  \bibfield  {author} {\bibinfo {author} {\bibfnamefont {T.}~\bibnamefont {de~Forest~Jr.}}\ and\ \bibinfo {author} {\bibfnamefont {J.}~\bibnamefont {Walecka}},\ }\bibfield  {title} {\bibinfo {title} {Electron scattering and nuclear structure},\ }\href {https://doi.org/10.1080/00018736600101254} {\bibfield  {journal} {\bibinfo  {journal} {Advances in Physics}\ }\textbf {\bibinfo {volume} {15}},\ \bibinfo {pages} {1} (\bibinfo {year} {1966})},\ \Eprint {https://arxiv.org/abs/https://doi.org/10.1080/00018736600101254} {https://doi.org/10.1080/00018736600101254} \BibitemShut {NoStop}%
\bibitem [{\citenamefont {Fitzpatrick}\ \emph {et~al.}(2013)\citenamefont {Fitzpatrick}, \citenamefont {Haxton}, \citenamefont {Katz}, \citenamefont {Lubbers},\ and\ \citenamefont {Xu}}]{Fitzpatrick:2012ix}%
  \BibitemOpen
  \bibfield  {author} {\bibinfo {author} {\bibfnamefont {A.~L.}\ \bibnamefont {Fitzpatrick}}, \bibinfo {author} {\bibfnamefont {W.}~\bibnamefont {Haxton}}, \bibinfo {author} {\bibfnamefont {E.}~\bibnamefont {Katz}}, \bibinfo {author} {\bibfnamefont {N.}~\bibnamefont {Lubbers}},\ and\ \bibinfo {author} {\bibfnamefont {Y.}~\bibnamefont {Xu}},\ }\bibfield  {title} {\bibinfo {title} {{The Effective Field Theory of Dark Matter Direct Detection}},\ }\href {https://doi.org/10.1088/1475-7516/2013/02/004} {\bibfield  {journal} {\bibinfo  {journal} {JCAP}\ }\textbf {\bibinfo {volume} {02}},\ \bibinfo {pages} {004}},\ \Eprint {https://arxiv.org/abs/1203.3542} {arXiv:1203.3542 [hep-ph]} \BibitemShut {NoStop}%
\bibitem [{\citenamefont {Donnelly}\ and\ \citenamefont {Haxton}(1979)}]{DONNELLY1979103}%
  \BibitemOpen
  \bibfield  {author} {\bibinfo {author} {\bibfnamefont {T.}~\bibnamefont {Donnelly}}\ and\ \bibinfo {author} {\bibfnamefont {W.}~\bibnamefont {Haxton}},\ }\bibfield  {title} {\bibinfo {title} {Multipole operators in semileptonic weak and electromagnetic interactions with nuclei: Harmonic oscillator single-particle matrix elements},\ }\href {https://doi.org/https://doi.org/10.1016/0092-640X(79)90003-2} {\bibfield  {journal} {\bibinfo  {journal} {Atomic Data and Nuclear Data Tables}\ }\textbf {\bibinfo {volume} {23}},\ \bibinfo {pages} {103} (\bibinfo {year} {1979})}\BibitemShut {NoStop}%
\bibitem [{\citenamefont {Brown}\ and\ \citenamefont {Rae}(2014)}]{Brown2014TheNuShellXMSU}%
  \BibitemOpen
  \bibfield  {author} {\bibinfo {author} {\bibfnamefont {B.~A.}\ \bibnamefont {Brown}}\ and\ \bibinfo {author} {\bibfnamefont {W.~D.}\ \bibnamefont {Rae}},\ }\bibfield  {title} {\bibinfo {title} {{The Shell-Model Code NuShellX@MSU}},\ }\href {https://doi.org/10.1016/j.nds.2014.07.022} {\bibfield  {journal} {\bibinfo  {journal} {Nuclear Data Sheets}\ }\textbf {\bibinfo {volume} {120}},\ \bibinfo {pages} {115} (\bibinfo {year} {2014})}\BibitemShut {NoStop}%
\bibitem [{\citenamefont {Abdel~Khaleq}\ \emph {et~al.}(2024)\citenamefont {Abdel~Khaleq}, \citenamefont {Busoni}, \citenamefont {Simenel},\ and\ \citenamefont {Stuchbery}}]{AbdelKhaleq:2023ipt}%
  \BibitemOpen
  \bibfield  {author} {\bibinfo {author} {\bibfnamefont {R.}~\bibnamefont {Abdel~Khaleq}}, \bibinfo {author} {\bibfnamefont {G.}~\bibnamefont {Busoni}}, \bibinfo {author} {\bibfnamefont {C.}~\bibnamefont {Simenel}},\ and\ \bibinfo {author} {\bibfnamefont {A.~E.}\ \bibnamefont {Stuchbery}},\ }\bibfield  {title} {\bibinfo {title} {{Impact of shell model interactions on nuclear responses to WIMP elastic scattering}},\ }\href {https://doi.org/10.1103/PhysRevD.109.075036} {\bibfield  {journal} {\bibinfo  {journal} {Phys. Rev. D}\ }\textbf {\bibinfo {volume} {109}},\ \bibinfo {pages} {075036} (\bibinfo {year} {2024})},\ \Eprint {https://arxiv.org/abs/2311.15764} {arXiv:2311.15764 [hep-ph]} \BibitemShut {NoStop}%
\bibitem [{\citenamefont {Anand}\ \emph {et~al.}(2014)\citenamefont {Anand}, \citenamefont {Fitzpatrick},\ and\ \citenamefont {Haxton}}]{Anand:2013yka}%
  \BibitemOpen
  \bibfield  {author} {\bibinfo {author} {\bibfnamefont {N.}~\bibnamefont {Anand}}, \bibinfo {author} {\bibfnamefont {A.~L.}\ \bibnamefont {Fitzpatrick}},\ and\ \bibinfo {author} {\bibfnamefont {W.~C.}\ \bibnamefont {Haxton}},\ }\bibfield  {title} {\bibinfo {title} {{Weakly interacting massive particle-nucleus elastic scattering response}},\ }\href {https://doi.org/10.1103/PhysRevC.89.065501} {\bibfield  {journal} {\bibinfo  {journal} {Phys. Rev. C}\ }\textbf {\bibinfo {volume} {89}},\ \bibinfo {pages} {065501} (\bibinfo {year} {2014})},\ \Eprint {https://arxiv.org/abs/1308.6288} {arXiv:1308.6288 [hep-ph]} \BibitemShut {NoStop}%
\bibitem [{\citenamefont {Akimov}\ \emph {et~al.}(2022)\citenamefont {Akimov} \emph {et~al.}}]{COHERENT:2021xmm}%
  \BibitemOpen
  \bibfield  {author} {\bibinfo {author} {\bibfnamefont {D.}~\bibnamefont {Akimov}} \emph {et~al.} (\bibinfo {collaboration} {COHERENT}),\ }\bibfield  {title} {\bibinfo {title} {{Measurement of the Coherent Elastic Neutrino-Nucleus Scattering Cross Section on CsI by COHERENT}},\ }\href {https://doi.org/10.1103/PhysRevLett.129.081801} {\bibfield  {journal} {\bibinfo  {journal} {Phys. Rev. Lett.}\ }\textbf {\bibinfo {volume} {129}},\ \bibinfo {pages} {081801} (\bibinfo {year} {2022})},\ \Eprint {https://arxiv.org/abs/2110.07730} {arXiv:2110.07730 [hep-ex]} \BibitemShut {NoStop}%
\bibitem [{\citenamefont {Ackermann}\ \emph {et~al.}(2024)\citenamefont {Ackermann} \emph {et~al.}}]{Ackermann:2024kxo}%
  \BibitemOpen
  \bibfield  {author} {\bibinfo {author} {\bibfnamefont {N.}~\bibnamefont {Ackermann}} \emph {et~al.},\ }\bibfield  {title} {\bibinfo {title} {{Final CONUS results on coherent elastic neutrino nucleus scattering at the Brokdorf reactor}},\ }\href@noop {} {\  (\bibinfo {year} {2024})},\ \Eprint {https://arxiv.org/abs/2401.07684} {arXiv:2401.07684 [hep-ex]} \BibitemShut {NoStop}%
\bibitem [{\citenamefont {Augier}\ \emph {et~al.}(2023)\citenamefont {Augier} \emph {et~al.}}]{Ricochet:2023yek}%
  \BibitemOpen
  \bibfield  {author} {\bibinfo {author} {\bibfnamefont {C.}~\bibnamefont {Augier}} \emph {et~al.} (\bibinfo {collaboration} {Ricochet}),\ }\bibfield  {title} {\bibinfo {title} {{Results from a prototype TES detector for the Ricochet experiment}},\ }\href {https://doi.org/10.1016/j.nima.2023.168765} {\bibfield  {journal} {\bibinfo  {journal} {Nucl. Instrum. Meth. A}\ }\textbf {\bibinfo {volume} {1057}},\ \bibinfo {pages} {168765} (\bibinfo {year} {2023})},\ \Eprint {https://arxiv.org/abs/2304.14926} {arXiv:2304.14926 [physics.ins-det]} \BibitemShut {NoStop}%
\bibitem [{\citenamefont {Angloher}\ \emph {et~al.}(2019)\citenamefont {Angloher} \emph {et~al.}}]{NUCLEUS:2019igx}%
  \BibitemOpen
  \bibfield  {author} {\bibinfo {author} {\bibfnamefont {G.}~\bibnamefont {Angloher}} \emph {et~al.} (\bibinfo {collaboration} {NUCLEUS}),\ }\bibfield  {title} {\bibinfo {title} {{Exploring $\hbox {CE}\nu \hbox {NS}$ with NUCLEUS at the Chooz nuclear power plant}},\ }\href {https://doi.org/10.1140/epjc/s10052-019-7454-4} {\bibfield  {journal} {\bibinfo  {journal} {Eur. Phys. J. C}\ }\textbf {\bibinfo {volume} {79}},\ \bibinfo {pages} {1018} (\bibinfo {year} {2019})},\ \Eprint {https://arxiv.org/abs/1905.10258} {arXiv:1905.10258 [physics.ins-det]} \BibitemShut {NoStop}%
\bibitem [{\citenamefont {Akimov}\ \emph {et~al.}(2023)\citenamefont {Akimov} \emph {et~al.}}]{COHERENT:dataCsI:2023aln}%
  \BibitemOpen
  \bibfield  {author} {\bibinfo {author} {\bibfnamefont {D.}~\bibnamefont {Akimov}} \emph {et~al.} (\bibinfo {collaboration} {COHERENT}),\ }\bibfield  {title} {\bibinfo {title} {{COHERENT Collaboration data release from the measurements of CsI[Na] response to nuclear recoils}},\ }\href@noop {} {\  (\bibinfo {year} {2023})},\ \Eprint {https://arxiv.org/abs/2307.10208} {arXiv:2307.10208 [physics.ins-det]} \BibitemShut {NoStop}%
\bibitem [{\citenamefont {Akimov}\ \emph {et~al.}(2020)\citenamefont {Akimov} \emph {et~al.}}]{COHERENT:dataAr:2020ybo}%
  \BibitemOpen
  \bibfield  {author} {\bibinfo {author} {\bibfnamefont {D.}~\bibnamefont {Akimov}} \emph {et~al.} (\bibinfo {collaboration} {COHERENT}),\ }\bibfield  {title} {\bibinfo {title} {{COHERENT Collaboration data release from the first detection of coherent elastic neutrino-nucleus scattering on argon}}\ }\href {https://doi.org/10.5281/zenodo.3903810} {10.5281/zenodo.3903810} (\bibinfo {year} {2020}),\ \Eprint {https://arxiv.org/abs/2006.12659} {arXiv:2006.12659 [nucl-ex]} \BibitemShut {NoStop}%
\bibitem [{\citenamefont {Nummela}\ \emph {et~al.}(2001)\citenamefont {Nummela}, \citenamefont {Baumann}, \citenamefont {Caurier}, \citenamefont {Dessagne}, \citenamefont {Jokinen}, \citenamefont {Knipper}, \citenamefont {Le~Scornet}, \citenamefont {Mieh{\'{e}}}, \citenamefont {Nowacki}, \citenamefont {Oinonen}, \citenamefont {Radivojevic}, \citenamefont {Ramdhane}, \citenamefont {Walter},\ and\ \citenamefont {{\"{A}}yst{\"{o}}}}]{Nummela2001SpectroscopyStates}%
  \BibitemOpen
  \bibfield  {author} {\bibinfo {author} {\bibfnamefont {S.}~\bibnamefont {Nummela}}, \bibinfo {author} {\bibfnamefont {P.}~\bibnamefont {Baumann}}, \bibinfo {author} {\bibfnamefont {E.}~\bibnamefont {Caurier}}, \bibinfo {author} {\bibfnamefont {P.}~\bibnamefont {Dessagne}}, \bibinfo {author} {\bibfnamefont {A.}~\bibnamefont {Jokinen}}, \bibinfo {author} {\bibfnamefont {A.}~\bibnamefont {Knipper}}, \bibinfo {author} {\bibfnamefont {G.}~\bibnamefont {Le~Scornet}}, \bibinfo {author} {\bibfnamefont {C.}~\bibnamefont {Mieh{\'{e}}}}, \bibinfo {author} {\bibfnamefont {F.}~\bibnamefont {Nowacki}}, \bibinfo {author} {\bibfnamefont {M.}~\bibnamefont {Oinonen}}, \bibinfo {author} {\bibfnamefont {Z.}~\bibnamefont {Radivojevic}}, \bibinfo {author} {\bibfnamefont {M.}~\bibnamefont {Ramdhane}}, \bibinfo {author} {\bibfnamefont {G.}~\bibnamefont {Walter}},\ and\ \bibinfo {author} {\bibfnamefont {J.}~\bibnamefont {{\"{A}}yst{\"{o}}}},\ }\bibfield  {title} {\bibinfo {title} {{Spectroscopy of 34,35Si by {$\beta$} decay: sd-fp
  shell gap and single-particle states}},\ }\bibfield  {journal} {\bibinfo  {journal} {Physical Review C - Nuclear Physics}\ }\href {https://doi.org/10.1103/PhysRevC.63.044316} {10.1103/PhysRevC.63.044316} (\bibinfo {year} {2001})\BibitemShut {NoStop}%
\bibitem [{\citenamefont {Nowacki}\ and\ \citenamefont {Poves}(2009)}]{Nowacki2009NewSpace}%
  \BibitemOpen
  \bibfield  {author} {\bibinfo {author} {\bibfnamefont {F.}~\bibnamefont {Nowacki}}\ and\ \bibinfo {author} {\bibfnamefont {A.}~\bibnamefont {Poves}},\ }\bibfield  {title} {\bibinfo {title} {{New effective interaction for 0 {$\omega$} shell-model calculations in the sd-pf valence space}},\ }\bibfield  {journal} {\bibinfo  {journal} {Physical Review C - Nuclear Physics}\ }\href {https://doi.org/10.1103/PhysRevC.79.014310} {10.1103/PhysRevC.79.014310} (\bibinfo {year} {2009})\BibitemShut {NoStop}%
\bibitem [{\citenamefont {Kaneko}\ \emph {et~al.}(2011)\citenamefont {Kaneko}, \citenamefont {Sun}, \citenamefont {Mizusaki},\ and\ \citenamefont {Hasegawa}}]{Kaneko2011Shell-modelNuclei}%
  \BibitemOpen
  \bibfield  {author} {\bibinfo {author} {\bibfnamefont {K.}~\bibnamefont {Kaneko}}, \bibinfo {author} {\bibfnamefont {Y.}~\bibnamefont {Sun}}, \bibinfo {author} {\bibfnamefont {T.}~\bibnamefont {Mizusaki}},\ and\ \bibinfo {author} {\bibfnamefont {M.}~\bibnamefont {Hasegawa}},\ }\bibfield  {title} {\bibinfo {title} {{Shell-model study for neutron-rich sd-shell nuclei}},\ }\bibfield  {journal} {\bibinfo  {journal} {Physical Review C - Nuclear Physics}\ }\href {https://doi.org/10.1103/PhysRevC.83.014320} {10.1103/PhysRevC.83.014320} (\bibinfo {year} {2011})\BibitemShut {NoStop}%
\bibitem [{\citenamefont {Utsuno}\ \emph {et~al.}(2012)\citenamefont {Utsuno}, \citenamefont {Otsuka}, \citenamefont {Brown}, \citenamefont {Honma}, \citenamefont {Mizusaki},\ and\ \citenamefont {Shimizu}}]{Utsuno2012ShapeEffect}%
  \BibitemOpen
  \bibfield  {author} {\bibinfo {author} {\bibfnamefont {Y.}~\bibnamefont {Utsuno}}, \bibinfo {author} {\bibfnamefont {T.}~\bibnamefont {Otsuka}}, \bibinfo {author} {\bibfnamefont {B.~A.}\ \bibnamefont {Brown}}, \bibinfo {author} {\bibfnamefont {M.}~\bibnamefont {Honma}}, \bibinfo {author} {\bibfnamefont {T.}~\bibnamefont {Mizusaki}},\ and\ \bibinfo {author} {\bibfnamefont {N.}~\bibnamefont {Shimizu}},\ }\bibfield  {title} {\bibinfo {title} {{Shape transitions in exotic Si and S isotopes and tensor-force-driven Jahn-Teller effect}},\ }\bibfield  {journal} {\bibinfo  {journal} {Physical Review C}\ }\href {https://doi.org/10.1103/physrevc.86.051301} {10.1103/physrevc.86.051301} (\bibinfo {year} {2012})\BibitemShut {NoStop}%
\bibitem [{\citenamefont {Stuchbery}\ and\ \citenamefont {Wood}(2022)}]{physics4030048}%
  \BibitemOpen
  \bibfield  {author} {\bibinfo {author} {\bibfnamefont {A.~E.}\ \bibnamefont {Stuchbery}}\ and\ \bibinfo {author} {\bibfnamefont {J.~L.}\ \bibnamefont {Wood}},\ }\bibfield  {title} {\bibinfo {title} {To shell model, or not to shell model, that is the question},\ }\href {https://doi.org/10.3390/physics4030048} {\bibfield  {journal} {\bibinfo  {journal} {Physics}\ }\textbf {\bibinfo {volume} {4}},\ \bibinfo {pages} {697} (\bibinfo {year} {2022})}\BibitemShut {NoStop}%
\bibitem [{\citenamefont {Chen}(2017)}]{Chen:2017ngq}%
  \BibitemOpen
  \bibfield  {author} {\bibinfo {author} {\bibfnamefont {J.}~\bibnamefont {Chen}},\ }\bibfield  {title} {\bibinfo {title} {{Nuclear Data Sheets for A=40}},\ }\href {https://doi.org/10.1016/j.nds.2017.02.001} {\bibfield  {journal} {\bibinfo  {journal} {Nuclear Data Sheets}\ }\textbf {\bibinfo {volume} {140}},\ \bibinfo {pages} {1} (\bibinfo {year} {2017})}\BibitemShut {NoStop}%
\bibitem [{\citenamefont {{Stone}}(2020)}]{INDC0816}%
  \BibitemOpen
  \bibfield  {author} {\bibinfo {author} {\bibfnamefont {N.~J.}\ \bibnamefont {{Stone}}},\ }\href@noop {} {\emph {\bibinfo {title} {Table of Recommended Nuclear Magnetic Dipole Moments - Part II, Short-lived States}}},\ \bibinfo {type} {Tech. Rep.}\ \bibinfo {number} {INDC(NDS)-0816}\ (\bibinfo {year} {2020})\BibitemShut {NoStop}%
\bibitem [{\citenamefont {Stone}(2021)}]{indc.nds.0833}%
  \BibitemOpen
  \bibfield  {author} {\bibinfo {author} {\bibfnamefont {N.~J.}\ \bibnamefont {Stone}},\ }\href@noop {} {\emph {\bibinfo {title} {Table of Nuclear Electric Quadrupole Moments}}},\ \bibinfo {type} {Tech. Rep.}\ \bibinfo {number} {INDC(NDS)-0833}\ (\bibinfo {year} {2021})\BibitemShut {NoStop}%
\bibitem [{\citenamefont {Honma}\ \emph {et~al.}(2009)\citenamefont {Honma}, \citenamefont {Otsuka}, \citenamefont {Mizusaki},\ and\ \citenamefont {Hjorth-Jensen}}]{Honma2009NewNuclei}%
  \BibitemOpen
  \bibfield  {author} {\bibinfo {author} {\bibfnamefont {M.}~\bibnamefont {Honma}}, \bibinfo {author} {\bibfnamefont {T.}~\bibnamefont {Otsuka}}, \bibinfo {author} {\bibfnamefont {T.}~\bibnamefont {Mizusaki}},\ and\ \bibinfo {author} {\bibfnamefont {M.}~\bibnamefont {Hjorth-Jensen}},\ }\bibfield  {title} {\bibinfo {title} {{New effective interaction for f5pg9-shell nuclei}},\ }\bibfield  {journal} {\bibinfo  {journal} {Physical Review C - Nuclear Physics}\ }\href {https://doi.org/10.1103/PhysRevC.80.064323} {10.1103/PhysRevC.80.064323} (\bibinfo {year} {2009})\BibitemShut {NoStop}%
\bibitem [{\citenamefont {Mukhopadhyay}\ \emph {et~al.}(2017)\citenamefont {Mukhopadhyay}, \citenamefont {Crider}, \citenamefont {Brown}, \citenamefont {Ashley}, \citenamefont {Chakraborty}, \citenamefont {Kumar}, \citenamefont {McEllistrem}, \citenamefont {Peters}, \citenamefont {Prados-Est{\'{e}}vez},\ and\ \citenamefont {Yates}}]{Mukhopadhyay2017NuclearCalculations}%
  \BibitemOpen
  \bibfield  {author} {\bibinfo {author} {\bibfnamefont {S.}~\bibnamefont {Mukhopadhyay}}, \bibinfo {author} {\bibfnamefont {B.~P.}\ \bibnamefont {Crider}}, \bibinfo {author} {\bibfnamefont {B.~A.}\ \bibnamefont {Brown}}, \bibinfo {author} {\bibfnamefont {S.~F.}\ \bibnamefont {Ashley}}, \bibinfo {author} {\bibfnamefont {A.}~\bibnamefont {Chakraborty}}, \bibinfo {author} {\bibfnamefont {A.}~\bibnamefont {Kumar}}, \bibinfo {author} {\bibfnamefont {M.~T.}\ \bibnamefont {McEllistrem}}, \bibinfo {author} {\bibfnamefont {E.~E.}\ \bibnamefont {Peters}}, \bibinfo {author} {\bibfnamefont {F.~M.}\ \bibnamefont {Prados-Est{\'{e}}vez}},\ and\ \bibinfo {author} {\bibfnamefont {S.~W.}\ \bibnamefont {Yates}},\ }\bibfield  {title} {\bibinfo {title} {{Nuclear structure of Ge 76 from inelastic neutron scattering measurements and shell model calculations}},\ }\bibfield  {journal} {\bibinfo  {journal} {Physical Review C}\ }\href {https://doi.org/10.1103/PhysRevC.95.014327} {10.1103/PhysRevC.95.014327} (\bibinfo {year}
  {2017})\BibitemShut {NoStop}%
\bibitem [{\citenamefont {Men{\'{e}}ndez}\ \emph {et~al.}(2009)\citenamefont {Men{\'{e}}ndez}, \citenamefont {Poves}, \citenamefont {Caurier},\ and\ \citenamefont {Nowacki}}]{Menendez2009DisassemblingDecay}%
  \BibitemOpen
  \bibfield  {author} {\bibinfo {author} {\bibfnamefont {J.}~\bibnamefont {Men{\'{e}}ndez}}, \bibinfo {author} {\bibfnamefont {A.}~\bibnamefont {Poves}}, \bibinfo {author} {\bibfnamefont {E.}~\bibnamefont {Caurier}},\ and\ \bibinfo {author} {\bibfnamefont {F.}~\bibnamefont {Nowacki}},\ }\bibfield  {title} {\bibinfo {title} {{Disassembling the nuclear matrix elements of the neutrinoless {$\beta$}{$\beta$} decay}},\ }\bibfield  {journal} {\bibinfo  {journal} {Nuclear Physics A}\ }\href {https://doi.org/10.1016/j.nuclphysa.2008.12.005} {10.1016/j.nuclphysa.2008.12.005} (\bibinfo {year} {2009})\BibitemShut {NoStop}%
\bibitem [{\citenamefont {McCormick}\ \emph {et~al.}(2019)\citenamefont {McCormick}, \citenamefont {Stuchbery}, \citenamefont {Brown}, \citenamefont {Georgiev}, \citenamefont {Coombes}, \citenamefont {Gray}, \citenamefont {Gerathy}, \citenamefont {Lane}, \citenamefont {Kib\'edi}, \citenamefont {Mitchell}, \citenamefont {Reed}, \citenamefont {Akber}, \citenamefont {Bignell}, \citenamefont {Dowie}, \citenamefont {Eriksen}, \citenamefont {Hota}, \citenamefont {Palalani},\ and\ \citenamefont {Tornyi}}]{McCormick2019}%
  \BibitemOpen
  \bibfield  {author} {\bibinfo {author} {\bibfnamefont {B.~P.}\ \bibnamefont {McCormick}}, \bibinfo {author} {\bibfnamefont {A.~E.}\ \bibnamefont {Stuchbery}}, \bibinfo {author} {\bibfnamefont {B.~A.}\ \bibnamefont {Brown}}, \bibinfo {author} {\bibfnamefont {G.}~\bibnamefont {Georgiev}}, \bibinfo {author} {\bibfnamefont {B.~J.}\ \bibnamefont {Coombes}}, \bibinfo {author} {\bibfnamefont {T.~J.}\ \bibnamefont {Gray}}, \bibinfo {author} {\bibfnamefont {M.~S.~M.}\ \bibnamefont {Gerathy}}, \bibinfo {author} {\bibfnamefont {G.~J.}\ \bibnamefont {Lane}}, \bibinfo {author} {\bibfnamefont {T.}~\bibnamefont {Kib\'edi}}, \bibinfo {author} {\bibfnamefont {A.~J.}\ \bibnamefont {Mitchell}}, \bibinfo {author} {\bibfnamefont {M.~W.}\ \bibnamefont {Reed}}, \bibinfo {author} {\bibfnamefont {A.}~\bibnamefont {Akber}}, \bibinfo {author} {\bibfnamefont {L.~J.}\ \bibnamefont {Bignell}}, \bibinfo {author} {\bibfnamefont {J.~T.~H.}\ \bibnamefont {Dowie}}, \bibinfo {author} {\bibfnamefont {T.~K.}\ \bibnamefont {Eriksen}}, \bibinfo
  {author} {\bibfnamefont {S.}~\bibnamefont {Hota}}, \bibinfo {author} {\bibfnamefont {N.}~\bibnamefont {Palalani}},\ and\ \bibinfo {author} {\bibfnamefont {T.}~\bibnamefont {Tornyi}},\ }\bibfield  {title} {\bibinfo {title} {First-excited state $g$ factors in the stable, even ge and se isotopes},\ }\href {https://doi.org/10.1103/PhysRevC.100.044317} {\bibfield  {journal} {\bibinfo  {journal} {Phys. Rev. C}\ }\textbf {\bibinfo {volume} {100}},\ \bibinfo {pages} {044317} (\bibinfo {year} {2019})}\BibitemShut {NoStop}%
\bibitem [{\citenamefont {G\"urdal}\ \emph {et~al.}(2013)\citenamefont {G\"urdal} \emph {et~al.}}]{Gurdal:2013oma}%
  \BibitemOpen
  \bibfield  {author} {\bibinfo {author} {\bibfnamefont {G.}~\bibnamefont {G\"urdal}} \emph {et~al.},\ }\bibfield  {title} {\bibinfo {title} {{Measurements of $g(4^+_1,2^+_2)$ in $^{70,72,74,76}Ge$: Systematics of low-lying structures in $30\le Z \le 40$ and $30\le N \le 50$ nuclei}},\ }\href {https://doi.org/10.1103/PhysRevC.88.014301} {\bibfield  {journal} {\bibinfo  {journal} {Phys. Rev. C}\ }\textbf {\bibinfo {volume} {88}},\ \bibinfo {pages} {014301} (\bibinfo {year} {2013})}\BibitemShut {NoStop}%
\bibitem [{\citenamefont {G\"urdal}\ and\ \citenamefont {McCutchan}(2016)}]{Gurdal:2016opi}%
  \BibitemOpen
  \bibfield  {author} {\bibinfo {author} {\bibfnamefont {G.}~\bibnamefont {G\"urdal}}\ and\ \bibinfo {author} {\bibfnamefont {E.~A.}\ \bibnamefont {McCutchan}},\ }\bibfield  {title} {\bibinfo {title} {{Nuclear Data Sheets for A = 70}},\ }\href {https://doi.org/10.1016/j.nds.2016.08.001} {\bibfield  {journal} {\bibinfo  {journal} {Nucl. Data Sheets}\ }\textbf {\bibinfo {volume} {136}},\ \bibinfo {pages} {1} (\bibinfo {year} {2016})}\BibitemShut {NoStop}%
\bibitem [{\citenamefont {Abriola}\ and\ \citenamefont {Sonzogni}(2010)}]{Abriola:2010ome}%
  \BibitemOpen
  \bibfield  {author} {\bibinfo {author} {\bibfnamefont {D.}~\bibnamefont {Abriola}}\ and\ \bibinfo {author} {\bibfnamefont {A.~A.}\ \bibnamefont {Sonzogni}},\ }\bibfield  {title} {\bibinfo {title} {{Nuclear Data Sheets for A = 72}},\ }\href {https://doi.org/10.1016/j.nds.2009.12.001} {\bibfield  {journal} {\bibinfo  {journal} {Nucl. Data Sheets}\ }\textbf {\bibinfo {volume} {111}},\ \bibinfo {pages} {1} (\bibinfo {year} {2010})}\BibitemShut {NoStop}%
\bibitem [{\citenamefont {Singh}\ and\ \citenamefont {Farhan}(2006)}]{Singh:2006spf}%
  \BibitemOpen
  \bibfield  {author} {\bibinfo {author} {\bibfnamefont {B.}~\bibnamefont {Singh}}\ and\ \bibinfo {author} {\bibfnamefont {A.~R.}\ \bibnamefont {Farhan}},\ }\bibfield  {title} {\bibinfo {title} {{Nuclear Data Sheets for A = 74}},\ }\href {https://doi.org/10.1016/j.nds.2006.05.006} {\bibfield  {journal} {\bibinfo  {journal} {Nucl. Data Sheets}\ }\textbf {\bibinfo {volume} {107}},\ \bibinfo {pages} {1923} (\bibinfo {year} {2006})}\BibitemShut {NoStop}%
\bibitem [{\citenamefont {Singh}\ \emph {et~al.}(2024)\citenamefont {Singh}, \citenamefont {Chen},\ and\ \citenamefont {Farhan}}]{Singh:2024adz}%
  \BibitemOpen
  \bibfield  {author} {\bibinfo {author} {\bibfnamefont {B.}~\bibnamefont {Singh}}, \bibinfo {author} {\bibfnamefont {J.}~\bibnamefont {Chen}},\ and\ \bibinfo {author} {\bibfnamefont {A.~R.}\ \bibnamefont {Farhan}},\ }\bibfield  {title} {\bibinfo {title} {{Nuclear Structure and Decay Data for A=76 Isobars}},\ }\href {https://doi.org/10.1016/j.nds.2024.02.002} {\bibfield  {journal} {\bibinfo  {journal} {Nucl. Data Sheets}\ }\textbf {\bibinfo {volume} {194}},\ \bibinfo {pages} {3} (\bibinfo {year} {2024})}\BibitemShut {NoStop}%
\bibitem [{\citenamefont {Brown}\ \emph {et~al.}(2005)\citenamefont {Brown}, \citenamefont {Stone}, \citenamefont {Stone}, \citenamefont {Towner},\ and\ \citenamefont {Hjorth-Jensen}}]{Brown2005Magnetic132Sn}%
  \BibitemOpen
  \bibfield  {author} {\bibinfo {author} {\bibfnamefont {B.~A.}\ \bibnamefont {Brown}}, \bibinfo {author} {\bibfnamefont {N.~J.}\ \bibnamefont {Stone}}, \bibinfo {author} {\bibfnamefont {J.~R.}\ \bibnamefont {Stone}}, \bibinfo {author} {\bibfnamefont {I.~S.}\ \bibnamefont {Towner}},\ and\ \bibinfo {author} {\bibfnamefont {M.}~\bibnamefont {Hjorth-Jensen}},\ }\bibfield  {title} {\bibinfo {title} {{Magnetic moments of the 21+ states around 132Sn}},\ }\href {https://doi.org/10.1103/PhysRevC.71.044317} {\bibfield  {journal} {\bibinfo  {journal} {Physical Review C}\ }\textbf {\bibinfo {volume} {71}},\ \bibinfo {pages} {044317} (\bibinfo {year} {2005})}\BibitemShut {NoStop}%
\bibitem [{\citenamefont {Baldridge}\ and\ \citenamefont {Dalton}(1978)}]{Baldridge1978Shell-modelCases}%
  \BibitemOpen
  \bibfield  {author} {\bibinfo {author} {\bibfnamefont {W.~J.}\ \bibnamefont {Baldridge}}\ and\ \bibinfo {author} {\bibfnamefont {B.~J.}\ \bibnamefont {Dalton}},\ }\bibfield  {title} {\bibinfo {title} {{Shell-model studies for the Sn132 region. II. Exact and statistical results for multi-proton cases}},\ }\href {https://doi.org/10.1103/PhysRevC.18.539} {\bibfield  {journal} {\bibinfo  {journal} {Physical Review C}\ }\textbf {\bibinfo {volume} {18}},\ \bibinfo {pages} {539} (\bibinfo {year} {1978})}\BibitemShut {NoStop}%
\bibitem [{\citenamefont {{Stone}}(2019)}]{NSR2019STZV}%
  \BibitemOpen
  \bibfield  {author} {\bibinfo {author} {\bibfnamefont {N.~J.}\ \bibnamefont {{Stone}}},\ }\href@noop {} {\emph {\bibinfo {title} {Table of Recommended Nuclear Magnetic Dipole Moments: Part I - Long-lived States}}},\ \bibinfo {type} {Tech. Rep.}\ \bibinfo {number} {INDC(NDS)-0794}\ (\bibinfo {year} {2019})\BibitemShut {NoStop}%
\bibitem [{\citenamefont {Campbell}\ \emph {et~al.}(1977)\citenamefont {Campbell}, \citenamefont {Montet},\ and\ \citenamefont {Perlow}}]{Campbell1977}%
  \BibitemOpen
  \bibfield  {author} {\bibinfo {author} {\bibfnamefont {L.~E.}\ \bibnamefont {Campbell}}, \bibinfo {author} {\bibfnamefont {G.~L.}\ \bibnamefont {Montet}},\ and\ \bibinfo {author} {\bibfnamefont {G.~J.}\ \bibnamefont {Perlow}},\ }\bibfield  {title} {\bibinfo {title} {Anisotropy of the debye-waller factor in cesium-graphite intercalation compounds by m\"ossbauer spectroscopy, and the quadrupole moment of the 81-kev state in $^{133}\mathrm{Cs}$},\ }\href {https://doi.org/10.1103/PhysRevB.15.3318} {\bibfield  {journal} {\bibinfo  {journal} {Phys. Rev. B}\ }\textbf {\bibinfo {volume} {15}},\ \bibinfo {pages} {3318} (\bibinfo {year} {1977})}\BibitemShut {NoStop}%
\bibitem [{\citenamefont {Khazov}\ \emph {et~al.}(2011)\citenamefont {Khazov}, \citenamefont {Rodionov},\ and\ \citenamefont {Kondev}}]{Khazov:2011sug}%
  \BibitemOpen
  \bibfield  {author} {\bibinfo {author} {\bibfnamefont {Y.}~\bibnamefont {Khazov}}, \bibinfo {author} {\bibfnamefont {A.}~\bibnamefont {Rodionov}},\ and\ \bibinfo {author} {\bibfnamefont {F.~G.}\ \bibnamefont {Kondev}},\ }\bibfield  {title} {\bibinfo {title} {{Nuclear Data Sheets for A = 133}},\ }\href {https://doi.org/10.1016/j.nds.2011.03.001} {\bibfield  {journal} {\bibinfo  {journal} {Nucl. Data Sheets}\ }\textbf {\bibinfo {volume} {112}},\ \bibinfo {pages} {855} (\bibinfo {year} {2011})}\BibitemShut {NoStop}%
\end{thebibliography}%

\end{document}